\documentclass[12pt,a4paper]{amsart}
\usepackage{latexsym,amssymb,amsthm,amsmath,amscd}
\usepackage{graphicx,graphics}
\usepackage[all]{xy}

\usepackage{xypic}
\usepackage{mathrsfs}

\title[Dimensional reduction over the quantum sphere]
{Dimensional reduction over the quantum sphere \\[5pt] and
  non-abelian $\mbf q$-vortices} 
\date{ \ v2: October 2011 \hfill HWM--10--7 \ , \ EMPG--10--06 \ }

\author{Giovanni Landi }
\address{ Dipartimento di Matematica e Informatica,
  Universit\`{a} di Trieste, 
Via A.~Valerio~12/1, I-34127 Trieste, Italy, and INFN, Sezione di
Trieste, Trieste, Italy} 
\email{landi@univ.trieste.it}
\author{Richard J. Szabo}
\address{ Department of Mathematics, Heriot-Watt University,
  Colin Maclaurin Building, Riccarton, Edinburgh EH14 4AS, U.K., and
  Maxwell Institute for Mathematical Sciences, Edinburgh, U.K.}
\email{R.J.Szabo@ma.hw.ac.uk}

\numberwithin{equation}{section}

\newtheorem{theo}[equation]{Theorem}
\newtheorem{lemm}[equation]{Lemma}
\newtheorem{prop}[equation]{Proposition}

\newtheorem{coro}[equation]{Corollary}

\newcommand{\nn}{\nonumber}
\newcommand{\ce}{\mathcal{E}}
\newcommand{\ceX}{{\,\underline{\mathcal{E}}\,}}
\newcommand{\man}{{\,\underline{M}\,}}
\newcommand{\nablaCP}{\widehat\nabla}
\newcommand{\nablaX}{\,\underline{\nabla}\,}
\newcommand{\DeltaX}{\,\underline{\Delta}\,}
\newcommand{\betaX}{\,\underline{\beta}\,}
\newcommand{\omegaX}{\,\underline{\omega}\,}
\newcommand{\dd}{{\rm d}}
\newcommand{\ddX}{\,\underline{\dd}\,}
\newcommand{\cob}{{\rm b}}
\newcommand{\ca}{\mathcal{A}}
\newcommand{\cc}{\mathcal{C}}

\newcommand{\cl}{\mathcal{L}}

\newcommand{\cf}{\mathcal{F}}
\newcommand{\scrc}{\mathscr{C}}
\newcommand{\scru}{\mathscr{U}}
\newcommand{\scrr}{\mathscr{R}}

\newcommand{\asf}{\,{\sf a}}
\newcommand{\fsf}{\,{\sf f}}
\newcommand{\cfX}{\,\underline{\mathcal{F}}\,}
\newcommand{\hX}{\,\underline{h}\,}
\newcommand{\AX}{{\,\underline{A}\,}}
\newcommand{\uX}{\,\underline{u}\,}
\newcommand{\fX}{\,\underline{f}\,}
\newcommand{\starX}{\,\underline{\star}\,}

\newcommand{\slc}{{{\rm SL}(2,\IC)}}

\newcommand{\pq}{{\IC\mathrm{P}^1_{q}}}  
\newcommand{\Apq}{\ca(\IC\mathrm{P}^1_{q})}  
\newcommand{\cu}{\mathcal{U}}        
\newcommand{\SU}{\mathrm{SU}_q(2)}  
\newcommand{\ASU}{\ca(\mathrm{SU}_q(2))}  
\newcommand{\sq}{\mathrm{S}^2_{q}}  
\newcommand{\Asq}{\ca(\mathrm{S}^2_{q})}  
\newcommand{\su}{\cu_q(\mathrm{su}(2))}  

\newcommand{\cop}{\Delta}           
\newcommand{\co}[2]{#1_{(#2)}}      
\newcommand{\hs}[2]{\left\langle #1\,,\,#2\right\rangle}  

\newcommand{\lt}{\triangleright}    
\newcommand{\rt}{\triangleleft}
\newcommand{\IC}{{\mathbb C}} 
\newcommand{\IR}{{\mathbb R}} 
\newcommand{\IN}{{\mathbb N}} 
\newcommand{\IZ}{{\mathbb Z}} 
\DeclareMathOperator{\id}{id}       
\DeclareMathOperator{\Mat}{Mat}       
\DeclareMathOperator{\End}{End}       
\DeclareMathOperator{\Hom}{Hom}       
\DeclareMathOperator{\U}{U}       
\DeclareMathOperator{\tr}{tr}       
\newcommand{\qtr}{{{\rm tr}_q}}       
\DeclareMathOperator{\chern}{ch}

\DeclareMathOperator{\Aut}{Aut}       
\DeclareMathOperator{\YM}{YM}         
\DeclareMathOperator{\YMH}{YMH}       
\DeclareMathOperator{\Top}{Top}       
\DeclareMathOperator{\vol}{vol}       
\DeclareMathOperator{\rank}{rank}     
\DeclareMathOperator{\Lie}{Lie}

\newcommand{\abs}[1]{\left|#1\right|}

\newcommand{\figureheight}{8cm}
\newcommand{\putfig}[2]{\begin{figure}[htp]
        \special{isoscale c:/itex/texfig/#1.wmf, \the\hsize \figureheight}
        \vspace{\figureheight}
        \caption{#2}\label{fig:#1}
        \end{figure}}
\newcommand{\pictureheight}{4cm}
\newcommand{\putpicture}[2]{\begin{figure}[htp]
        \special{isoscale c:/itex/texfig/#1.wmf, \the\hsize \pictureheight}
        \vspace{\pictureheight}
        \caption{#2}\label{fig:#1}
        \end{figure}}
\newcommand{\mbf}[1]{{\boldsymbol {#1} }}

\newcommand{\beqa}{\begin{eqnarray}}
\newcommand{\eeqa}{\end{eqnarray}}
\newcommand{\beq}{\begin{equation}}
\newcommand{\eeq}{\end{equation}}

\newcommand{\dol}{\partial}
\newcommand{\dolb}{\overline{\dol}}

\newcommand{\K}{{\rm K}}

\newcommand{\mn}{\abs{n}}
\newcommand{\qP}{\mathfrak{p}}
\newcommand{\qpp}{\mathfrak{p}^{\left(n\right)}}

\newcommand{\bz}{B_{0}}
\newcommand{\bp}{B_{+}}
\newcommand{\bm}{B_{-}}
\def\={\ =\ }

\begin{document}
\maketitle

\thispagestyle{empty}

\maketitle

\begin{abstract}
We extend equivariant dimensional reduction techniques to the case of
quantum spaces which are the product of a K\"ahler manifold $M$ with
the quantum two-sphere. We work out the reduction of bundles which are
equivariant under the natural action of the quantum group $\SU$, and
also of invariant gauge connections on these bundles. The reduction of
Yang--Mills gauge theory on the product space leads to a
$q$-deformation of the usual quiver gauge theories on $M$. We
formulate generalized instanton equations on the quantum space and
show that they correspond to $q$-deformations of the usual holomorphic
quiver chain vortex equations on $M$. We study some topological
stability conditions for the existence of solutions to these
equations, and demonstrate that the corresponding vacuum moduli spaces
are generally better behaved than their undeformed counterparts, but
much more constrained by the $q$-deformation. We work out several explicit examples, including new examples of non-abelian vortices on Riemann surfaces, and $q$-deformations of instantons whose moduli spaces admit the standard hyper-K\"ahler quotient construction.
\end{abstract}

\tableofcontents

\section*{Introduction}

Let $M$ be a smooth manifold. In this paper we define and characterize
vector bundles over the quantum space $\man:=\pq\times M$ which are
equivariant under an action of the quantum group $\SU$. Here $\pq$ is
the quantum projective line which is defined in \S\ref{qdct}. The
vector bundles will be given as (finitely-generated and projective)
$\SU$-equivariant modules over the algebra of functions $\ca(\man)=\Apq
\otimes \ca(M)$. We will describe the dimensional reduction of
invariant connections on the $\SU$-equivariant modules over the
algebra $\ca(\man)$. In particular, we will reduce Yang--Mills gauge
theory on $\ca(\man)$ to a type of Yang--Mills--Higgs theory on the
manifold $M$. The vacuum equations of motion for this model give
$q$-deformations of some known vortex equations, whose solutions
possess, as we shall see, some remarkable properties.

In the $q=1$ case, a general and systematic treatment of ${\rm SU}(2)$-equivariant
dimensional reduction over the product $\IC\mathrm{P}^1\times M$ of
the ordinary complex projective line $\IC\mathrm{P}^1$ with a K\"ahler manifold $M$ was first
carried out in~\cite{A-CG-P1}. Here ${\rm SU}(2)$ acts in the standard
way by isometries of the homogeneous space $\IC\mathrm{P}^1$ and
trivially on $M$. It was shown in~\cite{A-CG-P1} that there is a
one-to-one correspondence between ${\rm SU}(2)$-equivariant vector
bundles over $\IC\mathrm{P}^1\times M$ and ${\rm U}(1)$-equivariant
vector bundles over $M$, with ${\rm U}(1)$ acting trivially on $M$. The
reduced vector bundle has the structure of a quiver bundle, in this
case a representation of the linear ${\rm A}_{m+1}$ quiver chain in
the category of complex vector bundles over $M$. Moreover, certain
natural first order gauge theory equations on $\IC\mathrm{P}^1\times
M$ reduce to generalizations of vortex equations called holomorphic
chain vortex equations, which contain a multitude of BPS-type integrable
equations as special cases~\cite{LPS1}. These include standard abelian and non-abelian vortex equations in two dimensions, and the self-duality and perturbed abelian Seiberg--Witten monopole equations in four dimensions. With suitable notions of stability for
holomorphic bundles over $\IC\mathrm{P}^1\times M$ and the
corresponding quiver bundles over $M$, a variant of the
Hitchin--Kobayashi correspondence identifies the necessary and
sufficient conditions for the existence of solutions to these
equations~\cite{A-CG-P1}. This particular reduction has been further developed
in~\cite{PS,LPS1,BS}, where some physical applications are also considered. The reduction generalizes to the product of $M$
with any homogeneous space which is a flag manifold;
see~\cite{A-CG-P3,LPS2} for the general theory.

In this paper we will study how equivariant dimensional reduction and
the ensuing vortex equations are modified when the `internal' sphere
$\IC\mathrm{P}^1$ is replaced with a particular noncommutative
deformation. Dimensional reduction over the \emph{fuzzy} sphere
$\IC\mathrm{P}_F^1$ was
considered in~\cite{Aschieri2,Aschieri1,Harland}, where it was shown
that the deformation significantly alters the vacuum structure of the
induced Yang--Mills--Higgs theory, which in some instances may not
coincide with the standard vortex models in the commutative limit. In
particular, solutions of abelian vortex
equations are studied in~\cite{Harland} which correspond to instantons
in the original Yang--Mills theory on $\IC\mathrm{P}_F^1\times M$ but
are nevertheless non-BPS states of the dimensionally reduced field
theory. In the following we will demonstrate that a similar vacuum
structure emerges when the dimensional reduction is performed over a
\emph{quantum} sphere $\pq$. As discussed in~\cite{Bradlow}, a basic
problem with standard vortex equations is that it is not
possible to reach the zeroes of the corresponding Yang--Mills--Higgs
action functional by means of non-trivial vortex solutions, due to
topological obstructions. In~\cite{A-CG-P2} it was shown that one can
improve this functional by using the formalism of twisted quiver
bundles, which yields zeroes of the action for bundles admitting
flat connections. In the present paper we show that, in contrast to the usual
quiver gauge theories that arise through dimensional reduction, the same is true for the
Yang--Mills--Higgs models which are systematically obtained via
$\SU$-equivariant dimensional reduction over $\pq$.

In order to rigorously carry out the dimensional reduction in parallel
to the commutative case, it is necessary to extend the equivariant
decompositions of~\cite{A-CG-P1,PS} within the algebraic framework
of noncommutative geometry and in a Hopf algebraic framework
appropriate to the action of the quantum group $\SU$. This is the
content of \S1--\S3 of the present paper. In \S1 and \S2 we extend the
requisite geometry of the projective line $\IC\mathrm{P}^1$ to the
$q$-deformed case $\pq$, using the fact that there are finitely-generated projective modules over the quantum sphere that correspond to the canonical line bundles on the Riemann sphere in the $q\to1$ limit.
In \S3 we generalize the decompositions
of~\cite{A-CG-P1,PS} to invariant gauge fields for the action of $\SU$
on $\man=\pq\times M$. In \S4 we study the reduction of Yang--Mills
theory on $\man$. In particular, we formulate a suitable notion of
generalized instanton on the quantum space $\man$ which coincides with
solutions of the vortex equations associated to minima of the induced
$q$-deformed Yang--Mills--Higgs action functional on $M$; we call the
(gauge equivalence classes of) solutions to these equations
`$q$-vortices'. We also examine in detail the structure of the corresponding vacuum
moduli spaces and the topological stability conditions for the existence of
solutions to the $q$-vortex equations, finding in general that these
moduli spaces are much more constrained than their classical $q\to1$ limits. In \S5 we study some explicit
examples and compare with analogous results in the literature for the
case $q=1$, showing that the $q$-deformation generically improves the geometrical structure of the associated moduli spaces. In particular, we analyse moduli spaces of $q$-vortices on Riemann surfaces giving new examples of non-abelian vortices, and show that our $q$-deformations of instantons on K\"ahler surfaces are analogous to those of some previous noncommutative deformations of the self-duality equations.

\subsubsection*{Conventions} \ \ 
In the following we shall use the terminology {\it covariance} to mean
both covariance for an action and `co-covariance' for a coaction. The $q$-number
\begin{equation}
[s]  =  [s]_q := \frac{q^s - q^{-s}}{q - q^{-1}} \ ,
\label{eq:q-integer}
\end{equation}
is defined for $q \neq 1$ and any $s \in \IR$. For a coproduct
$\Delta$ we use the conventional Sweedler notation
$\Delta(x)=x_{(1)}\otimes x_{(2)}$ (with implicit summation). This
convention is iterated to give $(\id\otimes\Delta)\circ\Delta(x) = (\Delta\otimes\id)\circ\Delta(x) = x_{(1)}\otimes x_{(2)}\otimes x_{(3)}$, and so on. 

\bigskip

\begin{center}
\textsc{Acknowledgments}
\end{center}

\vskip 4pt

GL was partially supported by the Italian Project
`Cofin08 -- Noncommutative Geometry, Quantum Groups and
Applications'. RJS was partially supported by the Grant ST/G000514/1 `String Theory Scotland' from the UK Science and Technology Facilities Council. We thank T. Brzezinski for an email exchange. 

\section{$\SU$-equivariant bundles on the quantum projective
  line }\label{se:qhb} 

The quantum projective line $\pq$ is defined as a quotient of the
sphere $\mathrm{S}^3_q \simeq\SU$ with respect to an action of the
group $\U(1)$. It is the standard Podle\'s sphere $\sq$ of \cite{Po87}
with additional structure. The construction we need is the well-known
quantum principal $\U(1)$-bundle over $\sq$, whose total space is the
manifold of the quantum group $\SU$. 

\medskip

\subsection{Quantum projective line $\pq$}\label{qdct} ~\\[5pt]
We begin with the algebras of $\mathrm{S}^3_q$ and
$\pq$. The manifold of $\mathrm{S}^3_q$ is identified with the
manifold of the quantum group $\SU$. The deformation parameter
$q\in\IR$ can be restricted to the interval $0<q<1$ without loss of
generality.  The coordinate algebra $\ASU$ is
the $*$-algebra generated by elements $a$ and~$c$ with the relations
\beqa
a\,c&=&q\,c\,a \quad \mbox{and} \quad c^*\,a^* \= q\,a^*\,c^*  \ ,     \quad a\,c^* \= q\,c^*\,a \ 
\quad \mbox{and} \quad c\,a^* \= q\,a^*\,c  \ , \nn \\[4pt] \ \ 
\ \ \ \ c\,c^* &=& c^*\,c \quad \mbox{and} \quad a^*\,a+c^*\,c \=
a\,a^*+q^{2}\,c\,c^* \= 1 \ . 
\label{derel}\eeqa
These relations are equivalent to requiring that the `defining' matrix
$$
U = \left(
\begin{array}{cc} a & -q\,c^* \\ c & a^*
\end{array}\right)
$$
is unitary, $U \,U^* = U^*\, U = 1$. The Hopf algebra structure for $\ASU$ is given by the
coproduct
$$
\Delta\,\left(
\begin{array}{cc} a & -q\,c^* \\ c & a^*
\end{array}\right)=\left(
\begin{array}{cc} a & -q\,c^* \\ c & a^*
\end{array}\right)\otimes\left(
\begin{array}{cc} a & -q\,c^* \\ c & a^*
\end{array}\right) \ ,
$$
with a `tensor product' of rows by columns, e.g. $\Delta(a)= a \otimes
a - q\, c^* \otimes c$, etc., the antipode
$$
S\,\left(
\begin{array}{cc} a & -q\,c^* \\ c & a^*
\end{array}\right)=\left(
\begin{array}{cc} a^* & c^* \\ -q\,c & a
\end{array}\right) \ ,
$$
and the counit
$$\epsilon \left(
\begin{array}{cc} a & -q\,c^* \\ c & a^*
\end{array}\right)=\left(
\begin{array}{cc} 1 & 0 \\ 0 & 1
\end{array}\right) \ .
$$

The quantum universal enveloping algebra $\su$ is the Hopf $*$-algebra
generated as an algebra by four elements $K,K^{-1},E,F$ with 
$K\,K^{-1}=1=K^{-1}\,K$ and relations
\beq \label{relsu} 
K^{\pm\,1}\,E = q^{\pm\,1}\,E\,K^{\pm\,1} \ , \quad 
K^{\pm\,1}\,F = q^{\mp\,1}\,F\,K^{\pm\,1} \quad \mbox{and} \quad
[E,F]  = \frac{K^{2}-K^{-2}}{q-q^{-1}} \ . 
\eeq 
The $*$-structure is simply
$$
K^* = K\ , \qquad E^* = F \qquad \mbox{and} \qquad F^* = E \ , 
$$
and the Hopf algebra structure is provided by the coproduct $\Delta$, 
the antipode $S$, and the counit $\epsilon$ defined by
$$
\begin{array}{cc}
\Delta(K^{\pm\,1})=K^{\pm\,1}\otimes K^{\pm\,1} \ ,
\quad \Delta(E)=E\otimes K+K^{-1}\otimes E \ , \quad
\Delta(F)=F\otimes K+K^{-1}\otimes F \ , 
\\~ \\
S(K)  = K^{-1} \ ,
\quad S(E)  = -q\,E \ , \quad S(F)  = -q^{-1}\,F \ , 
\\~\\
\epsilon(K) = 1 \ , \quad \epsilon(E) = \epsilon(F) = 0 \ .
\end{array}
$$
There is a bilinear pairing between $\su$ and $\ASU$ given on
generators by
\begin{align*}
&\langle K,a\rangle = q^{-1/2} \ , \quad \big\langle
K^{-1},a\big\rangle = q^{1/2} \ , \quad
\langle K,a^*\rangle = q^{1/2} \quad \mbox{and} \quad 
\big\langle K^{-1},a^*\big\rangle = q^{-1/2} \ , \nn\\[4pt]
&\langle E,c\rangle = 1 \quad \mbox{and} \quad \langle
F,c^*\rangle = -q^{-1} \ ,
\end{align*}
with all other couples of generators pairing to~0. One regards $\su$
as a subspace of the linear dual of~$\ASU$ via this pairing. 
There are canonical left and right $\su$-module algebra structures
on~$\ASU$ such that~\cite{wor87}
$$
\hs{g}{h \lt x} := \hs{g\,h}{x}\qquad \mbox{and} \qquad
\hs{g}{x \rt  h} := \hs{h\,g}{x}
$$
for all $g,h \in \su,\ x \in \ASU$. They are given by $h \lt x :=
\hs{(\id \otimes h)}{\cop (x)}$ and $x \rt  h := \hs{(h \otimes
  \id)}{\cop (x)}$, or equivalently
$$
h \lt x := \co{x}{1} \,\hs{h}{\co{x}{2}} \qquad \mbox{and} \qquad
x \rt  h := \hs{h}{\co{x}{1}}\, \co{x}{2}   
$$
in the Sweedler notation.
These right and left actions
mutually commute,
\begin{eqnarray*}
(h \lt x) \rt  g  \=  \left(\co{x}{1} \,\hs{h}{\co{x}{2}}\right) \rt
g &=& \hs{g}{\co{x}{1}} \,\co{x}{2}\, \hs{h}{\co{x}{3}} \nn\\[4pt]
&=& h \lt \left(\hs{g}{\co{x}{1}}\, \co{x}{2}\right)  \=  h \lt (x \rt
g) \ ,
\end{eqnarray*}
and since the pairing satisfies
$$
\hs{S(h)^*}{x} = \overline{\hs{h}{x^*}}
$$
for all $h \in \su,\ x \in \ASU$, the $*$-structure is compatible with 
both actions,
$$
h \lt x^*  =  \big(S(h)^* \lt x\big)^*  \qquad \mbox{and} \qquad
x^* \rt  h  =  \big(x \rt S(h)^*\big)^*
$$
for all $h \in \su, \ x \in \ASU$. The left action for any $s\in\IN_0$ 
is given explicitly by
\beqa
K^{\pm\,1}\triangleright a^{s}  = q^{\mp\,\frac{s}{2}}\,a^{s}, \qquad
&\mbox{and}& \qquad
K^{\pm\,1}\triangleright a^{*}\,^{ s}  =
q^{\pm\,\frac{s}{2}}\,a^{*}\,^{ 
  s} \ , \nn\\[4pt] 
K^{\pm\,1}\triangleright c^{s}  = q^{\mp\,\frac{s}{2}}\,c^{s} \qquad
&\mbox{and}& \qquad
K^{\pm\,1}\triangleright c^{*}\,^{ s}  =
q^{\pm\,\frac{s}{2}}\,c^{*}\,^{ 
  s} \ , \nn\\[4pt]
F\triangleright a^{s}  = 0 \qquad  &\mbox{and}& \qquad F\triangleright
a^{*}\,^{s}  = q^{(1-s)/2} \,[s]\, c\, a^{*}\,^{ s-1} \ , \nn\\[4pt]
F\triangleright c^{s}  = 0 \qquad &\mbox{and}& \qquad 
F\triangleright c^{*}\,^{s}  = -q^{-(1+s)/2}\, [s]\, a\, c^{*}\,^{s-1}
\ , \nn \\[4pt]
E\triangleright a^{s}  = -q^{(3-s)/2}\, [s]\, a^{s-1}\, c^{*}
\qquad&\mbox{and}& \qquad E\triangleright a^{*}\,^{ s}  = 0 \ ,
\nn\\[4pt] 
E\triangleright c^{s}  = q^{(1-s)/2}\, [s] \, c^{s-1}\, a^*
\qquad&\mbox{and}& \qquad E\triangleright c^{*}\,^{ s}  = 0 \ .
\label{lact}\eeqa
The right action is given explicitly by
\beqa
a^{s}\triangleleft K^{\pm\,1}  = q^{\mp\,\frac{s}{2}}\,a^{s} \qquad
&\mbox{and}& \qquad  
a^{*}\,^{ s}\triangleleft K^{\pm\,1}  =
q^{\pm\,\frac{s}{2}}\,a^{*}\,^{ s} 
\ , \nn\\[4pt]
c^{s}\triangleleft K^{\pm\,1}  = q^{\pm\,\frac{s}{2}}\,c^{s}
\qquad&\mbox{and}& \qquad   
c^{*}\,^{ s}\triangleleft K^{\pm\,1}  =
q^{\mp\,\frac{s}{2}}\,c^{*}\,^{ s} 
\ , \nn\\[4pt]
a^{s}\triangleleft F  = q^{(s-1)/2}\, [s]\, c\, a^{s-1}
\qquad&\mbox{and}& \qquad 
a^{*}\,^{ s}\triangleleft F  = 0 \ , \nn\\[4pt]
c^{s}\triangleleft F  = 0 \qquad&\mbox{and}& \qquad 
c^{*}\,^{ s}\triangleleft F  = -q^{-(s-3)/2}\, [s]\,
a^{*}\,c^{*}\,^{s-1} \ , \nn\\[4pt]
a^{s}\triangleleft E  = 0 \qquad&\mbox{and}& \qquad 
a^{*}\,^{s}\triangleleft E  = -q^{(3-s)/2}\, [s]\,
c^{*}\,a^{*}\,^{s-1} \ , \nn\\[4pt]
c^{s}\triangleleft E  = q^{(s-1)/2}\, [s]\, c^{s-1}\, a
\qquad&\mbox{and}& \qquad c^{*}\,^{ s}\triangleleft E  = 0 \ .
\label{ract}\eeqa

Now we describe the $\U(1)$-principal bundle over $\sq$, whose total
space is the manifold of the quantum group $\SU$. It is an example of
a quantum homogeneous space~\cite{BM93} constructed as follows. If  
$\ca(\U(1)):=\IC[\zeta,\zeta^*] \big/ \langle \zeta\,\zeta^*
-1\rangle$ denotes the (commutative) algebra of coordinate functions
on the group $\U(1)$, the map
\beq  \label{qprp}
\pi\,:\, \ASU ~ \longrightarrow~ \ca(\U(1)) \ ,
\qquad \pi\,\left(
\begin{array}{cc} a & -q\,c^* \\ c & a^*
\end{array}\right) = 
\left(
\begin{array}{cc} \zeta & 0 \\ 0 & \zeta^*
\end{array}\right)
\eeq 
is a surjective Hopf $*$-algebra homomorphism, so that $\ca(\U(1))$
becomes a quantum subgroup of ${\rm SU}_{q}(2)$ with a right coaction
\beq 
\Delta_{R}:= (\id\otimes\pi) \circ \Delta \, : \, \ASU
~\longrightarrow~\ASU \otimes
\ca(\U(1)) \ . \label{cancoa} 
\eeq 
The coinvariant elements for this coaction, i.e. elements 
$\{ x \in\ASU  \; | \; \Delta_{R}(x)=x\otimes 1\}$, generate a
subalgebra of $\ASU$ which is the coordinate algebra $\Asq$ of the
standard Podle\'s sphere $\sq$ first described in \cite{Po87}.

For the purposes of the present paper, it will be useful to also have
an equivalent description of the bundle by taking an action (irrelevantly right or left)
  of the abelian group $\U(1)= \{ z\in \IC \, | \, z \,z^* = 1\}$ on
the algebra $\ASU$, i.e. we consider the map
\beq\label{canact}
\alpha \, : \, \U(1) ~ \longrightarrow ~ \Aut\big(\ASU\big)
\eeq
defined on generators by  
\begin{eqnarray}
\alpha_z(a) = a \, z  \qquad &\mbox{and}& \qquad
\alpha_z(a^*) = a^*\, z^{*} \, , \nn\\[4pt] 
\alpha_z(c) = c\,  z \qquad &\mbox{and}& \qquad
\alpha_z(c^*) = c^*\, z^{*} \ , 
\label{rco}\end{eqnarray}
and extended as an algebra map, $\alpha_z(x\, y) =
\alpha_z(x)\,\alpha_z(y)$ for $x,y\in \ASU$ and $z\in \U(1)$. Here the complex number
$z$ is the evaluation of the function $\zeta \in \ca(\U(1))$. The
coordinate algebra $\Asq$ is then regarded as the
subalgebra of \emph{invariant} elements in $\ASU$, 
\beq\label{qsp-ps}
\Asq := \ASU^{\U(1)} := \big\{ x \in\ASU  \; \big| \; \alpha_z(x)= x
\big\} \ .
\eeq
As a set of generators for $\Asq$ we may take 
\beq \label{podgens}
B_{-} := a\,c^* \ , \qquad 
B_{+} := c\,a^* \qquad \mbox{and} \qquad 
B_{0} := c\,c^* \ ,
\eeq 
for which one finds relations
\begin{align*}
B_{-}\,B_{0} &= q^{2}\, B_{0}\,B_{-} \qquad \mbox{and} \qquad
B_{+}\,B_{0} = q^{-2}\, B_{0}\,B_{+} \ , \\[4pt]
B_{-}\,B_{+} &= q^2\, B_{0} \,\big( 1 - q^2\, B_{0} \big) \qquad
\mbox{and} \qquad B_{+}\,B_{-}= B_{0} \,\big( 1 - B_{0} \big) \ ,
\end{align*}
and $*$-structure $(B_{0})^*=B_{0}$ and $(B_{+})^*= B_{-}$.
The algebra inclusion $\Asq\hookrightarrow\ASU$ is a quantum principal
bundle and can be endowed with compatible calculi~\cite{BM93}, a
construction that we shall illustrate later on.

In \S\ref{se:cals2} we will describe a natural complex structure on
the quantum two-sphere $\sq$ for the unique two-dimensional covariant
calculus on it. This will transform the sphere $\sq$ into a quantum
riemannian sphere or quantum projective line $\pq$. Having this in
mind, with a slight abuse of `language' we will speak of $\pq$ rather
than $\sq$ from now on.

The sphere $\sq$ (and hence the quantum projective line $\pq$) is a
quantum homogeneous space of $\SU$ and the coproduct of $\ASU$
restricts to a left coaction of $\ASU$ on $\Asq$ (or $\Apq$):
$$
\Delta_L \,:\, \Apq ~ \longrightarrow ~ \ASU \otimes \Apq \ .
$$
In particular, the elements
$$
Y_{-} :=- a\,c^* \, , \qquad Y_{+} := q\, c\,a^* \, \qquad
\mathrm{and} \qquad Y_{0} := q^{2}\,\big(1+q^{2}\big)^{-1} - q^{2}\,
c\,c^* 
$$ 
transform according to the fundamental `vector corepresentation' of
$\SU$ given by
\begin{eqnarray}
\Delta_L(Y_{-})&=&a^{2}\otimes Y_{-}-\big(1+q^{-2}\big)\,Y_{-}\otimes
Y_{0}+ c^{*}\,^{ 2}\otimes Y_{+} \ , \nn\\[4pt]
\Delta_L(Y_{0})&=& q\, a\,c\otimes Y_{-}+\big(1+q^{-2}\big)\,
Y_{0}\otimes Y_{0}- c^*\,a^* \otimes Y_{+} \ , \nn\\[4pt]
\Delta_L(Y_{+})&=&q^{2}\, c^{2}\otimes
Y_{-}+\big(1+q^{-2}\big)\,Y_{+}\otimes 
Y_{0}+a^{*}\,^{ 2}\otimes Y_{+} \ .
\label{cosq}
\end{eqnarray}
The following result is evident.
\begin{prop}\label{prop:1invCPq1}
The element $1\in\Apq$ is the only coinvariant element for this
coaction, i.e. the only    $x\in\Apq$ for which $\Delta_{L}(x)= 1
\otimes x$.
\end{prop}

\medskip

\subsection{Equivariant line bundles on $\pq$}\label{se:avb} ~\\[5pt]
Let $\rho : \U(1) \to V$ be a representation of $\U(1)$ on a
finite-dimensional complex vector space $V$. The corresponding space
of $\rho$-equivariant elements is given by
\begin{equation}\label{coeq}
\ASU \boxtimes_\rho V := \big\{ \varphi  \in  \ASU \otimes V \, \big| \,  
(\alpha \otimes \id)\varphi = \big((\id \otimes \rho^{-1} )\big)
\varphi \big\} \ ,
\end{equation}
where $\alpha$ is the action (\ref{canact}) of $\U(1)$ on $\ASU$.
The space (\ref{coeq})
is an $\ca(\pq)$-bimodule. We shall think of
it as the module of sections of the vector bundle associated with the
quantum principal $\U(1)$-bundle on $\pq$ via the representation
$\rho$. There is a natural $\SU$-equivariance, in that the left
coaction $\Delta$ of $\ASU$ on itself extends in a natural way to a
left coaction on $\ASU \boxtimes_\rho V$ given by
\begin{equation}\label{co-mod}
\Delta^\rho= \Delta \otimes \id\,:\, \ASU \boxtimes_\rho V
~\longrightarrow~ \ASU \otimes \big(\ASU \boxtimes_\rho V\big) \ .
\end{equation}

The irreducible representations of $\U(1)$ are labelled by an integer
$n\in\IZ$. If $C_n\simeq\IC$ is the irreducible one-dimensional left
$\U(1)$-module of weight $n$, they are given by  
\begin{equation}\label{ircore}
\rho_n \,:\, \U(1) ~\longrightarrow~ \Aut(C_n) \ , \qquad C_n \ni v
~\longmapsto~ z^n\, v \in C_n \ .
\end{equation}
The corresponding spaces of equivariant elements are well-known and
amount to a vector space decomposition~\cite[eq.~(1.10)]{maetal}
\beq
\ASU=\bigoplus_{n\in\IZ}\, \cl_n \ ,
\label{SUlndecomp}\eeq
where
\beq\label{libu} 
\cl_n := \ASU \boxtimes_{\rho_n} \IC \simeq \big\{x \in \ASU ~\big|~
\alpha_z(x) = x \, (z^*)^n \big\} \ .
\eeq 
In particular, $\cl_0 = \Apq$. One has $\cl_n^* =
\cl_{-n}$ and $\cl_n\,\cl_m = \cl_{n+m}$. Each $\cl_n$ is
clearly a bimodule over $\Apq$ and is naturally isomorphic to $\ASU
\boxtimes_{\rho_n} C_n$. It was shown in~\cite[Prop.~6.4]{SWPod} that
each $\cl_n$ is a finitely-generated projective left (and right)
$\Apq$-module of rank one. They give the modules of $\SU$-equivariant
elements or of sections of line bundles over the quantum projective
line $\pq$ with monopole charges $-n$. One has the following results (cfr. \cite[Prop.~3.1]{{KLvS}}).
\begin{lemm}\label{masu}
\begin{enumerate}
\item Each $\cl_n$ is the bimodule of equivariant elements associated 
with the irreducible representation of $\U(1)$ with weight
$n$.
\item The natural map $\cl_n \otimes \cl_m \to \cl_{n+m}$ defined by
multiplication induces an isomorphism of $\Apq$-bimodules 
$$
\cl_n \otimes_{\Apq} \cl_m \simeq \cl_{n+m} \ ,
$$
and in particular $\Hom_{\Apq}(\cl_m, \cl_n) \simeq\cl_{n-m} $.
\end{enumerate}
\end{lemm}
\begin{proof}
These results follow by using the representation theory of $\U(1)$ as
well as the relations  
$$
a \otimes_{\Apq} c = q\, c \otimes_{\Apq} a \ , \quad 
a \otimes_{\Apq} c^* = q\, c^* \otimes_{\Apq} a \ , \quad 
c \otimes_{\Apq} c^*= c^*\otimes_{\Apq} c \ , 
$$ 
and so on, which are easily established. 
\end{proof}

{}From the transformations in \eqref{rco}, it follows that an
$\Apq$-module generating set for $\cl_n$ is given by elements
\beq
\big|\Psi^{(n)}\big\rangle_{\mu}  =  \left\{
\begin{array}{l}
~ c^{*}\,^{ \mu}\,a^{*}\,^{ n-\mu} \qquad
\mathrm{for }\quad n\geq0 ~,~ \mu = 0,1, \dots,  n \ , \\[4pt]
~ c^{\mn-\mu}\,a^{\mu}
\qquad \mathrm{for }\quad n\leq0 ~,~ \mu = 0,1, \dots,  \mn \ .
\end{array}
\right.
\label{qpro}\eeq
Then one writes equivariant elements as
\beq
\varphi_f  =  \left\{
\begin{array}{l} 
\displaystyle{ \, \sum_{\mu=0}^{n}\, c^{*}\,^{\mu}\, a^{*}\,^{n-\mu} \,
f_{\mu}  
 =  \sum_{\mu=0}^{n}\, \tilde f_{\mu}~c^{*}\,^{\mu}\, a^{*}\,^{n-\mu} 
\qquad \mathrm{for} \quad n \geq 0} \ , \\ ~ \\
\displaystyle{ \,  \sum_{\mu=0}^{\mn}\, c^{\mn-\mu}\, a^{\mu}\, f_{\mu}
 =  \sum_{\mu=0}^{\mn}\, \tilde f_{\mu}~c^{\mn-\mu}\, a^{\mu}
\qquad \mathrm{for} \quad n \leq 0} \ , 
\end{array} \right.
\label{eqmap}\eeq
with $f_\mu$ and $\tilde f_\mu$ generic elements in $\Apq$. The
elements in \eqref{qpro} are not independent over $\Apq$ since the
bimodules $\cl_n$ are not free modules.

A generic finite-dimensional representation $(V,\rho)$ for $\U(1)$ is
given by a weight decomposition 
\begin{equation}\label{repdeco}
V  =  \bigoplus_{n\in W(V)}\, C_n  \otimes V_n \ , \qquad \rho  = 
\bigoplus_{n\in W(V)}\, \rho_n \otimes  \id \ .
\end{equation}
Here $(C_n, \rho_n)$ is the one-dimensional irreducible representation
of $\U(1)$ with weight $n\in\IZ$ given in \eqref{ircore}, the spaces
$V_n=\Hom_{\U(1)}(C_n,V)$ are the multiplicity spaces, and the set
$W(V) = \{n\in\IZ ~|~V_n \not= 0\}$ is the set of weights of
$V$. For the corresponding space of $\rho$-equivariant elements we
have a corresponding decomposition 
\begin{equation}\label{su2deco}
\ASU \boxtimes_\rho V = \bigoplus_{n\in W(V)}\, \cl_n  \otimes V_n \ ,
\end{equation}
with $\cl_n$ the irreducible modules in \eqref{libu} giving sections
of line bundles over $\pq$.

The left action of the group-like element $K$ on $\ASU$ allows one to
give a dual presentation of the line bundles $\cl_n$ as 
\beq\label{libudual} 
\cl_n = \big\{x \in \ASU ~\big|~ K \lt x = q^{n/2}\, x \big\} \ .
\eeq 
Indeed, if $H$ is the infinitesimal generator of the $\U(1)$-action $\alpha$, 
the group-like element $K$ can be written as $K = q^{-H/2}$.
Then from the relations \eqref{relsu} of $\su$ one finds
\beq\label{rellb}
E \lt \cl_n ~\subset~ \cl_{n+2} \qquad \mbox{and} \qquad
F \lt \cl_n ~\subset~ \cl_{n-2} \ .
\eeq
On the other hand, commutativity of the left and right actions of
$\su$ yields 
\beq\label{lellb}
\cl_n \rt  h \subset \cl_n
\eeq
for all $h\in \su$. It was shown
in~\cite[Thm.~4.1]{SWPod} that there is also a decomposition
\beq
\label{decoln}
\cl_n=\bigoplus_{J=\tfrac{|n|}{2}, \tfrac{|n|}{2} +1,
\tfrac{|n|}{2} +2, \dots}\,V_{J}^{\left(n\right)} \ ,
\eeq 
with $V^{(n)}_{J}$ the spin~$J$ re\-pre\-sen\-ta\-tion space (for the
right action) of $\su$. Combined with (\ref{SUlndecomp}), we get a
Peter-Weyl decomposition for $\ASU$~\cite{wor87}. A PBW-basis for
$\ASU$ is given by monomials $a^m\, c^k\, c^{*\,l}$ for $k,l =0,1,
\dots$ and $m\in\IZ$, with the convention that $a^{-m}$ is 
short-hand notation for $a^{*\,m}$ when $m>0$. Furthermore, a similar
basis for $\cl_n$ is given by the monomials $a^{l-k}\, c^k\,
c^{*\,l+n}$, since from \eqref{lact} it follows that $K \lt (a^m
\,c^k\, c^{*\,l}) = q^{(-m-k+l)/2}\, a^m\, c^k\, c^{*\,l}$ and the
requirement that $-m-k+l = n$ is met by redefining $l \to l+n$ forcing
in turn $m=l-k$. In particular, the monomials $a^{l-k}\, c^k\,
c^{*\,l}$ are the only $K$-invariant elements, thus providing a
PBW-basis for $\cl_0=\Apq$.

\section{$\SU$-invariant gauge fields on the quantum projective
  line}\label{se:cotqpb}

We will now describe connections on the quantum projective
line. For this, we will need an explicit description of
the calculi on the quantum principal bundle over $\pq$. The principal
bundle $(\ASU,\Apq,\ca(\U(1)))$ is endowed with compatible
non-universal calculi~\cite{BM93,BM97} obtained from the
three-dimensional left-covariant calculus on $\SU$~\cite{wor87},
which we present first. We then describe the unique left-covariant
two-dimensional calculus on the quantum projective line
$\pq$~\cite{pod89} obtained by restriction, and also the projected
calculus on the structure group $\U(1)$. The calculus on $\pq$ can be
canonically decomposed into a holomorphic and an anti-holomorphic
part. All the calculi are compatible in a natural sense. These
constructions will produce a connection on the quantum principal
bundle over $\pq$ with respect to the left-covariant calculus
$\Omega^\bullet(\pq)$, also with a natural holomorphic structure. This
connection will determine a covariant derivative on the module of
equivariant elements $\cl_n$, which can be shown~\cite{LRZ} to
correspond to the canonical Grassmann connection on the associated
projective modules over $\Apq$. We also briefly recall how to compute
the monopole number $n\in\IZ$ by means of a Fredholm module. On the
other hand, to integrate the gauge curvature one needs a `twisted
integral' and the result is no longer an integer but rather its
$q$-analogue.

\medskip

\subsection{Left-covariant forms on $\SU$}\label{se:lcc}~\\[5pt]
The first differential calculus we take on the quantum group $\SU$
is the left-covariant calculus developed in~\cite{wor87}. It is
three-dimensional, such that its quantum tangent space
is generated by the three elements
$$
X_{z}  = \frac{1-K^{4}}{1-q^{-2}} \ , \qquad X_{-}
 = q^{-1/2}\,F\,K  \qquad \mbox{and} \qquad X_{+}  = q^{1/2}\,E\,K =
 X_{-}^* \ .
$$
Their coproducts and antipodes are easily found to be
\begin{equation}\label{cotb}
\cop(X_z)  =  1\otimes X_z + X_z \otimes K^4 \qquad \mbox{and} \qquad
\cop(X_\pm)  =  1\otimes X_\pm + X_\pm \otimes K^2 \ ,
\end{equation}
\begin{equation}\label{antb}
S(X_z) = - X_z \, K^{-4} \qquad \mbox{and} \qquad S(X_\pm) = - X_\pm
\, K^{-2} \ .
\end{equation}

The dual space of one-forms $\Omega^1(\SU)$ has a basis 
\beq
\beta_{z}  = a^*~\dd a+c^*~\dd c \ , \qquad \beta_{-}  = c^*~\dd
a^*-q\,a^*~\dd c^* \qquad \mbox{and} \qquad
\beta_{+}  = a~\dd c-q\,c~\dd a 
\label{q3dom}\eeq 
of left-invariant forms. The differential $\dd: \ASU \to
\Omega^1(\SU)$ is given by
\beq\label{exts3} 
\dd f =
(X_{-}\triangleright f) \,\beta_{-} + (X_{+}\triangleright f)
\,\beta_{+} + (X_{z}\triangleright f) \,\beta_{z} 
\eeq 
for all $f\in\ASU$. If $\Delta^{(1)}$ is the (left) coaction of $\ASU$
on itself extended to forms, the left-coinvariance of the basis forms 
is the statement that 
\beq\label{in-cl}
\Delta^{(1)}(\beta_{s})=1\otimes\beta_{s} \ , 
\eeq
while the left-covariance of the calculus is stated as
$$
(\Delta \otimes \id)\circ \Delta^{(1)} = \big(\Delta^{(1)} \otimes
\id\big)\circ \Delta^{(1)} \qquad \mathrm{and} \qquad  
(\epsilon \otimes \id)\circ \Delta^{(1)} = 1 \ .
$$
The requirement that it is a $*$-calculus, i.e. $\dd (f^*) = (\dd f)^*
$, yields
$$
\beta_{-}^*=-\beta_{+}  \qquad \mbox{and} \qquad
\beta_{z}^*=-\beta_{z} \ .
$$
The bimodule structure is given by
\begin{align}\label{bi1}
\beta_{z}\,a = q^{-2}\,a\,\beta_{z} \ , \quad
\beta_{z}\,a^* = q^{2}\,a^*\,\beta_{z} \ , \quad 
\beta_{\pm}\,a = q^{-1}\,a\,\beta_{\pm} \quad \mbox{and} \quad
\beta_{\pm}\,a^* = q\,a^*\,\beta_{\pm} \ , \nn \\[4pt]
\beta_{z}\,c = q^{-2}\,c\,\beta_{z} \ , \quad
\beta_{z}\,c^* = q^{2}\,c^*\,\beta_{z} \ , \quad 
\beta_{\pm}\,c = q^{-1}\,c\,\beta_{\pm} \quad \mbox{and} \quad
\beta_{\pm}\,c^* = q\,c^*\,\beta_{\pm} \ .
\end{align}

Higher degree forms can be defined in a natural way by requiring
compatibility with the commutation relations (the bimodule structure
(\ref{bi1})) and that $\dd^2=0$. One has
\beq \label{dformc3} \ 
\dd\beta_{z}  = -\beta_{-}\wedge\beta_{+} \ , \quad \dd \beta_{+}
 = q^{2}\,\big(1+q^{2}\big)\,\beta_{z}\wedge\beta_{+} \quad
\mbox{and} \quad 
\dd \beta_{-} = -q^{-2}\,\big(1+q^{2}\big)\,\beta_{z}\wedge\beta_{-}
\eeq 
together with the commutation relations
\begin{eqnarray}
\beta_{+}\wedge\beta_{+} \= \beta_{-}\wedge\beta_{-} \= \beta_{z}
\wedge\beta_{z}&=&0 \ , \nn\\[4pt]
\beta_{-}\wedge\beta_{+}+q^{-2}\,\beta_{+}\wedge\beta_{-}&=&0 \ ,
\nn\\[4pt]
\beta_{z}\wedge\beta_{-}+q^{4}\,\beta_{-}\wedge\beta_{z}&=&0 \ ,
\nn\\[4pt] 
\beta_{z}\wedge\beta_{+}+q^{-4}\,\beta_{+}\wedge\beta_{z}&=&0 \ .  
\label{commc3}\end{eqnarray}
Finally, there is a unique top form
$\beta_{-}\wedge\beta_{+}\wedge\beta_{z}$.
We may summarize the above results as follows.
\begin{prop}\label{3dsu}
For the three-dimensional left-covariant differential calculus on
$\SU$, the bimodules of forms are all trivial (left) $\ASU$-modules
given explicitly as 
\begin{align*}
\Omega^0(\SU)&=\ASU \ , \\[4pt]
\Omega^1(\SU)&=\ASU \langle \beta_-\,,\, \beta_+\,,\, \beta_z \rangle
\ , \\[4pt]
\Omega^2(\SU)&=\ASU \langle \beta_-\wedge\beta_+\,,\,
\beta_-\wedge\beta_z\,,\, \beta_+\wedge \beta_z \rangle \ , \\[4pt]
\Omega^3(\SU)&=\ASU \, \beta_{-}\wedge\beta_{+}\wedge\beta_{z} \ .
\end{align*}
The exterior differential and commutation relations are obtained from
\eqref{dformc3} and \eqref{commc3}, whereas the bimodule structure is
obtained from \eqref{bi1}.
\end{prop}

\medskip

\subsection{Holomorphic forms on $\pq$}\label{se:cals2}~\\[5pt]
The restriction of the three-dimensional calculus of \S\ref{se:lcc} to
the quantum projective line $\pq$ yields the unique left-covariant
two-dimensional calculus on $\pq$~\cite{maj05}. Further development of
this approach has led to a description of this calculus in terms of a
Dirac operator~\cite{SW04}. The `cotangent bundle' $\Omega^1(\pq)$ is
shown to be isomorphic to the direct sum $\cl_{-2}\oplus\cl_{+2}$ of the
line bundles with degree (monopole charge) $\pm\, 2$. Since the
element $K$ acts as the identity on $\Apq$, the differential
\eqref{exts3} restricted to $\Apq$ becomes
$$
\dd f  =  (X_{-}\triangleright f) \,\beta_{-} + (X_{+}\triangleright
f) \,\beta_{+} = q^{-1/2} (F \triangleright f) \,\beta_{-} + q^{1/2} (E \triangleright f) \,\beta_{+}
$$
for $f\in\Apq$. This leads to a decomposition of the exterior
differential into a holomorphic and an anti-holomorphic part, $\dd
=\dolb + \dol$, with 
$$ 
\dolb f = \left(X_{-}\triangleright f\right)\,\beta_{-} \qquad
\mbox{and} \qquad
\dol f = \left(X_{+}\triangleright f\right)\,\beta_{+} 
$$ 
for $f\in\Apq$. An explicit computation on the generators
\eqref{podgens} of $\pq$ yields 
\begin{eqnarray*}
\dolb \bm =  -q^{-1} \, a^{2} \, \beta_{-} \ , \qquad
\dolb\bz = -q^{-1} \, c\,a  \, \beta_{-} \qquad &\mbox{and}& \qquad
\dolb\bp =  c^{2} \, \beta_{-} \ , \nn\\[4pt]
\dol\bp =  q \, a^{*}\,^{2} \, \beta_{+} \ , \qquad
\dol\bz =  c^{*}\,a^{*} \, \beta_{+} \qquad &\mbox{and}&
\qquad \dol\bm =  -q^{2}\, c^{*}\,^{2} \, \beta_{+} \ .
\end{eqnarray*}

It follows that
$$
\Omega^1(\pq)=\Omega^{0,1}(\pq) \oplus
\Omega^{1,0}(\pq) \ ,
$$ 
where $\Omega^{0,1}(\pq)\simeq\cl_{-2} \beta_- \simeq\dolb(\Apq)$ is
the $\Apq$-bimodule generated by
$$
\big\{\,\dolb\bm\,,\,\dolb\bz\,,\,\dolb\bp
\big\} = \big\{a^{2}\,,\,c\,a\,,\,c^{2}\big\}\,\beta_{-}  = 
q^{2}\,\beta_{-}\,\big\{a^{2}\,,\,c\,a\,,\,c^{2}\big\}
$$
and $\Omega^{1,0}(\pq)\simeq\cl_{+2} \beta_+ \simeq\dol(\Apq)$ is the
$\Apq$-bimodule generated by
$$
\big\{\dol\bp\,,\,\dol\bz\,,\,\dol\bm\big
\} = \big\{a^{*}\,^{2}\,,\,c^{*}\,a^{*}\,,\,c^{*}\,^{2}\big\}\,
\beta_{+} =  q^{-2}\, \beta_{+}
\,\big\{a^{*}\,^{2}\,,\,c^{*}\,a^{*}\,,\,c^{*}\,^{2}\big\} \ .
$$
That these two modules of forms are not free is also expressed by the
existence of relations among the differentials given by
$$
\dol\bz - q^{-2}\, \bm~\dol\bp + q^{2}\, \bp~\dol\bm = 0 \qquad
\mbox{and} 
\qquad \dolb\bz - \bp~\dolb\bm + q^{-4}\, \bm~\dolb\bp = 0 \ .
$$
The two-dimensional calculus on $\pq$ has then quantum tangent space generated by the two elements $X_+$ and 
$X_-$ (or, equivalently $F$ and $E$). It has a unique (up to scale) top invariant form
$\beta$, which is central, $\beta \,f = f\, \beta$ for all $f\in\Apq$, and
$\Omega^2(\pq)$ is the free $\Apq$-bimodule generated by $\beta$,
i.e. one has $\Omega^2(\pq)=\beta \Apq = \Apq \beta$. Both
$\beta_{\pm}$ commute with elements of $\Apq$ and so does
$\beta_{-}\wedge\beta_{+}$, which may be taken as the natural
generator $\beta=\beta_{-}\wedge\beta_{+}$ of
$\Omega^2(\pq)$ (cfr. \cite{maj05} or \cite[App.]{SW04}). 
Writing any one-form as $\alpha = x\,
\beta_{-} + y\, \beta_{+} \in \cl_{-2} \beta_{-} \oplus \cl_{+2}
\beta_{+}$, the product of one-forms is given by
$$
(x \,\beta_{-} + y \,\beta_{+} ) \wedge (t \,\beta_{-} + z\, \beta_{+}
) = \big( x\, z - q^{ 2}\, y\, t \big)\, \beta_{-}\wedge\beta_{+} \ .
$$

By \eqref{dformc3} it is natural (and consistent) to demand $\dd \beta_-=\dd
\beta_+=0$ when restricted to $\pq$. Then the exterior derivative
of any one-form  $\alpha = x\, \beta_{-} + y\, \beta_{+} \in \cl_{-2}
\beta_{-} \oplus \cl_{+2} \beta_{+}$ is 
\beqa
\dd \alpha & =& \dd (x \,\beta_{-} + y\, \beta_{+}) \nn\\[4pt]
&=& \dol x \wedge \beta_{-} + \dolb y \wedge \beta_{+} \=
 \big( X_- \lt y - q^{2}\, X_+ \lt x \big) \,
\beta_{-}\wedge\beta_{+} \ ,
\label{d1f}\eeqa
since $K$ acts as $q^{\pm\,1}$ on $\cl_{\pm\,2}$. Notice that in
(\ref{d1f}), both $X_+ \lt x$ and $X_- \lt y$ belong to $\Apq$, as they
should. We may summarize these results as follows.
\begin{prop}\label{2dsph}
The two-dimensional differential calculus on the quantum projective
line $\pq$ is given by
$$
\Omega^{\bullet}(\pq) \= \Apq ~\oplus~ \left(\cl_{-2} \beta_{-}
\oplus \cl_{+2} \beta_{+} \right)~ \oplus~ \Apq
\beta_{-}\wedge\beta_{+} \ .
$$
Moreover, the splitting $\Omega^1(\pq) = \Omega^{1,0}(\pq)
\oplus\Omega^{0,1}(\pq)$, together with the two maps $\dol$ and
$\dolb$ given above, constitute a complex structure for the
differential calculus.
\end{prop}

A Hodge operator at the level of one-forms is constructed in
\cite{maj05} via a left-covariant map $\hat\star:\Omega^{1}(\pq) \to
\Omega^{1}(\pq)$ which squares to the identity $\id$. In the
description of the calculus as given in Proposition~\ref{2dsph}, it is defined
by
\beq\label{hod1} 
\hat\star (\dol f) = \dol f  \qquad \mbox{and} \qquad \hat\star
\big(\,\dolb f\big) = -\dolb f 
\eeq 
for all $f\in \Apq$. One then demonstrates its compatibility with the
bimodule structure, i.e. the map $\hat\star$ is a bimodule map. Thus
$\hat\star$ has values $\pm\, 1$ on holomorphic or anti-holomorphic
one-forms respectively, i.e. 
one has $\hat\star = \pm \id$ on $\Omega^{1,0}(\pq)$ or 
$\Omega^{0,1}(\pq)$ respectively.
In particular, $\hat\star 
\beta_\pm = \pm\, \beta_\pm$. The calculus has one central top
two-form and the Hodge operator is naturally extended by requiring
\beq\label{hod2} 
\hat\star 1 = \beta_{-}\wedge\beta_{+} \qquad \mbox{and} \qquad
\hat\star \left(\beta_{-}\wedge\beta_{+}\right) = 1 \ .
\eeq

We conclude this section by mentioning the calculus on $\U(1)$ which
makes all three calculi compatible from the quantum principal bundle
point of view. The strategy~\cite{BM93} consists in defining the
calculus on the coordinate algebra $\ca(\U(1))$ via the Hopf
projection $\pi$ in \eqref{qprp}. One finds that the projected
calculus is one-dimensional and bicovariant. Its quantum tangent space
is generated by
\beq\label{vvf} 
X = X_{z}  = \frac{1-K^{4}}{1-q^{-2}}
\eeq 
with dual one-form given by $\beta_{z}$. Explicitly, one finds
$$ 
\beta_{z}  =  z^{*}~ \dd z \ , \qquad 
\dd z = z\,\beta_{z} \qquad \mbox{and} \qquad
\dd z^{*}  = -q^{2}\,z^{*} \,\beta_{z} 
$$
along with the noncommutative commutation relations
$$
\beta_{z} \,z  =  q^{-2}\, z\,\beta_{z} \ , \qquad
\beta_{z}\, z^{*}  =  q^{2}\,z^{*}\,\beta_{z} \qquad \mbox{and} \qquad
z ~\dd z = q^{2}~\dd z~z \ .
$$
The data $(\ASU,\Apq,\ca(\U(1)))$ defines a `topological' quantum
principal bundle. There are differential calculi both on the total
space $\ASU$ (the three-dimensional left-covariant calculus) and on
$\ca(\U(1))$ (obtained from it via the same projection $\pi$ in
\eqref{qprp} giving the bundle structure). Moreover, from the calculus 
on $\ASU$ one also obtains by restriction a calculus on the base space
$\Apq$. The three calculi are compatible with the bundle
structure~\cite{BM93} (see also~\cite{LRZ}), thus constructing a
quantum principal bundle with non-universal calculi. The vector field
$X_z$ is vertical for the fibration.

\medskip

\subsection{Connections on equivariant line bundles over $\pq$
}\label{se:con}~\\[5pt]
The most efficient way to define a connection on a quantum principal
bundle (with given calculi) is by decomposing the one-forms on the
total space into horizontal and vertical forms~\cite{BM93,BM97}. Since 
horizontal one-forms are given in the structure group of the principal 
bundle, one needs a projection onto forms whose range is the subspace
of vertical one-forms. The projection is required to be covariant with
respect to the right coaction of the structure Hopf algebra.

For the principal bundle over the quantum projective line $\pq$ that
we are considering, a principal connection is a covariant left module 
projection $\Pi : \Omega^1(\SU) \to \Omega^1_{\mathrm{ver}}(\SU)$, i.e.
$\Pi^2=\Pi$ and $\Pi(x \,\alpha) = x\, \Pi(\alpha)$ for
$\alpha\in\Omega^1(\SU)$ and $x \in\mathcal{A}(\SU)$. Equivalently,
it is a covariant splitting $\Omega^1(\SU)=\Omega^1_{\mathrm{ver}}(\SU)
\oplus\Omega^1_{\mathrm{hor}}(\SU)$. The covariance of the connection
is the requirement that 
$$
\alpha_{R}^{(1)}\circ \Pi= \Pi \circ \alpha_{R}^{(1)} \ , 
$$ 
with $\alpha_{R}^{(1)}$ the extension to one-forms of the action
$\alpha_{R}$ in \eqref{canact}--\eqref{rco} of the structure Hopf
algebra $\U(1)$. It is not difficult to see that with the left-covariant
three-dimensional calculus on $\ASU$, a basis for
$\Omega^1_{\mathrm{hor}}(\SU)$ is given by
$\beta_{-},\beta_{+}$. Furthermore, one has
$$
\alpha_{R}^{(1)}(\beta_{z}) = \beta_{z}  \ , \qquad
\alpha_{R}^{(1)}(\beta_{-}) = \beta_{-}\ z^{*}\,^{2} \qquad
\mbox{and} \qquad
\alpha_{R}^{(1)}(\beta_{+}) = \beta_{+}\ z^{2} \ ,
$$
and so a natural choice of connection $\Pi=\Pi_z$ is to define
$\beta_{z}$ to be vertical~\cite{BM93,maj05}, whence
$$
\Pi_{z}(\beta_{z}):=\beta_{z} \qquad \mbox{and} \qquad
\Pi_{z}(\beta_{\pm}):=0 \ . 
$$

With a connection, one has a covariant derivative acting on right
$\Apq$-modules $\ce$ of equivariant elements,
$$
\nabla:= (\id - \Pi_{z}) \circ \dd\,:\,
\ce ~\longrightarrow~ \ce\otimes_{\Apq} \Omega^1(\pq)  \ , 
$$
and one readily proves the Leibniz rule $\nabla(\varphi\cdot f)=
(\nabla\varphi)\cdot f+\varphi\otimes \dd f$ for all $\varphi\in\ce$
and $f\in\Apq$. We shall take for $\ce$ the line bundles $\cl_n$ of
\eqref{libu}. Then with the left-covariant two-dimensional calculus
on $\Apq$ (coming from the left-covariant three-dimensional calculus
on $\ASU$ as described in \S\ref{se:cals2}), we have
\beq
\nabla \varphi= \left(X_{+}\triangleright\varphi\right)\,\beta_{+}
+\left(X_{-}\triangleright\varphi\right)\,\beta_{-}
\label{coder2d}\eeq
with $X_\pm\lt\varphi\in\cl_{n\pm2}$ for $\varphi\in\cl_n$. Using
Lemma~\ref{masu} we conclude that
$$
\nabla\varphi \ \in \ \cl_{n-2}   \, \beta_- \oplus \cl_{n+2}   \,
\beta_+ \simeq  \cl_{n} \otimes_{\Apq}  \Omega^1(\pq)
$$
as required.

A generic covariant derivative on the module $\cl_n$ is of the form
$\nabla_\alpha = \nabla + \alpha$, with $\alpha$ an element in
$\mathrm{Hom}_{\Apq}(\cl_n,\cl_n\otimes_{\Apq} \Omega^1(\pq) )$. For
later use it is helpful to characterize this space. More generally,
from Lemma~\ref{masu} we can infer the following results.
\begin{lemm} For any $n\in\IZ$ one has
\begin{equation}\label{homcon}
\mathrm{Hom}_{\Apq}\big(\cl_n\,,\, \cl_n\otimes_{\Apq} \Omega^1(\pq)
\big) 
\simeq  \cl_{-2} \beta_- \oplus \cl_{+2} \beta_+ = \Omega^1(\pq) \ ,
\end{equation}
while for any two distinct integers $n,m\in\IZ$ one has
\begin{align}\label{homhom}
\mathrm{Hom}_{\Apq}\big(\cl_n\,,\, \cl_m\otimes_{\Apq} \Omega^1(\pq)
\big) & \simeq
\cl_{m-n-2}   \, \beta_- \oplus \cl_{m-n+2}   \, \beta_+ \nn \\[4pt]
& \simeq \cl_{m-n} \otimes_{\Apq} \Omega^1(\pq) \ .
\end{align}
\end{lemm}

Given the connection, we can work out an explicit expression for its
curvature, defined to be the $\Apq$-linear (by construction) map 
$$ 
\nabla^2:=\nabla\circ\nabla \, : \, \cl_n ~ \longrightarrow ~ \cl_n 
\otimes_{\Apq}\Omega^2(\pq) \ .
$$
\begin{prop}
Let $\nablaCP_n$ be the connection on the line bundle $\cl_n$ defined
in \eqref{libu}, given in \eqref{coder2d} for the canonical
left-covariant two-dimensional calculus on $\Apq$. Then, with
$\varphi\in\cl_n$, its curvature is given by
\beq
\nablaCP_n^2 \varphi = \left(X_z \lt
  \varphi\right)\,\beta_-\wedge\beta_+\ .  
\label{gcurphi}\eeq
As an element in 
$\mathrm{Hom}_{\Apq}(\cl_n, \cl_n \otimes_{\Apq}\Omega^2(\pq) )$, one
has
\begin{equation}\label{gcur}
\nablaCP_n^2 = -q^{n+1}\, [n] \, \beta_-\wedge\beta_+ \ .
\end{equation}
\end{prop}
\begin{proof}
Using (\ref{coder2d}), (\ref{commc3}) and the fact that
$\dd\beta_\pm=0$ on $\pq$, by the Leibniz rule we have
\begin{eqnarray*}
\nablaCP_n\big(\nablaCP_n\varphi
\big)&=&(X_-\,X_+\lt\varphi)\,\beta_-\wedge\beta_++
(X_+\,X_-\lt\varphi)\,\beta_+\wedge\beta_- \\[4pt]
&=&  
\big((X_-\,X_+-q^{2}\,X_+\,X_-)\lt\varphi\big)\,\beta_-\wedge\beta_+ \
,
\end{eqnarray*}
and (\ref{gcurphi}) follows from the relation
$X_-\,X_+-q^{2}\,X_+\,X_-=X_z$. Since
$X_z\lt\varphi=-q^{n+1}\,[n]\,\varphi$ 
for $\varphi\in\cl_n$, one has (\ref{gcur}). Since $X_z\lt\Apq=0$,
the curvature is $\Apq$-linear.
\end{proof}

We can also derive an explicit expression for the corresponding gauge
potential $\asf_n$ defined by $\varphi\asf_n = \nabla \varphi - \dd
\varphi$ for $\varphi\in\cl_n$. With $X_z$ the vertical vector field
in \eqref{vvf}, using (\ref{exts3}) and (\ref{coder2d}) we find
$ 
\varphi\asf_n =  -\left(X_z \lt \varphi\right) \,\beta_z 
 =  q^{n+1}\, [n] \,\varphi\, \beta_z  
$,
or 
\beq\label{gpcan}
\asf_n=q^{n+1}\, [n] \, \beta_z \ . 
\eeq
As usual, $\asf_n$ is \emph{not} defined on $\pq$ but rather on the
total space $\SU$ of the bundle, i.e. $ \asf_n \in
\mathrm{Hom}_{\Apq}(\cl_n,\cl_n\otimes_{\Apq} \Omega^1(\SU) )$. In
terms of the gauge potential, the curvature is given by
\beq\label{gpot}
\fsf_n:=\nablaCP_n^2 = \dd\asf_n
\eeq
as a direct consequence of the first identity in (\ref{dformc3}).

\medskip

\subsection{Holomorphic structures }\label{se:hol}~\\[5pt]
The connection given in \S\ref{se:con} can be naturally decomposed
into a holomorphic and an anti-holomorphic part, $\nabla =
\nabla^{\dol} +\nabla^{\dolb}$. They are given by
\begin{equation}\label{coher}
\nabla^{\dol} \varphi  =  \left(X_{+}\lt \varphi\right)\,\beta_{+}
\qquad \mbox{and} \qquad
\nabla^{\dolb} \varphi  =
\left(X_{-}\triangleright\varphi\right)\,\beta_{-}
\end{equation}
with the corresponding Leibniz rules 
$$
\nabla^{\dol}(\varphi\cdot f) = \big(\nabla^{\dol}\varphi\big)\cdot f
+\varphi\otimes\dol f \qquad \mbox{and} \qquad
\nabla^{\dolb}(\varphi\cdot f) = \big(\nabla^{\dolb}\varphi\big)\cdot
f+ \varphi\otimes \dolb f \ ,
$$
for all $\varphi\in\cl_n$ and $f\in\Apq$. They are both flat,
i.e. $(\nabla^{\dol})^2=0=(\nabla^{\dolb}\,)^2$, and so the connection
$\nabla$ is integrable.

Holomorphic `sections' are elements $\varphi\in\cl_n$ which satisfy
$$
\nabla^{\dolb} \varphi = 0 \ . 
$$
{}From the actions given in \eqref{lact} we see that $F\lt a^{s} =0$ and
$F\lt c^{s} =0$ for any $s\in\IN_0$, while $F\lt a^{*}\,^{s} \not= 0$
and $F\lt c^{*}\,^{s} \not= 0$ for any $s\in\IN$. Then, from the
expressions \eqref{eqmap} for generic equivariant elements, we see that
there are no holomorphic elements in $\cl_n$ for $n>0$. On the other
hand, for $n\leq 0$ the elements 
$c^{\mn-\mu}\,a^{\mu},~ \mu = 0,1, \dots,\mn$ are holomorphic,
$$
\nabla^{\dolb} \big( c^{\mn-\mu}\,a^{\mu} \big) = 0 \ . 
$$
Since $\ker \dolb= \IC$ (as only the constant functions on $\pq$ do
not contain the generator $a^*$ or $c^*$), so that the only
holomorphic functions on $\pq$ are the constants, these are the only
invariants in degree $n$. We may conclude that holomorphic equivariant
elements are all polynomials in two variables $a,c$ with the
commutation relation $a\,c = q\,c\,a$, which defines the coordinate
algebra of the quantum plane. Further aspects of these
holomorphic structures are reported in~\cite{KLvS}.

\medskip

\subsection{Unitarity and gauge transformations }~\\[5pt]
On each line bundle $\cl_n$, $n\geq0$ there is an $\Apq$-valued
hermitian structure 
$$
\hat h_{n}\,:\,  \cl_n  \times \cl_n ~\longrightarrow~ \Apq
$$
defined by
\beq
\hat h_{n} \Big(\,\mbox{$\sum\limits_{\mu=0}^{n}\, c^{*}\,^{\mu}\,
  a^{*}\,^{n-\mu}~f_{\mu} \,,\,
  \sum\limits_{\nu=0}^{n}\,c^{*}\,^{\nu}\, a^{*}\,^{n-\nu}~g_{\nu}$}\,
\Big) = \sum_{\mu=0}^{n}\, f_{\mu}^* \, a^{n-\mu}\, c^{\mu}
~c^{*}\,^{\mu}\, a^{*}\,^{n-\mu} ~ g_{\mu}
\label{herstr}\eeq
in the $\Apq$-module basis \eqref{qpro}--\eqref{eqmap}. Having taken
the right $\Apq$-module structure for $\cl_n$, the hermitian structure 
(\ref{herstr}) is right $\Apq$-linear and left $\Apq$-antilinear. It
is covariant under the natural left coaction of $\ASU$ on $\cl_n$
induced by the inclusion $\cl_n \subset \ASU$. There is an analogous 
formula for $n\leq0$. By composing $ \hat h_{n}$ with the Haar
functional of $\ASU$ restricted to $\Apq$, one obtains a $\IC$-valued
inner product on $\cl_n$. Since the Haar functional of $\ASU$ is
invariant under the coaction of $\ASU$ on itself~\cite[\S4.2.6]{KS97},
we get an $\SU$-invariant inner product on each $\cl_n$. If we
write elements $\varphi\in\cl_n$ as vector-valued functions
$\varphi=(\varphi_\mu \ , \ \mu=0,1, \dots, \mn)$, the hermitian
structure is simply $\hat h_{n}(\varphi, \psi) = \sum_\mu\,
\varphi_\mu^* \, \psi_\mu$.

\begin{lemm}
The connection $\nablaCP_n$ is unitary, i.e. it is compatible with the
hermitian structure $\hat h_{n}$,
$$
\hat h_{n}\big(\nablaCP_n \varphi\,,\, \psi\big) + \hat
h_{n}\big(\varphi\,,\, \nablaCP_n \psi\big) = \dd \big(\, \hat
h_{n}(\varphi, \psi)\, \big) \qquad \mbox{for any} \quad
\varphi, \psi \in \cl_n \ .
$$
\end{lemm}
\begin{proof}
On the one hand,
$ 
\dd \big(\, \hat h_{n}(\varphi, \psi)\, \big) = \big(X_+ \lt \hat
h_{n}(\varphi, \psi)\big)\, \beta_+ + \big(X_- \lt \hat h_{n}(\varphi,
\psi)\big)\, \beta_- 
$. Using the coproducts \eqref{cotb} we have  
\begin{align*}
\big(X_\pm \lt \hat h_{n}(\varphi, \psi)\big) = \sum_{\mu=0}^{|n|}\,
X_\pm \lt \big( \varphi_\mu^*\, \psi_\mu \big) & =
\sum_{\mu=0}^{|n|}\, \Big(\varphi_\mu^*\, \big(X_\pm \lt
  \psi_\mu\big) + \left(X_\pm \lt \varphi_\mu^*\right)\, \left(K^2
  \lt \psi_\mu\right)\Big) \\[4pt]
& = \sum_{\mu=0}^{|n|}\,\Big( \varphi_\mu^*\, \big(X_\pm \lt
  \psi_\mu\big) + q^n\, \left(X_\pm \lt \varphi_\mu^*\right)\,
\psi_\mu \Big) \ ,
\end{align*}
and in turn
$$ 
\dd \big(\, \hat h_{n}(\varphi, \psi)\, \big) = \sum_\pm ~
\sum_{\mu=0}^{|n|} \, \Big( \varphi_\mu^*\, \big(X_\pm \lt
  \psi_\mu\big) 
+ q^n\, \left(X_\pm \lt \varphi_\mu^*\right)\, \psi_\mu \Big)\,
\beta_\pm \ . 
$$ 
On the other hand, using the antipodes \eqref{antb} and
$\beta_\pm^*=-\beta_\pm$ we have
\begin{align*}
\hat h_{n}\big(\nablaCP_n \varphi\,,\, \psi\big) = \sum_\pm~
\sum_{\mu=0}^{|n|}\,\beta_\pm^* \, \big(X_\pm \lt \varphi_\mu\big)^*\,
\psi_\mu & = q^n\, \sum_\pm~ \sum_{\mu=0}^{|n|}\, q^{\mp\,2}\,
\beta_\pm \, \big(X_\pm \lt \varphi_\mu^*\big) \,\psi_\mu \\[4pt]
& = q^n \, \sum_\pm~ \sum_{\mu=0}^{|n|}\, \big(X_\pm \lt
\varphi_\mu^*\big) \, \psi_\mu\, \beta_\pm \ ,
\end{align*}
and in turn
$$
\hat h_{n}\big(\nablaCP_n \varphi\,,\, \psi\big) + \hat h_{n}\big(
\varphi\,,\, \nablaCP_n \psi \big) 
= \sum_\pm~ \sum_{\mu=0}^{|n|}\, \Big( q^n\, \big(X_\pm \lt
\varphi_\mu^*\big) \, \psi_\mu  +
\varphi_\mu^*\, \big( X_\pm \lt \psi_\mu  \big) \Big)\, \beta_\pm \ .
$$
A direct comparison now gives the result. 
\end{proof}

We already know that any other connection is written as $\nabla_\alpha =
\nabla + \alpha$ with $\alpha$ a generic element in
$\mathrm{Hom}_{\Apq}(\cl_n, \cl_n\otimes_{\Apq} \Omega^1(\pq) )$
which, for a unitary connection $\nabla$, is necessarily
anti-hermitian,
$$
\hat h_{n}(\alpha\varphi, \psi) + \hat h_{n}(\varphi, \alpha\psi)=0
\qquad \textup{for} \quad \varphi, \psi \in \cl_n \ .
$$
\begin{lemm}\label{uni-gp}
Unitary elements $\alpha\in\mathrm{Hom}_{\Apq}(\cl_n,
\cl_n\otimes_{\Apq} \Omega^1(\pq) )$ are of the form 
$$
\alpha = x\, \beta_- + q^2\, x^*\, \beta_+ = x\, \beta_-
- (x\, \beta_-)^* \ , 
$$ 
with $x$ a generic element in $\cl_{-2}$.
\end{lemm}
\begin{proof}
{}From the identification \eqref{homcon}, we seek elements in
$\Omega^1(\pq) = \cl_{-2} \beta_- \oplus \cl_{+2} \beta_+$ which are
unitary. It is straightforward to verify that a generic one-form $\alpha
= x_- \,\beta_- + x_+\, \beta_+$ with
$x_\mp\in\cl_{\mp\,2}$ is unitary with respect to the hermitian
structure $\hat h_{n}$ if and only if it is written as claimed.
\end{proof}

The group $\cu(\cl_n)$ of gauge transformations consists of unitary
elements in $\End_{\Apq}(\cl_n)$ (with respect to the hermitian
structure $\hat h_{n}$). It acts on a connection $\nabla$ by
$$
(u, \nabla) ~\longmapsto~ \nabla^u = u \circ \nabla \circ u^* \ .
$$
An arbitrary connection $\nabla_\alpha=\nabla+\alpha$ will then transform
to $(\nabla_\alpha)^u=\nabla+\alpha^u$ with
$$
\alpha^u=u\,(\nabla u^*) + u \,\alpha\, u^* \ . 
$$
We know from Lemma~\ref{masu} that
$\End_{\Apq}(\cl_n)\simeq\cl_0=\Apq$. Thus $\cu(\cl_n)$ consists of
unitary elements in the coordinate algebra $\Apq$.
Of these there are none which are nontrivial. Indeed, in the
  coordinate algebra of $\ASU$ there are no nontrivial invertible
  elements \cite[App.]{HM99}. Since $\Apq$ is a subalgebra of the latter, it cannot contain any nontrivial invertible (hence unitary) elements either.

\medskip

\subsection{$\SU$-invariant connections and gauge transformations
}\label{incogt}~\\[5pt]
Recall that there is a coaction \eqref{co-mod} of $\ASU$ on modules of
sections. Let us denote by $\Delta_{(n)}$ the coaction on $\cl_n$,
\beq\label{rec-co}
\Delta_{(n)} \, : \, \cl_n ~ \longrightarrow ~ \ASU \otimes \cl_n \ ,
\qquad \Delta_{(n)}(\varphi) = \varphi_{(-1)} \otimes \varphi_{(0)} \
, 
\eeq
with implicit summation as usual. By combining it with the coaction
$\Delta_L^{(1)}$ of $\ASU$ on the bimodule of one-forms
$\Omega^1(\pq)$ we get an analogous coaction
\begin{align}\label{rec-co-for}
\Delta_{(n)}^{(1)} \, : \, \cl_n \otimes_{\Apq}\Omega^1(\pq) & ~
\longrightarrow ~ \ASU \otimes \big( \cl_n \otimes_{\Apq}
\Omega^1(\pq) \big) \ , \nn \\ \Delta_{(n)}^{(1)}(\omega) & \=
\omega_{(-1)} \otimes \omega_{(0)} \ .
\end{align}

Next we give an `adjoint' coaction of $\ASU$ on the space $\cc(\cl_n)$
of unitary connections,
$$
\Delta^{\cc} \, : \, \cc(\cl_n) ~ \longrightarrow ~ \ASU \otimes
\cc(\cl_n) \ ,
$$
defined by
$$
\Delta^{\cc} (-)= m_{12} \circ \big(\id \otimes
\Delta_{(n)}^{(1)}\big) \circ \big( \id \otimes (-) \big) \circ \big(S
\otimes \id\big) \circ \Delta_{(n)}
$$
with $m_{12}$ the multiplication in the first two factors of the
tensor product and $S$ the antipode. Thinking of $\Delta^{\cc} (-)$ as
acting on $1 \otimes \varphi$ with $\varphi\in\cl_n$, and using
\eqref{rec-co}, we get the `explicit' expression
$$
\Delta^{\cc} (\nabla_\alpha) (\varphi) =
S\big(\varphi_{(-1)}\big)\,\big(\nabla_\alpha(\varphi_{(0)})
\big)_{(-1)} \otimes \big(\nabla_\alpha (\varphi_{(0)}) \big)_{(0)} \ .
$$
\begin{lemm}\label{invcancon}
The canonical connection $\nablaCP_n$ in \eqref{coder2d} is the unique
invariant connection for this coaction, i.e. the unique element
$\nabla\in\cc(\cl_n)$ for which 
$$
\Delta^{\cc} (\nabla ) = 1 \otimes \nabla \ .
$$
In particular, there is no non-trivial element in
$\mathrm{Hom}_{\Apq}(\cl_n,\cl_n\otimes_{\Apq} \Omega^1(\pq) )$ which
is invariant.
\end{lemm}
\begin{proof}
The left-coinvariance of the canonical connection $\nablaCP_n$ is most
easily seen from the corresponding gauge potential in
\eqref{gpcan}. This is clearly left-coinvariant from the properties
\eqref{in-cl} of the basis one-forms, and in particular of
$\beta_z$. Since a unitary element $\alpha\in\mathrm{Hom}_{\Apq}(\cl_n,
\cl_n\otimes_{\Apq} \Omega^1(\pq) )$ is of the form given in
Lemma~\ref{uni-gp}, it is evident that $\alpha=0$ is the only such
left-invariant element.
\end{proof}

An `adjoint' coaction of $\ASU$ on the group $\cu(\cl_n)$ of gauge transformations, 
$$
\Delta^{\cu} \, : \, \cu(\cl_n) ~ \longrightarrow ~ \ASU \otimes
\cu(\cl_n) \ ,
$$
would be defined analogously as above by
$$
\Delta^{\cu} (-)= m_{12} \circ \big(\id \otimes  \Delta_{(n)}\big)
\circ \big( \id \otimes (-) \big) \circ \big(S \otimes \id\big) \circ
\Delta_{(n)} \ ,
$$
and thinking of $\Delta^{\cu} (-)$ as acting on $1 \otimes \varphi$,
with $\varphi\in\cl_n$, one has.
$$
\Delta^{\cu} (u) (\varphi) =
S\big(\varphi_{(-1)}\big)\,\big(u(\varphi_{(0)})\big)_{(-1)} \otimes
\big(u(\varphi_{(0)})\big)_{(0)} \ . 
$$
In fact, we already know that $\cu(\cl_n)$ consists of unitary elements
in $\Apq$. Then, the $\ASU$-coaction $\Delta^{\cu}$ is just the restriction
to $\cu(\cl_n)$ of the canonical $\ASU$-coaction on $\Apq$ given in
\eqref{cosq}. Also, as $\cu(\cl_n)$ is made only of complex numbers of
modulus one, the following result is immediate.

\begin{lemm}\label{invcangt}
The element $1\in\cu(\cl_n)$ is the unique invariant gauge
transformation for this coaction, i.e. the unique element $u
\in\cu(\cl_n)$ for which 
$$
\Delta^{\cu} (u ) = 1 \otimes u\ .
$$
\end{lemm}

\noindent
This  also follow from Proposition~\ref{prop:1invCPq1} giving $1$ as the only $\SU$-invariant
element in the algebra $\Apq$.

\medskip

\subsection{K-theory charges }\label{se:wn}~\\[5pt]
The line bundles on the sphere $\pq$ described in \S\ref{se:avb} are
classified by their monopole number $n\in\IZ$. One writes $\cl_n=\qpp
(\Apq)^{\mn+1}$ with suitable projections $\qpp$ in
$\Mat_{\mn+1}(\Apq)$. They are given explicitly by
\beq\label{qproj}
\qpp_{\mu\nu} = 
\begin{cases}
\; \sqrt{\alpha_{n,\mu}\,\alpha_{n,\nu}} \; c^{n-\mu}\,a^{\mu}\,
a^{*\,\nu}\,c^{*\,n-\nu} \ , \qquad n\geq 0\\
~\\ 
\; \sqrt{\beta_{n,\mu}\,\beta_{n,\nu}} \;
c^{*\,\mu}\,a^{*\,\mn-\mu}\,a^{\mn-\nu}\, c^{\nu} \ , \qquad n\leq 0 \ ,
\end{cases}
\eeq
with $\mu, \nu = 0, 1, \dots,  \mn $ and the numerical coefficients
$$
\alpha_{n,\mu}=\prod_{j=0}^{n-\mu-1}\, \frac{1-q^{2\left(n-j\right)}}
{1-q^{2\left(j+1\right)}} \qquad \mbox{and} \qquad
\beta_{n,\mu}=q^{2\mu}\,\prod_{j=0}^{\mu-1}\,
\frac{1-q^{-2\left(\mn-j\right)}}{1-q^{-2\left(j+1\right)}} \ .
$$
We use the convention $\prod_{j=0}^{-1}\,(-):=1$.

The projections in \eqref{qproj} are representatives of classes in the
K-theory of $\pq$, i.e. $[\qpp]\in \K_0(\pq)$. One computes the
corresponding monopole number by pairing them with a non-trivial
element in the dual K-homology, i.e. with (the class of) a non-trivial
Fredholm module $[\mu]\in  \K^0(\pq)$. For this, one first calculates
the corresponding Chern characters in the cyclic homology
$\chern_\bullet(\qpp) \in \mathrm{HC}_\bullet(\pq)$ and cyclic
cohomology $\chern^\bullet(\mu)\in \mathrm{HC}^\bullet(\pq)$
respectively, and then uses the pairing between cyclic homology and
cohomology.

The Chern character of the projections $\qpp$ has a non-trivial
component in degree zero $\chern_0(\qpp)\in \mathrm{HC}_0(\pq)$ given
simply by a (partial) matrix trace
$$
\chern_0(\qpp) := \tr (\qpp) = 
\begin{cases}
\; \displaystyle \sum_{\mu=0}^{n}\, \alpha_{n,\mu}\,(c^*\,c)^{n-\mu} ~
\prod_{j=0}^{\mu-1}\, \big(1-q^{2j}\, c^*\,c\big) \ , \qquad n\geq 0
\\ ~ \\ \displaystyle 
\; \sum_{\mu=0}^{\mn}\, \beta_{n,\mu}\,(c^*\,c)^\mu ~
\prod_{j=0}^{\mn-\mu-1}\, \big(1-q^{-2j}\, c^*\,c\big) \ , \qquad
n\leq 0 \ ,
\end{cases}
$$
and $\chern_0(\qpp)\in\Apq$. Dually, one needs a cyclic zero-cocycle,
i.e. a trace on $\Apq$. This was obtained in~\cite{MNW} and it is a
trace on $\Apq / \IC$, i.e. it vanishes on $\IC \subset \Apq$. On the
other hand, its values on powers of the element $c^*\,c$ is given by 
$$
\mu\left((c^*\,c)^k\right) = \big(1-q^{2k}\big)^{-1} \ , \qquad k>0 \
.
$$
The pairing was computed in~\cite{H00} and results in  
\beq\label{ind}
\hs{[\mu]}{[\qpp]}:= \mu\left(\chern_0(\qpp)\right)= -n \ .
\eeq

This integer is a topological quantity that depends only on the
bundle, both over the quantum sphere and over its classical limit
which is an ordinary two-sphere. In this limit it could also be
computed by integrating the curvature two-form of any
connection. However, in order to integrate the gauge curvature on the
quantum sphere $\pq$ one requires a `twisted integral', and the result
is no longer an integer but rather a $q$-integer. We recall here the
main facts, refering to~\cite{{LRZ}} for additional details.

It is known~\cite[Prop.~4.15]{KS97} that the modular automorphism
associated with the Haar state $H$ on the algebra $\ASU$ when
restricted to the subalgebra $\Apq$ yields a faithful, invariant state
on $\Apq$, i.e. $H(a\rt X)=H(a)\, \epsilon(X)$ for $a\in\Apq$ and
$X\in\su$, with modular automorphism
\beq\label{masp}
\vartheta(g) = g \rt K^2 \qquad \mathrm{for} \quad g\in\Apq \ ,
\eeq
such that
\beq
H(a \, b) = H\big(\vartheta(b)\, a\big)
\eeq
for $a,b \in \Apq$. With $\beta_{-}\wedge\beta_{+}$ the central
generator of $\Omega^2({\pq})$, $H$ the Haar state on $\Apq$, and
$\vartheta$ its modular automorphism in \eqref{masp}, it was proven in
\cite{SW04} that the linear functional
\beq\label{int}
\int_\pq \, : \;\; \Omega^2({\pq}) ~ \longrightarrow ~ \IC \ , \qquad
\int_\pq\, a\, \beta_{-}\wedge\beta_{+} := H(a)
\eeq
defines a non-trivial $\vartheta$-twisted cyclic two-cocycle $\tau$ on
$\Apq$ given by
\beq\label{twcc}
\tau(a_0, a_1, a_2):= \frac{1}{2}\, \int_\pq\, a_0~ \dd a_1 \wedge \dd
a_2 \ .
\eeq
This means that $\cob_\vartheta \tau=0$ and $\lambda_\vartheta
\tau=\tau$, where $\cob_\vartheta$ is the $\vartheta$-twisted
coboundary operator
\beqa
(\cob_\vartheta\tau)(f_0,f_1,f_2,f_3) &:=&
\tau(f_0\,f_1,f_2,f_3)-\tau(f_0,f_1\,f_2,f_3)+\tau(f_0,f_1,f_2\,f_3)
\nn \\ && -\, \tau\big(\vartheta(f_3)\,f_0\,,\,f_1\,,\,f_2\big) \ ,
\nn
\eeqa
and $\lambda_\vartheta$ is the $\vartheta$-twisted cyclicity operator
$$
(\lambda_\vartheta\tau)(f_0,f_1,f_2) :=
\tau\big(\vartheta(f_2)\,,\,f_0\,,\,f_1\big) \ .
$$
The non-triviality means that there is no twisted cyclic one-cochain
$\alpha$ on $\Apq$ such that $\cob_\vartheta \alpha = \tau$ and
$\lambda_\vartheta \alpha= \alpha$, where here the operators
$\cob_\vartheta$ and $\lambda_\vartheta$ are defined by formul{ae} like
those above (and directly generalize in any degree). Thus $\tau$ is a
class in $\mathrm{HC}^2_\vartheta(\pq)$, the degree two twisted cyclic
cohomology of the quantum space $\pq$.

In terms of the projections $\qpp$, the curvature \eqref{gcur} of the
connection \eqref{coder2d} is given by
\beq\label{curv}
F_{\nablaCP_n} := \qpp ~ \dd \qpp \wedge\, \dd \qpp = - q^{n+1}\, [n]~
\qpp\, \beta_{-}\wedge\beta_{+} \ .
\eeq
Using the normalization $H(1)=1$ for the Haar state on $\Apq$, its
integral (\ref{int}) is computed to be
\beq\label{intcurv}
q^{-1}\, \int_\pq\, \qtr \big(F_{\nablaCP_n} \big)  = - [n]\ .
\eeq
Here $\qtr$ stands for the twisted or `quantum' trace defined as
follows~\cite{wa07}. Given an element $M \in \Mat_{|n|+1}(\Apq)$, its
(partial) quantum trace is the element $\qtr(M) \in \Apq$ defined by 
$$
\qtr(M):= \tr\left(M \, \sigma_{|n|/2} (K^{2})\right)
=\sum_{j,l=0}^{|n|}\, 
M_{jl} \, \left(\sigma_{|n|/2}(K^{2})\right)_{lj} \ ,
$$
where $\sigma_{|n|/2} (K^{2})$ is the matrix form of the spin
$J=|n|/2$ representation of the modular element $K^2$. In particular,
$\tr_q\big(\qpp\big)=q^{-n}$. The $q$-trace is
`twisted' by the automorphism~$\vartheta$,
$$
\qtr(M_1 \,M_2 ) = \qtr\left( (M_2 \rt K^{2})\, M_1 \right) = 
\qtr\big( \vartheta(M_2)\, M_1 \big) \ .
$$
{}From the definition \eqref{twcc} of the $\vartheta$-twisted cyclic
two-cocycle $\tau$ and the expression \eqref{curv} of the curvature
$F_{\nablaCP_n}$, the integral \eqref{intcurv} is also found to
coincide with the coupling of the cocycle $\tau$ to the projection
$\qpp$ as
\beq\label{qind}
\big( 2 q^{-1} \, \tau\big) \circ  \qtr
\big(\qpp\,,\,\qpp\,,\,\qpp\big) = -[n] \ .
\eeq

The pairing in \eqref{ind} is the index of the Dirac operator on
$\pq$. In parallel, the pairing in \eqref{qind} can be
obtained~\cite{NT,wa07} as the $q$-index of the same Dirac operator,
i.e. the difference between the quantum dimensions of its kernel and
cokernel computed using $\qtr$. Thus the $q$-integer (\ref{intcurv})
may be naturally regarded as a quantum Fredholm index computed from
the pairing between the $\vartheta$-twisted cyclic cohomology and the
(Hopf algebraic) $\SU$-equivariant K-theory
$\K^{\su}_0(\pq)$~\cite{NT,wa07}.

\section{Dimensional reduction of invariant gauge
  fields\label{se:dimred}}

For a smooth manifold $M$, let $\man$ denote the quantum space
$\pq\times M$. By this we mean the family of quantum projective lines
$\pq\times\{p\}\simeq\pq$ parametrized by points $p\in M$. Let
$\ca(M) = C^\infty(M)$ be the commutative algebra of smooth functions on $M$. 
Then the algebra of $\man$ is given by 
$$
\ca(\man):=\Apq\otimes\ca(M) \ .
$$ 
Using the connections on the quantum principal bundle over $\pq$ given
in \S\ref{se:cotqpb}, we will now construct invariant
connections on $\SU$-equivariant modules over the algebra $\ca(\man)$
and describe their dimensional reduction over $\pq$.

\medskip

\subsection{Dimensional reduction of $\SU$-equivariant vector
  bundles}\label{se:dimredbun}~\\[5pt]
We start by giving a coaction of the quantum group $\SU$ on
$\ca(\man)$, by coacting trivially on $\ca(M)$ and with the canonical
coaction $\Delta_L$ on $\ca(\pq)$ given in \eqref{cosq}. This gives a
map defined by
\beqa
\DeltaX \,:\, \ca(\man)  &~\longrightarrow~& \ASU\otimes\ca(\man) \ ,
\nn \\  
b\otimes f &~\longmapsto~& m_{13}\big( \Delta_L(b) \otimes (1\otimes
f) \big) = b_{(-1)} \otimes b_{(0)} \otimes f
\label{coxsq}\eeqa
for $b\in\Apq$, $f\in\ca(M)$, where we use the Sweedler-like notation
$\Delta_L(b) = b_{(-1)} \otimes b_{(0)}$ (with implicit summation),
and $m_{13}$ denotes multiplication in the first and third factors of
the tensor product. 
In parallel with the description \eqref{qsp-ps} of the two-sphere
algebra $\Apq$ as the subalgebra of invariant elements in $\ASU$,
there is an analogous description of the algebra $\ca(\man)$ in terms
of invariant elements in  $\ASU \otimes \ca(M)$. For this, we let
$\U(1)$ act trivially on $\ca(M)$ with corresponding map 
\beq\label{canactbis}
\widetilde{\alpha}_z \, : \, \U(1) ~ \longrightarrow ~ \Aut\big(\ASU
\otimes \ca(M) \big) \ , \qquad \widetilde{\alpha}_z(x\otimes f) =
\alpha_z(x)\otimes f \ ,
\eeq
with $\alpha_z$ the $\U(1)$-action on $\ASU$ given in
\eqref{canact}--\eqref{rco}. It is then evident that
\begin{align}\label{axinv}
\ca(\man)  = \big( \ASU\otimes\ca(M) \big)^{\U(1)} :=
\big\{\fX\in\ASU\otimes\ca(M)~\big|~
\widetilde{\alpha}_z(\fX)=\fX\big\} \ . 
\end{align}
It is also useful to regard the algebra $\ca(M)$ itself as coming from $\ca(\man)$ via a projection related to the 
map $\pi$ in \eqref{qprp} that establishes the `quantum group' $\ca(\U(1))$ as a quantum subgroup of $\ASU$. 
Indeed, by restricting $\pi$ to the subalgebra $\Apq \subset \ASU$ one gets a one-dimensional representation 
\beq
\pi : \Apq  \longrightarrow \IC \ , \qquad \pi(B_{-}) = \pi(B_{+}) = \pi(B_{0}) = 0
\eeq
on the generators and $\pi(1)=1$, which is none other that the counit $\epsilon$ restricted to $\Apq$.
We then have a surjective algebra homomorphism
\beq\label{suralg}
\widetilde{\pi} = \pi \otimes \id \, : \, \ca(\man) ~ \longrightarrow ~ \ca(M) \ , 
\qquad x \otimes f ~ \longmapsto ~ \epsilon(x) f \ . 
\eeq
A right $\ca(\man)$-module $\ceX$ is said to be $\SU$-equivariant if it
carries a left coaction 
$$
\delta \, : \, \ceX ~ \longrightarrow ~ \ASU\otimes\ceX 
$$ 
of the Hopf algebra $\ASU$ which is compatible with the coaction
$\DeltaX$ of $\ASU$ on~$\ca(\man)$,
$$
\delta(\varphi\cdot\fX) = \delta(\varphi)\cdot\DeltaX(\fX) \qquad
\mbox{for all} \quad \varphi \in \ceX \ , \  \fX\in\ca(\man) \ .
$$
Similarly, one defines $\SU$-equivariant left $\ca(\man)$-modules. The
remainder of this section is devoted to relating $\ASU$-equivariant
bundles $\ceX$ on the quantum space $\man$ to $\U(1)$-equivariant
bundles $E$ over the manifold~$M$.

Let $E\to M$ be a smooth, $\U(1)$-equivariant complex vector bundle,
with $\U(1)$ acting trivially on $M$. This induces an action $\rho$ of
the group $\U(1)$ on the (right) $\ca(M)$-module 
$\ce = C^\infty(M,E)$ of smooth sections of the bundle $E$, making it
$\U(1)$-equivariant. By the classical Serre--Swan theorem, the module
$\ce$ is a finitely-generated (right) projective module over
$\ca(M)$. Consider now the space $\ceX$ of equivariant  elements,
generalizing those in \eqref{coeq}, given by
\begin{equation}\label{bundlecoeq}
\ceX=\ASU \boxtimes_\rho \ce :=\big\{ \varphi  \in  \ASU
\otimes \ce ~\big|~
(\alpha \otimes \id)\varphi = \big((\id \otimes \rho^{-1} )\big)
\varphi \big\} \ . 
\end{equation}
There is a natural $\SU$-equivariance. Again the left coaction
$\Delta$ of $\ASU$ on itself extends naturally to a left coaction on 
$\ASU \boxtimes_\rho \ce$ given by
$$
\Delta^\rho= \Delta \otimes \id\,:\, \ASU \boxtimes_\rho
\ce~\longrightarrow ~ \ASU 
\otimes \big(\ASU \boxtimes_\rho \ce\big) \ .
$$
This coaction is naturally compatible with the corresponding
$\SU$-coaction in \eqref{coxsq}. The space (\ref{bundlecoeq}) is an
$\ca(\man)$-bimodule. Any $\varphi\in \ASU 
\boxtimes_\rho \ce$ can be written as $\varphi = \varphi^{(1)}
\otimes \varphi ^{(2)}$ with $\varphi^{(1)}\in\ASU$ and
$\varphi^{(2)}\in\ce$ (and an implicit sum understood). Then
the bimodule structure is given as 
$$
(b \otimes f)\,\big(\varphi^{(1)} \otimes \varphi ^{(2)}\big)  = 
\big(b\, \varphi^{(1)}\big) \otimes \big(f\, \varphi ^{(2)}\big) \quad
\mbox{and} \quad  
\big(\varphi^{(1)} \otimes \varphi ^{(2)}\big) \,(b \otimes f)   = 
\big(\varphi^{(1)}\, b \big) \otimes \big(\varphi ^{(2)}\, f\big) 
$$
for $b \otimes f\in\Apq \otimes\ca(M)$. As a right (or left)
$\ca(\man)$-module, it is finitely-generated and projective when it is
defined with the tensor product of modules $\ce$ which are
finitely-generated and projective, respectively. 

Conversely, let $\ceX$ be a finitely-generated $\SU$-equivariant right
(or left) projective $\ca(\man)$-module. The surjective algebra
homomorphism $\widetilde{\pi}:\ca(\man)\to\ca(M)$ in \eqref{suralg} (together with
the quantum group surjection in \eqref{qprp}) induces a map sending
$\ca(\man)$-modules to $\ca(M)$-modules, with a residual coaction of
the `quantum group' $\ca(\U(1))$ which is trivial on $\ca(M)$. From
$\ceX$ we obtain one such module $\ce$, such that the coaction of
$\ca(\U(1))$ is an action of $\U(1)$ on $\ce$. Again by the Serre--Swan
theorem, $\ce$ is the $\ca(M)$-module of smooth sections 
$\ce=C^\infty(M,E)$
of a complex vector bundle $E\to M$ which is
equivariant with respect to the action of $\U(1)$ lifting the trivial
action on $M$.

An alternative way to understand this correspondence between
$\SU$-equivariant modules over $\ca(\man)$ and $\U(1)$-equivariant
bundles over $M$ is as follows. Given $p\in M$, consider the
evaluation map ${\rm ev}_p:\ca(M)\to\IC$ defined by ${\rm
  ev}_p(f)=f(p)$ for $f\in\ca(M)$. By $\U(1)$-equivariance, it induces
a surjective algebra homomorphism ${\rm ev}_p:\ca(\man)\to\Apq$. Let
$\ceX$ be a finitely-generated $\SU$-equivariant projective right (or
left) $\ca(\man)$-module. Then the surjection ${\rm ev}_p$ induces a
finitely-generated $\SU$-equivariant projective right (or left)
module $\ceX_p$ over $\Apq$. We may in this way regard $\ceX$ also as
a family of finitely-generated $\SU$-equivariant projective right
(or left) $\Apq$-modules $\ceX_p$ of the type described
in~\S\ref{se:avb}, parametrized by points $p\in M$. The module
$\ceX_p$ is in correspondence with the representations of $\U(1)$ 
via the construction of~\S\ref{se:avb}, and admits a decomposition
(\ref{su2deco}) into irreducible rank~one modules (\ref{libu}). 

We are now ready to formulate the fundamental statement of dimensional 
reduction, which will enable us to think of $\ceX=\ASU
\boxtimes_\rho\ce$ as the module of sections of an $\SU$-equivariant
vector bundle on $\pq \times M $. We begin with the following
preliminary decomposition. 
\begin{lemm}
Let $M$ be a smooth manifold with trivial $\U(1)$-action. Let $C_n$,
$n\in \IZ$, denote the irreducible $\U(1)$-module of weight $n$ as
given in \eqref{ircore}. Then every $\U(1)$-equivariant
$\ca(M)$-bimodule $\ce$ is isomorphic to a finite direct sum  
\begin{equation}\label{vb-dec} 
\ce \simeq \bigoplus_{n\in W(\ce)} \,C_n \otimes\ce_n  \ ,
\end{equation}
where $W(\ce) \subset \IZ$  is the set of eigenvalues for the
$\U(1)$-action on $\ce$, and $\ce_n$ are $\ca(M)$-bimodules with trivial $\U(1)$-coaction. If
$\ce$ is finitely-generated (resp. projective) then the modules $\ce_n$ are
also finitely-generated (resp. projective). 
\label{lem:vb-dec}\end{lemm}
\begin{proof}
Denote by $\cc_n$, with $n\in\IZ$, the $\ca(M)$-bimodule of sections
of the trivial bundle over $M$ with typical fibre $C_n$. It is
naturally $\U(1)$-equivariant. 
Using the decomposition \eqref{repdeco} of a generic
finite-dimensional representation $(V,\rho)$ for $\U(1)$, the dual formulation
of~\cite[Prop.~1.1]{A-CG-P1} then gives a finite isotopical
decomposition
$$ 
\ce \simeq \bigoplus_{n\in W(\ce)} \,\cc_n \otimes_{\ca(M)} \ce_n \ ,
$$ 
where $W(\ce) \subset \IZ$ is the set of eigenvalues for the
$\U(1)$-action on $\ce$, so that  
$$
W(\ce) = \big\{n\in\IZ ~\big|~\ce_n \not= 0\big\}
$$ 
are the weights of $\ce$, and $\ce_n=\Hom_{\U(1)}(\cc_n,\ce)$ are $\ca(M)$-bimodules with
trivial $\U(1)$-action. Since $\cc_n$ is associated to the trivial
bundle, it is of the form $\cc_n\simeq C_n \otimes\ca(M)$ and the
decomposition (\ref{vb-dec}) follows. 
\end{proof}
\begin{prop}
Every finitely-generated $\SU$-equivariant projective bimodule
$\ceX$ over $\ca(\man)$ can be equivariantly decomposed, uniquely up
to isomorphism, as
\beq
\ceX = \bigoplus_{i=0}^m\,\ceX_i = 
\bigoplus_{i=0}^m\,\cl_{m-2i}\otimes\ce_i
\label{decomp}\eeq
for some $m\in\IN_0$, where $ \ce_i$ are bimodules of sections of
smooth vector bundles $E_i$ over $M$ with trivial $\SU$ coactions and 
$\cl_{n}$ are bimodules \eqref{libu} of sections of the
$\SU$-equivariant line bundles over $\pq$, together with morphisms
$$
\Phi_i\in\Hom_{\ca(\man)}(\ceX_{i-1},\ceX_i) \ , \qquad
i=1,\dots,m
$$ 
of $\ca(\man)$-bimodules.
\label{prop:dimred}\end{prop}
\begin{proof}
Since the $\U(1)$-action on $\ca(M)$ is trivial, by Lemma~\ref{lem:vb-dec}
we have that every $\U(1)$-equivariant $\ca(M)$-bimodule $\cf$ is
isomorphic to a finite direct sum  
$$ 
\cf \simeq\bigoplus_{n\in W(\cf)}\,C_n \otimes \cf_n \ ,
$$ 
where $W(\cf) = \big\{n\in\IZ ~\big|~\cf_n \not= 0\big\}$ 
are the weights of $\cf$ for the $\U(1)$-action, 
and $\cf_n=\Hom_{\U(1)}(\cc_n,\cf)$ are
$\ca(M)$-bimodules with trivial $\U(1)$-action. Putting this together
with the decomposition \eqref{su2deco} in terms of the line bundles
$\cl_n$, we arrive at a decomposition for the corresponding induced
bimodule over $\ca(\man)$ given by 
$$
\cfX  =  \ASU \boxtimes_\rho \cf  =  \bigoplus_{n\in W(\cf)} \cl_n
\otimes \cf_n  \ .
$$
This decomposition describes the $\U(1)$-action on $\cfX$. The rest of
the left $\SU$-coaction is incorporated by using the dual right
$\su$-action. From (\ref{lellb}) the latter leaves each line bundle $\cl_n$ alone but this is not the case for the bimodules $\cf_n$. From relations 
(\ref{relsu}) the right action of $E$ sends  $\cl_n\otimes\cf_n$ to $\cl_n\otimes\cf_{n-2}$ with corresponding
$\varphi_n:\cf_n\to\cf_{n-2}$ that are $\ca(M)$-bimodule morphisms. In particular, every indecomposable
bimodule $\cfX$ has weight set of the form
$W(\cfX)=\{m_-,m_-+2,\dots,m_+-2,m_+\}$ consisting of consecutive even
or odd integers. By defining $m=\frac12\,(m_+-m_-)$,
$\ceX=\cl_{-m_--m}\otimes\cfX$, $\ce_i=\cf_{m_+-2i}$, and
$\ceX_i=\cl_{m-2i}\otimes\ce_i$, we find that the $K$-action is given
by (\ref{decomp}), while the $E$-action is determined by a
\emph{chain} of $\ca(M)$-bimodule morphisms
$$
\ce_0~\xrightarrow{\phi_1}~\ce_1~\xrightarrow{\phi_2}~
\cdots~\xrightarrow{\phi_m}~\ce_m
$$
with $\phi_i:=\varphi_{m_+-2i}$. By fixing $\ca(M)$-valued hermitian
structures $h_i:\ce_i\times\ce_i\to\ca(M)$ on the modules $\ce_i$, the
action of $F=E^*$ is given by the adjoint morphisms
$\phi_i^*$ in $\Hom_{\ca(M)}(\ce_i,\ce_{i-1})$.
\end{proof}

\medskip

\subsection{Covariant hermitian structures
}\label{se:dimredherm}~\\[5pt]
We will now give a gauge theory formulation of the equivalence
between the $\SU$-equivariant bundles over
$\ca(\man)=\ca(\pq)\otimes\ca(M)$ and the module chains over $\ca(M)$
described in Proposition~\ref{prop:dimred}. We first describe the reduction of
$\SU$-covariant hermitian structures on the $\SU$-equivariant
bimodules $\ceX$ of \S\ref{se:dimredbun}. On each line bundle $\cl_n$,
there is the $\Apq$-valued hermitian structure defined in
\eqref{herstr}. Since we require an element in $\Apq$, any two
modules $\cl_n$ and $\cl_m$ with $m\neq n$ are taken to be
orthogonal.

Let $\ceX$ be a finitely-generated $\SU$-equivariant projective
right module over the algebra $\ca(\man)$, with corresponding
equivariant decomposition (\ref{decomp}). On each $\ca(M)$-module
$\ce_i$ in this decomposition we fix an $\ca(M)$-valued hermitian
structure
$$
h_i\,:\,  \ce_i \times \ce_i ~\longrightarrow~\ca(M) \ .
$$
Combined with (\ref{herstr}) this gives an $\ca(\man)$-valued hermitian
structure on $\ceX_i$ defined by
$$
\hX_i = \hat h_{m-2i} \otimes h_i\,:\, \ceX_i  \times \ceX_i
~\longrightarrow~ \Apq \otimes\ca(M) \ ,
$$
and in turn a left $\SU$-covariant hermitian structure on $\ceX$ by
\begin{equation}\label{eq-metric}
\hX = \bigoplus_{i=0}^m\, \hX_{i}  \,:\, \ceX \times \ceX
~\longrightarrow~ \ca(\man) \ .
\end{equation}
By construction, the modules $\ceX_i$, $i=0,1,\dots,m$ are
$\SU$-covariantly mutually orthogonal, i.e. $\hX(\ceX_i,\ceX_j)=0$ for
$i\not=j$.

\medskip

\subsection{Decomposition of covariant  connections
}\label{se:dimredcon}~\\[5pt]
Denote the left-covariant calculus on $\Apq$ constructed in
\S\ref{se:cals2} by $(\Omega^1(\pq), \hat\dd)$. Let
$(\Omega^1(M),\dd)$ be the standard $*$-calculus on $\ca(M)$, with
$\Omega^1(M)$ the vector space of (complex) differential one-forms and
$\dd$ the usual de~Rham exterior derivative on the smooth
manifold~$M$. Then we define a calculus $(\Omega^1(\man),\,\underline{\dd}\,)$ on
$\ca(\man)=\Apq\otimes\ca(M)$ by 
$$
\Omega^1(\man) = \big(\Omega^1(\pq) \otimes\ca(M)\big)\oplus\big(
\Apq \otimes \Omega^1(M)\big) \qquad \mbox{and} \qquad
\,\underline{\dd}\, = \hat\dd \otimes \id + \id \otimes \dd \ .
$$
Let $\ceX$ be a finitely-generated $\SU$-equivariant projective
right $\ca(\man)$-module. Then we define
$$
\Omega^1(\ceX) = \ceX\otimes_{\ca(\man)} \Omega^1(\man) \ ,
$$
and from the equivariant decomposition \eqref{decomp}, 
$
\ceX = \bigoplus_{i}\,\ceX_i = \bigoplus_{i}\,\cl_{m-2i}\otimes\ce_i 
$, we get a
corresponding decomposition
$$
\Omega^1(\ceX)   =  \bigoplus_{i=0}^m\, \Omega^1(\ceX_i)
$$
with
$$
\Omega^1(\ceX_i) \= \ceX_i \otimes_{\ca(\man)} \Omega^1(\man) \
\simeq \ \big(\Omega^1(\cl_{m-2i}) \otimes \ce_i \big)
\oplus \big(\cl_{m-2i} \otimes \Omega^1(\ce_i)\big) \ ,
$$
and obvious notations $\Omega^1(\cl_{m-2i})=\cl_{m-2i} \otimes_{\Apq}
\Omega^1(\pq)$ and $\Omega^1(\ce_i)=\ce_i \otimes_{\ca(M)}
\Omega^1(M)$.

A connection on the right $\ca(\man)$-module $\ceX$ is given via a
covariant derivative
$$
\nablaX \,:\, \ceX ~\longrightarrow~ \Omega^1(\ceX)
$$
obeying the Leibniz rule
$$
\nablaX\big( \varphi\cdot(b\otimes f) \big) = (\nablaX
\varphi)\cdot(b\otimes 
f) + \varphi\otimes_{\ca(\man)} \,\underline{\dd}\,(b\otimes f) \ ,
$$ 
for $\varphi\in\ceX$ and $b\otimes f\in \Apq \otimes\ca(M)$. 
The connection is unitary if in addition it is compatible with the
hermitian structure $\hX$ of \S\ref{se:dimredherm}, so that
\beq\label{herconn}
\hX(\nablaX \varphi, \psi) + \hX(\varphi , \nablaX \psi) =
\,\underline{\dd}\, \big(\hX(\varphi,\psi) \big)
\eeq
for $\varphi,\psi \in\ceX$. Here the metric $\hX$ is naturally
extended to a map $\Omega^1(\ceX)\times \Omega^1(\ceX) \to
\Omega^2(\man)$ by the formulae
\begin{align*}
\hX(\varphi\otimes_{\ca(\man)} \eta  , \psi)  =  \eta^*\,
\hX(\varphi , \psi) \qquad \mbox{and} \qquad 
\hX(\varphi ,  \psi\otimes_{\ca(\man)}\xi)  =  \hX(\varphi , \psi)
\, \xi \ , 
\end{align*}
for $\varphi,\psi \in\ceX$ and $\eta,\xi\in\Omega^1(\man)$, which 
respectively define metrics $\Omega^1(\ceX)\times \ceX \to
\Omega^1(\man)$ and $\ceX \times\Omega^1(\ceX)\to \Omega^1(\man)$. For
any $p\geq0$, the connection $\nablaX$ is extended to a $\IC$-linear
map $\nablaX:\Omega^p(\ceX)\to \Omega^{p+1}(\ceX)$ by the
graded Leibniz rule, where
$$
\Omega^p(\ceX)=\bigoplus_{i=0}^m\,\Omega^p(\ceX_i)
$$
with
$$
\Omega^p(\ceX_i)=\big(\cl_{m-2i}\otimes\Omega^p(\ce_i)\big)\oplus
\big(\Omega^1(\cl_{m-2i})\otimes\Omega^{p-1}(\ce_i)\big)\oplus
\big(\Omega^2(\cl_{m-2i})\otimes\Omega^{p-2}(\ce_i)\big)
$$
and $\Omega^0(\ce_i):=\ce_i$.

As usual, for any two connections $\nablaX,\nablaX'$ their difference 
is an element
$$
\nablaX' - \nablaX = \AX \ \in \ \Hom_{\ca(\man)}\big(\ceX \,,\,
\Omega^1(\ceX) \big) \ ,
$$
and if the connections are unitary then the `matrix of one-forms'
$\AX$ is in addition anti-hermitian,
$$
 \hX(\AX \varphi, \psi) + \hX(\varphi, \AX \psi) =0 \, \qquad
 \mbox{for} \quad \varphi \, , \psi \in \ceX \ . 
$$
The collection of anti-hermitian elements in 
$\Hom_{\ca(\man)}(\ceX, \Omega^1(\ceX) )$ will be denoted by
$\Hom^a_{\ca(\man)}(\ceX, \Omega^1(\ceX) )$. The group $\cu(\ceX)$ of
gauge transformations consists of unitary elements in
$\End_{\ca(\man)}(\ceX)$, with respect to the hermitian structure
$\hX$. It acts on a connection $\nablaX$ by
$$
(u, \nablaX) ~ \longmapsto ~ \nablaX^u = u \circ \nablaX \circ u^* \
,
$$
where here $u$ acts implicitly as $u\otimes_{\ca(\man)}\id_{\Omega^1(\man)}$. A
connection $\nablaX_\AX=\nablaX+\AX$ will then transform to
$(\nablaX_\AX)^u=\nablaX+\AX^u$ with 
$$
\AX^u=u\,(\nablaX u^*) + u \, \AX \, u^* \ . 
$$
That each $\cu(\cl_n)\simeq S^1$, the complex numbers of modulus one, means that the part of a gauge transformation in 
$\cu(\ceX)$ acting on the bundles $\cl_{m-2i}$ is trivial. This fact
will be used in \S\ref{se:YMdimred} for the gauge invariance of the
Yang--Mills action functional.

\begin{lemm}
Any unitary connection $\nablaX$ on $(\ceX,\hX)$ decomposes as 
 $$
 \nablaX = \sum_{i=0}^m\,\Big( \nablaX_i + \sum_{j < i}\, \big(
 \betaX_{ji} - \betaX_{ji}^*\big) \Big) \ ,
 $$
where:
\begin{enumerate}
 \item
 Each $\nablaX_i$ is a unitary connection on $(\ceX_i, \hX_{i})$, i.e. 
 $$
 \hX_{i}(\nablaX_i \varphi, \psi) + \hX_{i}(\varphi, \nablaX_i \psi) =
 \,\underline{\dd}\, \big(\hX_{i}(\varphi,\psi) \big) \qquad \mbox{for} \quad
 \varphi,\psi \in \ceX_i \ .
 $$
  \item
 For $j\not=i$, $\betaX_{ji}\in
 \Hom_{\ca(\man)}(\ceX_i,\Omega^1(\ceX_j))$ is the adjoint of
 $-\betaX_{ij}$, i.e.
 $$
 \hX(\betaX_{ji} \varphi, \psi) + \hX(\varphi, \betaX_{ij} \psi) = 0 \,
 \qquad \mbox{for} \quad \varphi \in \ceX_i \ , \ \psi \in \ceX_j \ .
 $$
 \end{enumerate}
 \end{lemm}
 \begin{proof}
Decompose the connection as 
 $$ 
\nablaX=\sum_{i=0}^m\, \nablaX\big|_{\ceX_i} \qquad \mathrm{with} \quad  
 \nablaX\big|_{\ceX_i} \, : \, \ceX_i ~ \longrightarrow ~
 \Omega^1(\ceX)
$$ 
 and 
 $$ 
\nablaX\big|_{\ceX_i} = \sum_{j=0}^m\, \betaX_{j i} \qquad
\mathrm{with} \quad \betaX_{j i} \, : \, \ceX_i ~ \longrightarrow ~
\Omega^1(\ceX_j) \ .
 $$ 
 Then, since $\hX(\ceX_i,\ceX_j)=0$ when $i\not=j$, the unitarity
 condition \eqref{herconn} for $\nablaX$ breaks into pieces giving the
 claimed decomposition with $\nablaX_i = \betaX_{ii}$.
 \end{proof}

In a completely analogous way, one can decompose any given element
$\AX$ of the space $\Hom^a_{\ca(\man)}(\ceX, \Omega^1(\ceX) )$ as  
$$
 \AX = \sum_{i=0}^m\,\Big( \AX_i + \sum_{j < i} \, \big(
 \AX_{ji} - \AX_{ji}^* \big) \Big) \ ,
 $$
with $\AX_i\in\Hom^a_{\ca(\man)}(\ceX_i, \Omega^1(\ceX_i) )$ and
$\AX_{ji}\in\Hom_{\ca(\man)}(\ceX_i, \Omega^1(\ceX_j) )$, leading
to a decomposition
\beqa
\Hom^a_{\ca(\man)}\big(\ceX\,,\, \Omega^1(\ceX) \big) & \simeq &
\bigoplus_{i=0}^m \, \Big( \Hom^a_{\ca(\man)}\big( \ceX_i\,,\,
\Omega^1(\ceX_i) \big) \nn \\ && \qquad\qquad
\oplus \ \bigoplus_{j < i} \, \Hom_{\ca(\man)}\big( \ceX_i \,,\,
\Omega^1(\ceX_j) \big) \Big) \ . \nn
\eeqa

\medskip

\subsection{$\SU$-invariant connections and gauge transformations
}~\\[5pt]
On $\ceX = \bigoplus_i\,\ceX_i = \bigoplus_i\,\cl_{m-2i}\otimes\ce_i$
we denote by  $\Delta_{\ceX}$ the coaction of the Hopf algebra $\ASU$
which combines the natural coaction of $\ASU$ on the modules
$\cl_{m-2i}$ given in \eqref{rec-co} and the trivial coaction on the
modules $\ce_i$,
$$
\Delta_{\ceX} \, : \, \ceX ~ \longrightarrow ~ \ASU \otimes \ceX \ , 
$$
and by  $\Delta_{\ceX}^{(1)}$ its lift to $\Omega^1(\ceX)$, 
$
\Delta_{\ceX}^{(1)}: \Omega^1(\ceX) \to \ASU \otimes \Omega^1(\ceX). 
$
In complete parallel with \S\ref{incogt} there are `adjoint' coactions
of $\ASU$ on the space $\cc(\ceX)$ of unitary connections on $\ceX$,
on the group $\cu(\ceX)$ of gauge transformations, as well as on the
spaces $\Hom_{\ca(\man)}(\ceX_i, \ceX_j)$ and
$\Hom_{\ca(\man)}(\ceX_i, \Omega^1(\ceX_j))$. We shall denote by
$\cc(\ceX)^{\SU}$, etc. the corresponding space of coinvariant elements,
i.e.
$$
\cc(\ceX)^{\SU} = \big\{\nablaX \in \cc(\ceX) \ \big| \
\Delta^{\cc}(\nablaX) = 1 \otimes \nablaX \big\} \ ,
$$
and similarly for the other spaces and coactions. The spaces
$\cc(\ceX)^{\SU}$ and $\cu(\ceX)^{\SU}$ of invariant connections and
gauge transformations are described in terms of objects defined on $M$
and of canonical (and unique) objects defined on $\pq$. We begin with
the space~$\cc(\ceX)^{\SU}$.

\begin{lemm} One has
$$
\left( \Hom^a_{\ca(\man)}( \ceX_i, \Omega^1(\ceX_i) ) \right)^{\SU} =
\IC \otimes \Hom^a_{\ca(M)}( \ce_i , \Omega^1(\ce_i) ) \ ,
$$
while for $i\not= j$ one has
$$
\left( \Hom_{\ca(\man)}( \ceX_i, \Omega^1(\ceX_j) ) \right)^{\SU} =
\begin{cases}
\; 0 & \textup{if}  \quad  j\not= i \pm 1  \ , \\
\; \IC\beta_- \otimes \Hom_{\ca(M)}( \ce_i, \ce_{i-1}) & \textup{if}
\quad  j = i-1  \ , \\ 
\; \IC\beta_+ \otimes \Hom_{\ca(M)}( \ce_i, \ce_{i+1}) & \textup{if}
\quad  j = i+1 \ .
\end{cases}
$$
\label{lem:invformer}\end{lemm}
\begin{proof}
For any $i,j$ one has
\begin{multline*}
\Hom_{\ca(\man)}( \ceX_i , \Omega^1(\ceX_j) ) \simeq
\big(\Hom_{\Apq}( \cl_{m-2i} , \cl_{m-2j}  )  
\otimes \Hom_{\ca(M)}( \ce_i, \Omega^1(\ce_j)) \big) \\ \, \oplus\,
\big( \Hom_{\Apq}( \cl_{m-2i} , \Omega^1(\cl_{m-2j}) ) 
\otimes \Hom_{\ca(M)}( \ce_i, \ce_j) \big)    \,  
\end{multline*}
and, since $\SU$ coacts trivially on the bundles $\ce_i$, for the
coinvariant elements one finds
\begin{multline*}
\left( \Hom_{\ca(\man)}( \ceX_i, \Omega^1(\ceX_j) )\right)^{\SU} \\
\simeq \Big( \big(\Hom_{\Apq}( \cl_{m-2i} , \cl_{m-2j}  ) \big)^{\SU} 
\otimes \Hom_{\ca(M)}( \ce_i , \Omega^1(\ce_j)) \Big) \\
\, \oplus\, 
\Big( \big(\Hom_{\Apq}( \cl_{m-2i} , \Omega^1(\cl_{m-2j}) ) \big)^{\SU}
\otimes \Hom_{\ca(M)}( \ce_i, \ce_j) \Big) \ .  
\end{multline*}
In order to proceed we need only the fact that there are no
non-trivial $\SU$-invariant elements in the modules $\cl_n$ for
$n\not=0$, while $1$ is the only invariant element in the algebra
$\cl_0=\Apq$. Then by Lemma~\ref{masu} one has
\beq\label{text}
\big( \Hom_{\Apq}( \cl_{m-2i} , \cl_{m-2j}  ) \big)^{\SU} \simeq
\left( \cl_{2i-2j} \right)^{\SU} = 
\begin{cases}
\; \IC & \textup{if} \quad 2i=2j \ , \\
\; 0 & \textup{if} \quad 2i\not=2j \ .
\end{cases}
\eeq
On the other hand, using (\ref{homhom}) one finds
\begin{align*}
\big( \Hom_{\Apq}( \cl_{m-2i} , \Omega^1(\cl_{m-2j}) ) \big)^{\SU} &
\simeq \big(\cl_{2i-2j-2} \beta_- \oplus \cl_{2i-2j+2}
\beta_+\big)^{\SU} \\[4pt] & 
\simeq \begin{cases}
\; 0 & \textup{if}  \quad  2i-2j\not= \pm\, 2  \ , \\
\; \IC\beta_-  & \textup{if}  \quad  2i-2j = 2  \ , \\
\; \IC\beta_+ & \textup{if}  \quad  2i-2j = -2 \ ,
\end{cases}
\end{align*} 
and the results now follow.
\end{proof}

Using Lemma~\ref{lem:invformer} and $\beta_-^*=-\beta_+$, an element
$\AX\in \left( \Hom^a_{\ca(\man)}( \ceX, \Omega^1(\ceX) )
\right)^{\SU}$ can be written as
\beq\label{invaut}
\AX = \sum_{i=0}^{m} \, \big( 1 \otimes  A_i + \beta_+ \otimes
\phi_{i+1} +\beta_-\otimes \phi_{i+1}^* \big) \ , 
\eeq
where the `gauge potentials' $ A_i\in \Hom^a_{\ca(M)}( \ce_i,
\Omega^1(\ce_i) )$ and `Higgs fields' $\phi_{i+1}$ in the space
$\Hom_{\ca(M)}(\ce_i, \ce_{i+1})$ for $i=0,1,\dots,m$ with
$\ce_{m+1}:=0$. Now let $(\ceX, \hX)$ be an $\SU$-equivariant
hermitian $\ca(\man)$-module decomposed as in \eqref{decomp} with the
metric $\hX$ decomposed as in \eqref{eq-metric}. Let $(\ce_i, h_i)$
for $i=0,1,\dots,m$ be the hermitian $\ca(M)$-modules composing
$(\ceX,\hX) $, and let $\cc(\ce_i)$ be the corresponding spaces of
unitary connections. 
\begin{prop}\label{brconn}
There is a bijection between the spaces $\cc(\ceX)^{\SU}$ and
$$
\scrc(\ceX) := \prod_{i=0}^m\,\big(\cc(\ce_i) \times \Hom_{\ca(M)}(
\ce_i, \ce_{i+1})\big) 
$$
which associates to any element $(\nabla,\phi)$ of $\scrc(\ceX)$,
given by connections $\nabla_i\in\cc(\ce_i)$ and Higgs fields
$\phi_{i+1}\in \Hom_{\ca(M)}(\ce_i, \ce_{i+1} )$ for $i=0,1,\dots,m$,
the $\SU$-invariant unitary connection $\nablaX\in \cc(\ceX)^{\SU}$
given by
\beq\label{invcon}
\nablaX = \sum_{i=0}^{m}\,\big( \nablaX_i + \beta_+ \otimes \phi_{i+1} 
+ \beta_-\otimes \phi_{i+1}^* \big) \ .
\eeq
Here $\nablaX_i$ is the unitary connection on $(\ceX_i, \hX_i)$ given
by 
$$
\nablaX_i = \widehat{\nabla}_{m-2i} \otimes \id + \id \otimes \nabla_i
\ , 
$$
where $\widehat{\nabla}_{m-2i}$ is the unique (by Lemma~\ref{invcancon})
$\SU$-invariant unitary connection on the hermitian line bundle
$(\cl_{m-2i}, \hat{h}_{m-2i})$ given in \eqref{coder2d} and
\eqref{herstr}.
\begin{proof}
Fix a unitary connection $\nabla_i^{0}\in\cc(\ce_i)$ for each
$i=0,1,\dots,m$. Define unitary connections on each $(\ceX_i, \hX_i)$
by $\nablaX_i^{0} = \widehat{\nabla}_{m-2i} \otimes \id + \id \otimes
\nabla_i^{0}$. They are clearly $\SU$-invariant and give rise to an
$\SU$-invariant unitary connection $\nablaX^{0}$ on $(\ceX, \hX)$
defined by the sum $\nablaX^{0} = \sum_i\,\nablaX_i^{0}$. All
$\SU$-invariant unitary connections $\nablaX$ on $(\ceX, \hX)$ take
the form $\nablaX=\nablaX^{0}+\AX$ with $\AX\in \left(
  \Hom^a_{\ca(\man)}( \ceX, \Omega^1(\ceX) ) \right)^{\SU}$. The
general form of such an element is given in \eqref{invaut}, from which
the expression \eqref{invcon} follows. 
\end{proof}
\end{prop}

Next we give a similar characterization of the space
$\cu(\ceX)^{\SU}$.
\begin{lemm} One has
$$
\left( \Hom_{\ca(\man)}( \ceX_i, \ceX_j  ) \right)^{\SU} =
\begin{cases}
\; \IC \otimes \End_{\ca(M)}( \ce_i ) & \textup{if}  \quad i=j \ ,
\\ \; 0  & \textup{if}  \quad  i\not= j   \ .
\end{cases}
$$
\label{lem:HomA}\end{lemm}
\begin{proof}
For any $i,j$ one has
$$
\Hom_{\ca(\man)}( \ceX_i, \ceX_j  ) \simeq \Hom_{\Apq}( \cl_{m-2i} ,
\cl_{m-2j}  ) \otimes \Hom_{\ca(M)}( \ce_i,  \ce_j)
$$
and, since $\SU$ coacts trivially on the bundles $\ce_i$, for the
invariant elements one finds
$$
\left( \Hom_{\ca(\man)}( \ceX_i,  \ceX_j  )\right)^{\SU} \\ \simeq 
\big( \Hom_{\Apq}( \cl_{m-2i} , \cl_{m-2j}  ) \big)^{\SU} 
\otimes \Hom_{\ca(M)}( \ce_i,  \ce_j)     \ .  
$$
The result now follows from \eqref{text}.
\end{proof}

In the same setting as in Proposition~\ref{brconn}, let $\cu(\ce_i)$ be the
group of gauge transformations corresponding to the hermitian
$\ca(M)$-module $(\ce_i,h_i)$. 
\begin{prop}\label{brgt}
There is a bijection between the groups $\cu(\ceX)^{\SU}$ and
$$
\scru(\ceX):=\prod_{i=0}^m\,\cu(\ce_i) \ ,
$$
which associates to any element $u=(u_0, u_1, \dots,
u_m)\in\scru(\ceX)$ the $\SU$-invariant gauge transformation of
$(\ceX, \hX)$ given by
$$
\uX = \sum_{i=0}^m \, \uX_i 
$$
with $\uX_i = 1 \otimes  u _i \in \cu(\ceX_i)^{\SU} \simeq \IC \otimes
\cu(\ce_i)$.
\end{prop}  
\begin{proof}
This follows from Lemma~\ref{invcangt} and Lemma~\ref{lem:HomA}.
\end{proof}

The group $\cu(\ce_i)$ acts on both spaces
$\Hom_{\ca(M)}( \ce_i,\ce_{i+1})$ and $\Hom_{\ca(M)}( \ce_{i+1},
\ce_i)$ of Higgs fields by
$$
u_i( \phi_{i+1}) = \phi_{i+1} \circ u_i^{-1} \, \qquad \textup{and}
\qquad u_i\big( \phi_{i+1}^*\big) = u_i \circ \phi_{i+1}^* \ ,
$$
for $u_i\in \cu(\ce_i)$ and $\phi_{i+1}\in \Hom_{\ca(M)}( \ce_i,
\ce_{i+1})$, $\phi_{i+1}^*\in \Hom_{\ca(M)}( \ce_{i+1}, \ce_i)$. There
is also an induced action of the group $\scru(\ceX)$ on the space
$\scrc(\ceX)$ of connections. The following result is then immediate.
\begin{prop}\label{brorb}
The bijections between invariant connections and between
invariant gauge transformations of Proposition~\ref{brconn} and Proposition~\ref{brgt},
respectively, are compatible with the actions of the groups of
Proposition~\ref{brgt} on the connections of Proposition~\ref{brconn}, and there is a
bijection between gauge orbits
$$
\cc(\ceX)^{\SU} \, \big/ \, \cu(\ceX)^{\SU} \ \equiv \ \scrc(\ceX) \,
\big/ \, \scru(\ceX) \ .
$$
\end{prop}

\medskip

\subsection{Integrable connections }~\\[5pt]
In the sequel we will need to work with \emph{integrable} connections
as well. Let $M$ be a complex manifold, with the standard complex
structure for the complexified de~Rham differential calculus. Combined
with the complex structure for the differential calculus on $\Apq$
described in Proposition~\ref{2dsph}, we get a natural complex structure for the
calculus on the algebra $\ca(\man)=\Apq\otimes\ca(M)$. If $\nablaX$ is
a connection on the $\ca(\man)$-bimodule $\ceX$ with equivariant
decomposition (\ref{decomp}), then the $(0,2)$-component of its
curvature $F_{\nablaX}^{0,2}$ is an element of
$\Hom_{\ca(\man)}(\ceX,\Omega^{0,2}(\ceX))$, where
$$
\Omega^{0,2}(\ceX)= \bigoplus_{i=0}^m\,
\Big(\big(\Omega^{0,2}(\cl_{m-2i})
\otimes\ce_i\big)\oplus\big(\Omega^{0,1}(\cl_{m-2i})\otimes
\Omega^{0,1}(\ce_i) \big) 
\oplus \big(\cl_{m-2i}\otimes \Omega^{0,2}(\ce_i)\big)
\Big) \ .
$$
The connection $\nablaX$ is then integrable if
$F_{\nablaX}^{0,2}=0$. In this case the pair $(\ceX,\nablaX)$ is a
holomorphic vector bundle~\cite[\S2]{KLvS}.

By \eqref{invcon} an $\SU$-invariant unitary holomorphic connection on
$(\ceX,\hX)$ is of the form
\beq\label{invconhol}
\nablaX^{\dolb} = \sum_{i=0}^{m}\,\big( \nablaX_i^{\dolb}+ \beta_-
\otimes \phi_{i+1}^* \big) \ , 
\eeq
where $\nablaX^{\dolb}_i$ is the holomorphic connection on $(\ceX_i,
\hX_i)$ given by
$$
\nablaX^{\dolb}_i = \widehat{\nabla}\,^{\dolb}_{m-2i} \otimes \id
+ \id \otimes \nabla_i^{\dolb}
$$
with $\widehat{\nabla}^{\dolb}_{m-2i}$ the unique
$\SU$-invariant unitary holomorphic connection on the hermitian line
bundle $(\cl_{m-2i}, \hat{h}_{m-2i})$ given in \eqref{coher}, and
$\nabla^{\dolb}_i$ is a holomorphic unitary connection on $(\ce_i,
h_i)$. As before the Higgs fields $\phi^*_{i+1}\in \Hom_{\ca(M)}( \ce_{i+1},
\ce_i)$. Its curvature is readily found to be
\beq\label{curhol}
F_{\nablaX}^{0,2} := \big(\nablaX^{\dolb}\,\big)^2 = \sum_{i=0}^{m}\,
\Big( \id \otimes \big(\nabla^{\dolb}_i\,\big)^2 + \beta_- \otimes 
\big( \phi_{i+1}^* \circ \nabla^{\dolb}_{i+1} - \nabla_i^{\dolb}
\circ  \phi_{i+1}^* \big) \Big) \ , 
\eeq
where we have used
$\big(\,\widehat{\nabla}\,^{\dolb}_{m-2i}\big)^2=0$ and
$\beta_-\wedge\beta_-=0$.

There is a natural coaction of the quantum group $\SU$ on the
collection $\cc(\ceX)^{1,1}$ of integrable unitary connections on
$(\ceX,\hX)$, obtained by restricting the `adjoint' coaction
$\Delta^\cc$ of $\ASU$. Let $\big(\cc(\ceX)^{1,1}\big)^{\SU}$ be the
$\SU$-invariant subspace. This is the space of holomorphic structures
on $\ceX$~\cite{KLvS} for which the coaction of the Hopf algebra
$\ASU$ is holomorphic. Let  $\cc(\ce_i)^{1,1}$ be the collection of
integrable unitary connections on $(\ce_i, h_i)$.

\begin{prop}\label{brholconn}
Let $\scrc(\ceX)^{1,1}$ be the subspace of $\scrc(\ceX)$ consisting of
integrable connections $\nabla^{\dolb}_i \in\cc(\ce_i)^{1,1}$ and
Higgs fields $\phi^*_{i+1}\in\Hom_{\ca(M)}(\ce_{i+1}, \ce_i)$ for
$i=0,1,\dots,m$ on which the holomorphic connection
$\nabla_{i+1,i}^{\dolb}$ on $\Hom_{\ca(M)}( \ce_{i+1}, \ce_i)$
induced by $\nabla^{\dolb}_{i+1}$ and $\nabla^{\dolb}_{i}$
vanishes, 
$$
\nabla_{i+1,i}^{\dolb}(\phi_{i+1}^*) := \phi^*_{i+1} \circ
\nabla_{i+1}^{\dolb} - \nabla^{\dolb}_i \circ  \phi_{i+1}^* = 0 \ .
$$
Then the bijection of Proposition~\ref{brconn} defines a bijection between the
spaces $\big(\cc(\ceX)^{1,1}\big)^{\SU}$ and $\scrc(\ceX)^{1,1}$.
\end{prop}
\begin{proof}
This is a direct consequence of the expression \eqref{curhol} for the 
curvature.
\end{proof}

\section{Quiver gauge theory and non-abelian coupled $q$-vortex equations}

In the remainder of this paper we will assume that $M$ is a connected 
K\"ahler manifold of complex dimension $d$, with fixed K\"ahler form
$\omega\in\Omega^{1,1}(M)$. We then proceed to work out the equivariant
dimensional reduction of Yang--Mills theory defined on the quantum 
space $\man=\pq\times M$. This will produce a $q$-deformation of the
usual quiver gauge theories on
$M$ associated to the linear A$_{m+1}$ quiver~\cite{A-CG-P1,A-CG-P3,PS,LPS1,LPS2,BS}. The vacuum states of the resulting  quiver gauge theory are described by
$q$-deformations of chain vortex equations, which arise by dimensional
reduction of BPS-type gauge theory equations on $\man$ and whose solutions
we call `$q$-vortices'. The $q$-vortices on the manifold $M$ are in a
bijective correspondence with generalized instantons on the quantum
space $\man$. The data in the space $\scrc(\ceX)$ of Proposition~\ref{brconn}, defining a $q$-vortex will be
refered to as a `stable $q$-quiver bundle' over $M$. In contrast to
the $q=1$ case, the degree of a $q$-quiver bundle is generically
non-zero, and there are $q$-vortices which are realized as zeroes of
the quiver gauge theory action functional. In fact, we will find that the $q$-deformation of quiver
bundles over the manifold $M$ is analogous in some ways to the twistings of quiver
bundles considered in~\cite{A-CG-P2}. In particular, we will find
analogous constraints on the
characteristic classes of stable $q$-quiver bundles over $M$, which
can be used to naturally construct \emph{flat} connections on
$M$. Henceforth we fix a deformation parameter $0<q<1$ and an integer
$m\geq0$ parametrizing an $\SU$-equivariant decomposition as in~(\ref{decomp}).

\medskip

\subsection{Metrics on $\SU$-equivariant vector
  bundles }\label{se:metrics}~\\[5pt]
In the following we will make use of various $\SU$-invariant metrics
defined on the equivariant modules over $\ca(\man)$ considered in
\S\ref{se:dimred}. We start by defining
a natural Hodge duality operator on the forms
$$
\Omega^p(\man)=\big(\ca(\pq)\otimes\Omega^p(M)\big)\oplus\big(
\Omega^1(\pq)\otimes\Omega^{p-1}(M)\big)\oplus\big(
\Omega^2(\pq)\otimes\Omega^{p-2}(M)\big)
$$
with $\Omega^0(M)=\ca(M)$ and
$\Omega^{<0}(M):=0=:\Omega^{>2d}(M)$. Let $\star:\Omega^p(M)\to
\Omega^{2d-p}(M)$ be the Hodge operator corresponding to the K\"ahler
metric of $M$, with $\star1= \frac{\omega^d}{d!}$ and $\star^2=\id$. Using the left-covariant Hodge
operator $\hat\star$ on $\Omega^\bullet(\pq)$ defined in
\S\ref{se:cals2}, we then define the bimodule map
$$
\starX:=\hat\star\otimes\star\,:\,\Omega^p(\man)~\longrightarrow~
\Omega^{2(d+1)-p}(\man)
$$
with $\starX^2=\id$. Using the integration defined in (\ref{int}), we
define an integral
$$
\int_{\man}\,:=\int_{\pq}\,\otimes\int_M\,:\,
\Omega^2(\pq)\otimes\Omega^{2d}(M)~\longrightarrow~\IC
$$
when the integral over $M$ exists. We set $\int_{\man}\,\alpha:=0$
whenever $\alpha\notin\Omega^2(\pq)\otimes\Omega^{2d}(M)$. One then
introduces a complex inner product on $\Omega^p(\man)$ for each
$p\geq0$ by
\beq
(\alpha,\alpha'\,)_{\Omega^p(\man)}:=\int_{\man}\,
\alpha^*\wedge\starX\alpha'
\label{Ompmetric}\eeq
for $\alpha,\alpha'\in\Omega^p(\man)$. Forms of different degrees are
defined to be orthogonal. This is a natural generalization of analogous inner products
$(-,-)_{\Omega^p(M)}$ for each $p\geq0$ defined via the Hodge operator  $\star$ on $M$.

Let $\ceX$ be a finitely-generated $\SU$-equivariant projective
bimodule over the algebra $\ca(\man)$, with equivariant decomposition
(\ref{decomp}). Given hermitian structures on its components
$h_i:\ce_i\times\ce_i\to\ca(M)$, we define $L^2$-metrics and
$L^2$-norms on the modules of sections $\ce_i$ for each
$i=0,1,\dots,m$ by
$$
(\varphi_i,\psi_i)_{h_i}=\int_M\,h_i(\varphi_i,\psi_i)\,
\frac{\omega^d}{d!} \qquad \mbox{and} \qquad
\|\varphi_i\|_{h_i}=(\varphi_i,\varphi_i)_{h_i}^{1/2}
$$
for $\varphi_i,\psi_i\in\ce_i$. The hermitian structures $h_i$ also
induce a metric $h_{i,i+1}$ on each of the spaces of Higgs fields
$\Hom_{\ca(M)}(\ce_i,\ce_{i+1})$ as follows. Since each bimodule
$\ce_i$ is finitely-generated and projective, it is of the form
$\ce_i=\qP_i(\ca(M))^{n_i}$ for some $n_i\in\IN$ and a projection
$\qP_i\in\Mat_{n_i}(\ca(M))$; then
$\End_{\ca(M)}(\ce_i)\simeq\qP_i\Mat_{n_i}(\ca(M))\qP_i$. We
denote by $\tr$ the partial matrix trace over `internal indices' of
the endomorphism algebra $\End_{\ca(M)}(\ce_i)$ and define
$$
h_{i,i+1}(\phi_{i+1},\phi_{i+1}')=\tr\big(\phi_{i+1}^*\circ
\phi_{i+1}'\big)
$$
for $\phi_{i+1},\phi_{i+1}'\in\Hom_{\ca(M)}(\ce_i,\ce_{i+1})$, where
  $\phi_{i+1}^*:\ce_{i+1}\to\ce_i$ is the adjoint morphism of the
  Higgs field $\phi_{i+1}:\ce_i\to\ce_{i+1}$ with respect to the
  hermitian metrics $h_i$ on $\ce_i$ and $h_{i+1}$ on $\ce_{i+1}$. The
  associated $L^2$-inner products and $L^2$-norms are obtained by
  integrating the hermitian metrics over $M$ to get
$$
(\phi_{i+1},\phi_{i+1}')_{h_{i,i+1}}=\int_M\,
h_{i,i+1}(\phi_{i+1},\phi_{i+1}'\,)\, 
\frac{\omega^d}{d!} \qquad \mbox{and} \qquad
\|\phi_{i+1}\|_{h_{i,i+1}}=(\phi_{i+1},\phi_{i+1})_{h_{i,i+1}}^{1/2} \ .
$$

Using the hermitian structure $\hX$ in (\ref{eq-metric}), we can further define
a complex inner product on the bimodules $\ceX$ over $\ca(\man)$. Let
$\varphi,\psi\in\ceX$ with decompositions
$\varphi=\sum_i\,\hat\varphi_i\otimes\varphi_i$ and
$\psi=\sum_i\,\hat\psi_i\otimes\psi_i$, where
$\hat\varphi_i,\hat\psi_i\in\cl_{m-2i}$ and
$\varphi_i,\psi_i\in\ce_i$. Using the orthogonality of the direct sum
decomposition $\ceX=\bigoplus_i\,\ceX_i$ with respect to $\hX$, we
define an $L^2$-metric and $L^2$-norm on $\ceX$ by 
$$
(\varphi,\psi)_{\hX}=\sum_{i=0}^m\,H\big(\hat h_{m-2i}
(\hat\varphi_i,\hat\psi_i)\big)\,(\varphi_i,\psi_i)_{h_i}
\qquad \mbox{and} \qquad
\|\varphi\|_{\hX}=(\varphi,\varphi)_{\hX}^{1/2} \ ,
$$
where $H$ is the Haar functional on $\ca(\pq)$.

Finally, we define $\SU$-invariant metrics on the spaces
$\Hom_{\ca(\man)}(\ceX,\Omega^p(\ceX))$ for each $p\geq0$. Since
$\ceX$ is finitely-generated and projective, any $\ca(\man)$-linear
map $\ceX\to\Omega^p(\ceX)$ can be regarded as an element in
$\End_{\ca(\man)}(\ceX)\otimes_{\ca(\man)}\Omega^p(\man)$, i.e. as a
matrix with entries in $\Omega^p(\man)$. By composing the hermitian
structure $\hX$ on $\ceX$ with an ordinary (partial) matrix trace
$\tr$ over `internal indices', one constructs an inner product on
$\End_{\ca(\man)}(\ceX)$. By combining this product with the inner
product on $\Omega^p(\man)$ given in (\ref{Ompmetric}), one then
obtains a natural $L^2$-inner product $(-,-)_{\hX}$ on the space
$\Hom_{\ca(\man)}(\ceX,\Omega^p(\ceX))$ with associated $L^2$-norm
$\|-\|_{\hX}$. In an analogous way, one defines $L^2$-inner products
$(-,-)_{h_i}$ on the orthogonal components
$\Hom_{\ca(M)}(\ce_i,\Omega^p(\ce_i))$ with associated $L^2$-norms
$\|-\|_{h_i}$, and $L^2$-inner products and $L^2$-norms
$(-,-)_{h_{i,i+1}}$ and $\|-\|_{h_{i,i+1}}$ on the spaces
$\Hom_{\ca(M)}(\ce_i,\Omega^p(\ce_{i+1}))$.

\medskip

\subsection{Dimensional reduction of the Yang--Mills action functional}\label{se:YMdimred}~\\[5pt]
Let $\cc(\ceX)$ be the space of unitary connections on an
$\SU$-equivariant hermitian $\ca(\man)$-module $(\ceX,\hX)$. The
Yang--Mills action functional $\YM:\cc(\ceX)\to[0,\infty)$ is defined
by 
\beq
\YM(\nablaX)=\big\|F_{\nablaX}\big\|^2_{\hX} \ ,
\label{YMdef}\eeq
where $F_{\nablaX}=\nablaX^2$ is the curvature of the connection
$\nablaX$, regarded as an element of
$\Hom_{\ca(\man)}(\ceX,\Omega^2(\ceX))$. Under a gauge transformation,
i.e. the action of a unitary endomorphism $u\in\cu(\ceX)$ of the
module $\ceX$ on $\cc(\ceX)$, one has
$$
F_{\nablaX^u}=u\,F_{\nablaX}\,u^* \ ,
$$
and by construction the functional (\ref{YMdef}) is gauge invariant,
i.e. $\YM(\nablaX^u)=\YM(\nablaX)$ for all $\nablaX\in\cc(\ceX)$ and
$u\in\cu(\ceX)$. Consequently, the Yang--Mills action functional
descends to a map on gauge orbits $\YM:\cc(\ceX)\big/\,\cu(\ceX)\to
[0,\infty)$.
We have already remarked that the part of a gauge transformation that
acts on the bundles $\cl_{m-2i}$ is trivial. This entails that in proving the 
invariance of the Yang--Mills functional under gauge transformation, there is no problem 
coming from the integral over $\pq$ not being a trace but rather a twisted trace.

\medskip

In this section we compute the restriction of the
functional (\ref{YMdef}) to the corresponding $\SU$-invariant
subspaces $\cc(\ceX)^{\SU}$ and
$\cc(\ceX)^{\SU}\big/\,\cu(\ceX)^{\SU}$.
\begin{prop}
Under the bijection of Proposition~\ref{brconn}, the restriction of the
Yang--Mills action functional $\YM|_{\cc(\ceX)^{\SU}}$ on the
quantum space $\man$ to $\SU$-invariant unitary connections is equal
to the Yang--Mills--Higgs functional $\YMH_{q,m}:\scrc(\ceX)\to 
  [0,\infty)$ on the manifold $M$ defined by
\beqa
\YMH_{q,m}(\nabla,\phi)&=&
\sum_{i=0}^m\,\Big(\,\big\|F_{\nabla_i}\big\|_{h_i}^2
+\big(q^2+1\big)\,\big\|\nabla_{i-1,i}(\phi_i)\big\|^2_{h_{i-1,i}}
   \nn \\ && \qquad \qquad
    +\,\big\|\phi_{i+1}^*\circ\phi_{i+1} -q^2\,\phi_{i}\circ\phi_{i}^*
    -q^{m-2i+1}\,[m-2i]\,\id_{\ce_i}\big\|_{h_i}^2 \, \Big) \ ,
\label{YMHdef}\eeqa
with $\phi_0:=0=:\phi_0^*$ and $\phi_{m+1}:=0=:\phi_{m+1}^*$. Here
$F_{\nabla_i}=\nabla_i^2$ is the curvature of the connection
$\nabla_i\in\cc(\ce_i)$ on $M$, regarded as an element of
$\Hom_{\ca(M)}(\ce_i,\Omega^2(\ce_i))$, while $\nabla_{i-1,i}$ is the
connection on $\Hom_{\ca(M)}(\ce_{i-1},\ce_i)$ induced by
$\nabla_{i-1}$ on $\ce_{i-1}$ and $\nabla_i$ on $\ce_i$ with
\beq
\nabla_{i-1,i}(\phi_i)=\phi_i\circ\nabla_{i-1}-\nabla_i\circ\phi_i \ .
\label{indconn}\eeq
Under the bijection of Proposition~\ref{brorb}, the functional \eqref{YMHdef}
restricts to a map on gauge orbits
$\YMH_{q,m}:\scrc(\ceX)\big/\,\scru(\ceX)\to [0,\infty)$.
\label{prop:YMH}\end{prop}
\begin{proof}
From (\ref{invcon}), any $\SU$-invariant unitary connection
$\nablaX\in\cc(\ceX)^{\SU}$ is of the form
$\nablaX=\sum_i\,(\nablaX_i+\beta_+\otimes\phi_{i+1}+
\beta_-\otimes\phi_{i+1}^*)$ with
$\nablaX_i=\widehat\nabla_{m-2i}\otimes\id+ \id\otimes\nabla_i$. 
Thus its curvature $F_{\nablaX}=\nablaX\circ\nablaX$ is given by
\beqa
F_{\nablaX}&=&
\sum_{i=0}^m\,\Big(\nablaX_i^2+ \beta_+\otimes\nabla_{i-1,i}(\phi_i)-
\beta_-\otimes\nabla_{i+1,i}(\phi_{i+1}^*) \nn \\ && \qquad \qquad
+\, (\beta_+\wedge\beta_-)\otimes(\phi_{i}\circ\phi_{i}^*)+
(\beta_-\wedge\beta_+)\otimes(\phi_{i+1}^*\circ\phi_{i+1}) \nn \\ &&
\qquad \qquad \qquad \qquad +\, (\beta_+\wedge\beta_+)\otimes
(\phi_{i+1}\circ\phi_{i})-
(\beta_-\wedge\beta_-)\otimes(\phi_{i}^*\circ\phi_{i+1}^*) \Big) \
. 
\label{curvcalc}\eeqa
A straightforward computation gives
$$
\nablaX_i^2=\fsf_{m-2i}\otimes\id+\id\otimes F_{\nabla_i} \ ,
$$
where $\fsf_{m-2i}=\nablaCP_{m-2i}^2$ is the curvature (\ref{gpot}) of
the canonical connection on the $\SU$-equivariant line bundle
$\cl_{m-2i}$ over $\pq$. By (\ref{gcur}) one has
$$
\fsf_{m-2i}= -q^{m-2i+1}\,[m-2i]\,\beta_-\wedge\beta_+
$$
as an element of $\Hom_{\ca(\pq)}(\cl_{m-2i},
\Omega^2(\cl_{m-2i}))$. By (\ref{commc3})
one has $\beta_-\wedge\beta_-=0=\beta_+\wedge\beta_+$ and
$\beta_+\wedge\beta_-= -q^2\,\beta_-\wedge\beta_+$. Substituting
everything into (\ref{curvcalc}), we arrive at
\beqa
F_{\nablaX}&=&
\sum_{i=0}^m\,\Big(\id\otimes F_{\nabla_i}+
\beta_+\otimes\nabla_{i-1,i}(\phi_i)-
\beta_-\otimes\nabla_{i+1,i}(\phi_{i+1}^*) \nn \\ &&  \qquad \qquad
+\, (\beta_-\wedge\beta_+)\otimes\big(\phi^*_{i+1}\circ\phi_{i+1}-
q^2\,\phi_{i}\circ\phi_{i}^*- q^{m-2i+1}\,[m-2i]\,\id\big) \Big) \
.
\label{curvdimred}\eeqa

We now use the definition of the Hodge operator and integral on $\man$
from \S\ref{se:metrics}, together with $\hat\star\beta_\pm=
\pm\,\beta_\pm$ and (\ref{hod2}), and the $*$-structure $\beta_\pm^*=-\beta_\mp$, 
to compute the corresponding Yang--Mills action functional
(\ref{YMdef}). One finds
\beqa
\YM(\nablaX)&=&\int_{\man}\,\tr\big(F_{\nablaX}^*\wedge \starX
F_{\nablaX}\big) \nn \\[4pt] &=& \sum_{i,j=0}^m~
\int_{\pq}\,\otimes\int_M \,
\tr\bigg[~\Big( \id\otimes F_{\nabla_i}^*-
\beta_-\otimes\nabla_{i-1,i}(\phi_i)^*+ 
\beta_+\otimes\nabla_{i+1,i}(\phi_{i+1}^*)^* \nn \\ && +\,
(\beta_-\wedge\beta_+)\otimes\big(\phi^*_{i+1}\circ\phi_{i+1}-
q^2\,\phi_{i}\circ\phi_{i}^*- q^{m-2i+1}\,[m-2i]\,\id\big)^* \Big)
\nn \\ && \wedge \, \Big(
(\beta_-\wedge\beta_+)\otimes (\star F_{\nabla_j}) -
\beta_+\otimes\big(\star\nabla_{j-1,j}(\phi_j) \big) -
\beta_-\otimes\big( \star \nabla_{j+1,j}(\phi_{j+1}^*) \big) \nn \\ &&
+\, \id\otimes\big(\phi^*_{j+1}\circ\phi_{j+1}-
q^2\,\phi_{j}\circ\phi_{j}^*- q^{m-2j+1}\,[m-2j]\,\id\big)\,
\frac{\omega^d}{d!} \, \Big)~\bigg] \ .
\label{YMcalc}\eeqa
Now use orthogonality of the splitting $\ceX=\bigoplus_i\,\ceX_i$,
together with the fact that only the top two-form
$\beta_-\wedge\beta_+\in\Omega^2(\pq)$ has a non-zero integral over
$\pq$ in (\ref{YMcalc}). Using the identities (\ref{commc3}) once again and
the normalization $H(1)=1$ for the Haar state on $\ca(\pq)$, one finds
that (\ref{YMcalc}) coincides with the Yang--Mills--Higgs action
functional (\ref{YMHdef}). By construction, the functional
$\YMH_{q,m}$ is invariant under the action of the gauge group
$\scru(\ceX)$ of Proposition~\ref{brgt}, and hence descends to a well-defined
functional on the orbit space $\scrc(\ceX)\big/\,\scru(\ceX)\to [0,\infty)$. 
\end{proof}

We can rewrite the Yang--Mills--Higgs functional (\ref{YMHdef}) in a more suggestive way. 
For this, consider the direct sum of $\ca(M)$-modules with induced connection $(\ce, \nabla)$, 
$$
\ce= \bigoplus_{i=0}^m\,\ce_i \ , \qquad \nabla=\bigoplus_{i=0}^m\, \nabla_i \ .
$$ 
The induced hermitian structure $h=\bigoplus_i\,h_i$ on $\ce$ defines an inner product given by $(\varphi,\psi)_h=\sum_i\,(\varphi_i,\psi_i)_{h_i}$ for $\varphi,\psi\in\ce$, with corresponding norm $\|\varphi\|_h=(\varphi,\varphi)_h^{1/2}$ and extension to $\Hom_{\ca(M)}(\ce,\Omega^p(\ce))$ as described in~\S\ref{se:metrics}. The Higgs fields $\phi_i\in\Hom_{\ca(M)}(\ce_{i-1},\ce_i)$, with
$i=1,\dots,m$, induce an element
$$
\phi=\bigoplus_{i=1}^m\, \phi_i
$$
of the quiver representation module
$\scrr(\ceX)\subset\End_{\ca(M)}(\ce)$ 
given by
$$
\scrr(\ceX)=\bigoplus_{i=1}^m\, \Hom_{\ca(M)}(\ce_{i-1},\ce_i) \ ,
$$
with induced connection $\nabla=\bigoplus_i\,\nabla_{i,i-1}$. The induced hermitian structure $h=\bigoplus_i\, h_{i-1,i}$ defines an inner product given by $(\phi,\phi'\,)_h=\sum_i\,(\phi_i,\phi_i'\,)_{h_{i-1,i}}$ for $\phi,\phi'\in\scrr(\ceX)$, with associated $L^2$-norm $\|\phi\|_h=(\phi,\phi)_h^{1/2}$. Given 
$\phi,\phi'\in\scrr(\ceX)$, we
introduce the endomorphisms 
$\phi\circ\phi'\,^*=\bigoplus_i\,(\phi\circ\phi'\,^*)_i$ and
$\phi^*\circ\phi'= \bigoplus_i\,(\phi^*\circ\phi'\,)_i$ in
$\bigoplus_i\,\End_{\ca(M)}(\ce_i)\subset \End_{\ca(M)}(\ce)$ by
$$
\big(\phi\circ\phi'\,^*\big)_i = \phi_i \circ\phi_i'\,^* \qquad
\mbox{and} \qquad \big(\phi^*\circ\phi'\, \big)_i =
\phi_{i+1}^* \circ\phi_{i+1}^{\prime} \ .
$$
Using these maps we define the $q$-commutator $[\phi^*,\phi'\,]_q\in
\End_{\ca(M)}(\ce)$ by
$$
\big[\phi^*\,,\,\phi'\,\big]_q = \phi^*\circ\phi'-q^2\,
\phi'\circ\phi^* \ .
$$
Finally, define an endomorphism $\Sigma_{q,m}$ of $\ce$ by
$$
\Sigma_{q,m}=\bigoplus_{i=0}^m\, q^{m-2i+1}\,[m-2i]\, \id_{\ce_i} \ .
$$
Then the Yang--Mills--Higgs action functional (\ref{YMHdef}) can be succinctly rewritten in compact form as a functional on $\YMH_{q,m}:\cc(\ce)\times\scrr(\ceX)\to [0,\infty)$ given by
$$
\YMH_{q,m}(\nabla,\phi)=\YM(\nabla)+\big(q^2+1\big) \, \big\|\nabla(\phi)\big\|_h^2 +\big\|[\phi^*,\phi]_q-\Sigma_{q,m}\big\|_h^2 \ .
$$

\medskip

\subsection{Holomorphic chain $q$-vortex equations }~\\[5pt]
Given a unitary connection $\nablaX\in\cc(\ceX)$, there is a
well-defined map $[\nablaX,-]$ from the space
$\Hom_{\ca(\man)}(\ceX,\Omega^\bullet(\ceX))$ to itself, where
$\Omega^\bullet(\ceX)= \bigoplus_{p\geq0}\, \Omega^p(\ceX)$. On
homogeneous morphisms $T\in \Hom_{\ca(\man)}(\ceX,\Omega^p(\ceX))$ it
is defined by
$$
[\nablaX,T]:=\nablaX\circ T- (-1)^p\,T\circ\nablaX \ .
$$
For the curvature $F_{\nablaX}=\nablaX^2 \in
\Hom_{\ca(\man)}(\ceX,\Omega^2(\ceX))$, one then has the Bianchi
identity
$$
[\nablaX,F_{\nablaX}]=0 \ .
$$
As the space of connections $\cc(\ceX)$ is an affine space modeled on 
$\Hom_{\ca(\man)}(\ceX,\Omega^1(\ceX))$, as usual, for the critical points of the Yang--Mills action functional
(\ref{YMdef}) one needs to compute it on a one-parameter family $\nablaX +t\  \eta$, and equate to zero the linear term in $t$ for the corresponding expansion. 
By using properties of the complex inner product $(-,-)_{\hX}$ on
$\Hom_{\ca(\man)}(\ceX,\Omega^\bullet(\ceX))$, this 
variational problem for the Yang--Mills action functional
(\ref{YMdef}) shows that its critical points $\nablaX$ obey the
Euler--Lagrange equation
\beq
[\nablaX^*,F_{\nablaX}]=0 \ ,
\label{YMeom}\eeq
where the adjoint operator of $[\nablaX,-]$ is defined with respect to the
inner product as
\beq
\big([\nablaX^*,T]\,,\,T'\,\big)_{\hX} =
\big(T\,,\,[\nablaX,T'\,]\big)_{\hX}
\label{adjnablacomm}\eeq
for $T,T'\in\Hom_{\ca(\man)}(\ceX,\Omega^\bullet(\ceX))$. Using the
definition of the inner product given in \S\ref{se:metrics}, one
easily shows that $[\nablaX^*,T]=-\starX[\nablaX,\starX T]$.

The purpose of this section is to characterize stable critical points
of the Yang--Mills
action functional (\ref{YMdef}) on $\man$, and to study their
dimensional reduction to configurations on
$M$. For this, following~\cite[\S1.2]{Tian} we introduce the notion of
generalized instanton on the quantum space $\man$.
\begin{lemm}\label{geninst}
Let $\nablaX\in\cc(\ceX)$ be a unitary connection and
$\Xi\in\Omega^{2d-2}(\man)$ a closed form of degree $2d-2$, regarded
as the element $\id_{\ceX}\otimes_{\ca(\man)}\Xi$, such that
\beq
\starX F_{\nablaX} = -F_{\nablaX}\wedge \Xi \ .
\label{ASDeq}\eeq
Then $\nablaX$ solves the Yang--Mills equation (\ref{YMeom}) and
$\YM(\nablaX)= \Top_2(\ceX,\Xi)$, where
\beq
\Top_2(\ceX,\Xi) = - \big(F_{\nablaX}\,,\, \starX(F_{\nablaX}\wedge \Xi)
\big)_{\hX} \ .
\label{Topaction}\eeq
\end{lemm}
\begin{proof}
Using (\ref{ASDeq}) and the graded right Leibniz rule we compute
\begin{eqnarray*}
[\nablaX^*,F_{\nablaX}] &=& -\starX[\nablaX,\starX F_{\nablaX}]
\\[4pt] &=& \starX[\nablaX,F_{\nablaX}\wedge \Xi] \\[4pt]
&=& \starX\big([\nablaX,F_{\nablaX}]\wedge\Xi+ F_{\nablaX}\wedge
\ddX\Xi \big) \= 0 \ ,
\end{eqnarray*}
where in the last line we used the Bianchi identity and
$\ddX\Xi=0$. The second statement \eqref{Topaction} 
follows from direct substitution of the equation (\ref{ASDeq})
into the Yang--Mills action functional
$\YM(\nablaX)=(F_{\nablaX},F_{\nablaX})_{\hX}$.
\end{proof}

Using $\ddX\Xi=0$, the definition of the inner product $(-,-)_{\hX}$,
and the Bianchi identity $[\nablaX,F_{\nablaX}]=0$, one shows in the
standard way that the functional (\ref{Topaction}) does not depend on
the choice of connection $\nablaX$ on $\ceX$. It thus defines a
`topological action' which depends only on the $\ca(\man)$-module
$\ceX$ and the closed form $\Xi$, and hence provides an \emph{a
  priori} lower bound on the Yang--Mills action functional. We refer
to the gauge invariant equation (\ref{ASDeq}) as the
$\Xi$-anti-selfduality equation. Gauge equivalence classes in
$\cc(\ceX)\big/\,\cu(\ceX)$ of solutions to this first order
equation are called generalized instantons or $\Xi$-instantons.

In the sequel we will use the natural closed $(1,1)$-form
$$
\omegaX= (\beta_-\wedge\beta_+)\otimes 1 + 1\otimes\omega
$$
on $\ca(\man)$, and set
\beq
\Xi= \frac{\omegaX^{d-1}}{(d-1)!} =
1\otimes\frac{\omega^{d-1}}{(d-1)!}+ (\beta_-\wedge\beta_+)\otimes
\frac{\omega^{d-2}}{(d-2)!} \ ,
\label{Xiomega}\eeq
where the second term is absent when $d=1$. We write
$\Top_2(\ceX,\omegaX)$ for the corresponding topological action
functional (\ref{Topaction}). For simplicity, we also assume that $\tr(F_{\nablaX})=0$
for any connection $\nablaX\in\cc(\ceX)$ without loss of generality, for otherwise one can
consider $\tilde F_{\nablaX}=F_{\nablaX}-\frac1r\, \tr(F_{\nablaX})\,
\qP$ where $r$ is the rank of the projection $\qP$ defining the
bimodule $\ceX$.

We recall that on the algebra $\ca(\man)=\ca(\pq)\otimes \ca(M)$ there is a natural complex structure 
which combines the complex structure of $M$ with the complex structure for the differential calculus 
on $\Apq$ given in Proposition~\ref{2dsph}. 
Using it we write
$$
\Omega^p(\ceX)=\bigoplus_{i+j=p}\,\Omega^{i,j}(\ceX)
$$
for the corresponding splitting of the space of $\ceX$-valued $p$-forms into $(i,j)$-forms.
The $*$-involution maps $\Omega^{i,j}(\ceX)$ to $\Omega^{j,i}(\ceX)$,
while the Hodge operator $\starX$ maps $\Omega^{i,j}(\ceX)$ to
$\Omega^{d+1-j,d+1-i}(\ceX)$. Consider then the linear operator
$$
L_{\omegaX}\,:\, \Omega^{i,j}(\ceX)~ \longrightarrow ~
\Omega^{i+1,j+1}(\ceX) \ , \qquad L_{\omegaX}(\alpha):=\alpha\wedge
\omegaX \ .
$$
Let $L_{\omegaX}^*:\Omega^{i,j}(\ceX)\to \Omega^{i-1,j-1}(\ceX)$ be its
adjoint with respect to the $L^2$-inner product defined in
\S\ref{se:metrics}. If $\alpha\in\Omega^{1,1}(\ceX)$, then
$$
\alpha^{\omegaX}:= L_{\omegaX}^*(\alpha)=\starX(\alpha^* \wedge\starX
\omegaX)
$$
is the component of $\alpha$ (in $\Omega^{0,0}(\ceX)=\ceX$) along the closed $(1,1)$-form $\omegaX$. As in \S\ref{se:metrics}, this definition is extended to the modules $\End_{\ca(\man)}(\ceX)$ over $\ca(\man)$ and the corresponding modules over $\ca(M)$ in the obvious ways.

For a connection $\nablaX\in\cc(\ceX)$, we decompose the curvature
\beq
F_{\nablaX}=F_{\nablaX}^{2,0}+F_{\nablaX}^{1,1}+ F_{\nablaX}^{0,2}
\label{curvholcomps}\eeq
into its $(2,0)$, $(1,1)$, and $(0,2)$ components. Since $\nablaX$ is
unitary, one has $F_{\nablaX}^{2,0}=-\big(F_{\nablaX}^{0,2}\big)^*$
and $F_{\nablaX}^{1,1}=-\big(F_{\nablaX}^{1,1} \big)^*$.
We recall that the connection $\nablaX$ is called integrable if $F_{\nablaX}^{0,2}=0$. 
\begin{lemm}\label{HYMlemm}
A connection $\nablaX$ solves the generalized instanton equation \eqref{ASDeq} if and only if it is
an integrable unitary connection in $\cc(\ceX)^{1,1}$ which satisfies
the hermitian Yang--Mills equation
\beq
F_{\nablaX}^{\omegaX}= 0 \ .
\label{HYMeq}\eeq
\end{lemm}
\begin{proof}
The Hodge operator $\starX$ acts with a grading $(-1)^j$ on
$\Omega^{i,j}(\ceX)$. Substituting the decomposition
(\ref{curvholcomps}) into (\ref{ASDeq}) with the choice
(\ref{Xiomega}) we thus find
$$
F_{\nablaX}^{0,2}=0 \ ,
$$
whence $\nablaX\in\cc(\ceX)^{1,1}$. For the remaining
$(1,1)$-component, we note first the identity
$$
F_{\nablaX}\wedge\frac{\omegaX^d}{d!}= F_{\nablaX}\wedge
\starX\omegaX=- F_{\nablaX}^{\omegaX}~ \frac{\omegaX^{d+1}}{(d+1)!}  \ .
$$
But by the $\Xi$-anti-selfduality equation (\ref{ASDeq}), the left-hand
side is also equal to
$$
F_{\nablaX}\wedge\frac{\omegaX^d}{d!}= -\starX 
F_{\nablaX}\wedge\omegaX= +F_{\nablaX}^{\omegaX}~
\frac{\omegaX^{d+1}}{(d+1)!} \ ,
$$
and comparing the two expressions gives (\ref{HYMeq}).
\end{proof}

We next describe the $\SU$-equivariant dimensional reduction of the above
generalized instanton equations.
\begin{prop}\label{instredprop}
Under the bijection of Proposition~\ref{brholconn}, the subspace of
$\SU$ invariant connections
$\nablaX^{\dolb}\in\big(\cc(\ceX)^{1,1}\big)^{\SU}$ solving the
generalized instanton equation on the quantum
space $\man$ corresponds bijectively to the
subspace of $\scrc(\ceX)^{1,1}$ consisting of elements $(\nabla^{\dolb},\phi^*)$
which satisfy the holomorphic chain $q$-vortex equations on the
manifold $M$ given by
\beq
F_{\nabla_i}^{\omega}=
q^2\,\phi_{i}\circ\phi_{i}^*- \phi_{i+1}^*\circ\phi_{i+1}
+ q^{m-2i+1}\,[m-2i]\,\id_{\ce_i}
\label{qchaineqs}\eeq
for each $i=0,1,\dots,m$. Here 
$F_{\nabla_i}^{\omega} =\star\big((F_{\nabla_i}^{1,1})^* \wedge\star\omega\big)$ is
the component (in $\End_{\ca(M)}(\ce_i)$) of the curvature of $\nabla_i$ along the K\"ahler form
of $M$.
\end{prop}
\begin{proof}
By Proposition~\ref{brholconn}, the $\SU$-invariant subspace of
generalized instanton connections on $\man$ described by Lemma~\ref{HYMlemm} consists of
connections and Higgs fields on $M$ satisfying
$$
F_{\nabla_i}^{0,2}=0 \qquad \mbox{and} \qquad
\nabla_{i+1,i}^{\dolb}(\phi_{i+1}^*)=0
$$
for $i=0,1,\dots,m$. Whence the $(1,1)$-component of the curvature
(\ref{curvdimred}) for a connection
$\nablaX^{\dolb}\in\big(\cc(\ceX)^{1,1}\big)^{\SU}$ is given by
$$
F_{\nablaX}^{1,1} =
\sum_{i=0}^m\,\Big(\id\otimes F_{\nabla_i}^{1,1}+
(\beta_-\wedge\beta_+)\otimes\big(\phi^*_{i+1}\circ\phi_{i+1}-
q^2\,\phi_{i}\circ\phi_{i}^*- q^{m-2i+1}\,[m-2i]\,\id\big) \Big) \
.
$$
We now use (\ref{hod2}) and multiply this form with
$$
\starX\omegaX=1\otimes\frac{\omega^d}{d!}+(\beta_-\wedge\beta_+)
\otimes (\star\omega)
$$
to get
$$
F_{\nablaX}^{\omegaX} = \sum_{i=0}^m\, \id \otimes
\big(F_{\nabla_i}^{\omega}+ \phi^*_{i+1}\circ\phi_{i+1}-
q^2\,\phi_{i}\circ\phi_{i}^*- q^{m-2i+1}\,[m-2i]\,\id\big) \ .
$$
Using orthogonality of the direct sum
$\ceX=\bigoplus_i\,\ceX_i$, the hermitian Yang--Mills equation
(\ref{HYMeq}) then corresponds to the set of equations
(\ref{qchaineqs}).
\end{proof}

The equations (\ref{qchaineqs}) are naturally invariant under the
action of the group $\scru(\ceX)$. Under the bijection of
Proposition~\ref{brorb}, an $\SU$-invariant generalized instanton
reduces to a gauge equivalence class of solutions to (\ref{qchaineqs}) in
$\scrc(\ceX)\big/\,\scru(\ceX)$. We call such a class a $q$-vortex on
the manifold $M$. We can rewrite the system of $q$-vortex equations
collectively in a more suggestive form. Using the definitions given at the end of 
\S\ref{se:YMdimred}, they can be
succinctly rewritten
in a compact form as the single equation in $\End_{\ca(M)}(\ce)$ given
by
$$
F_\nabla^\omega+ \big[\phi^*\,,\,\phi\big]_q = \Sigma_{q,m} \ .
$$
This equation illustrates the crucial feature of our deformation of standard
quiver vortex equations, in that the commutator of Higgs fields is
replaced with the $q$-commutator. We shall find various interesting
consequences of this deformation below.

\medskip

\subsection{Vacuum structure}~\\[5pt]
In order to establish a correspondence between the generalized
instanton equations on $\man$ and the non-abelian $q$-vortex equations
on $M$, we will now show directly how the latter equations describe stable
critical points of the Yang--Mills--Higgs functional (\ref{YMHdef}).
For this, we will assume in the following that, for each
$i=1,\dots,m$, the hermitian endomorphism $F_{\nabla_i}^\omega$
of $\ce_i$ is non-negative, i.e. it defines a non-negative
sesquilinear form on $\Hom_{\ca(M)}(\ce_{i-1},\ce_i)$ given by
$(F_{\nabla_i}^\omega\circ \phi_{i},\phi_{i}'\,)_{h_{i-1,i}}$ for
$\phi_{i},\phi_{i}'\in \Hom_{\ca(M)}(\ce_{i-1},\ce_i)$. Summing
these forms we then get a non-negative sesquilinear form on
$\scrr(\ceX)$ defined by
$$
(\phi,\phi'\,)_{\scrr(\ceX)}:=\sum_{i=1}^{m}\,
\big(F_{\nabla_i}^\omega\circ\phi_{i} 
\,,\, \phi_{i}'\,\big)_{h_{i-1,i}} \ ,
$$
for $\phi,\phi'\in\scrr(\ceX)$, with corresponding norm
$\|\phi\|_{\scrr(\ceX)}^2:=(\phi,\phi)_{\scrr(\ceX)}\geq0$ for each
$\phi\in\scrr(\ceX)$.

\begin{theo}\label{thm:YMHrestr}
The restriction of the Yang--Mills--Higgs functional $\YMH_{q,m}$ to
elements $(\nabla^{\dolb},\phi^*)\in\scrc(\ceX)^{1,1}$, for which
$F_{\nabla_i}^\omega\in \End_{\ca(M)}(\ce_i)$ is non-negative for
$i=1,\dots,m$, is given by
\begin{align}
\YMH_{q,m}\big(\nabla^{\dolb}\,,\,\phi^*\big)=&\sum_{i=0}^m\,
\big(2q^{m-2i+1}\,[m-2i]\, \Top_1(\ce_i,\omega)
-\Top_2(\ce_i,\omega)\big) \nonumber\\ & -\, 2\big(1-q^2\big)\, \|\phi\|_{\scrr(\ceX)}^2 \nonumber\\
& +\, \sum_{i=0}^m\, \big\|F_{\nabla_i}^\omega+\phi^*_{i+1}\circ\phi_{i+1} -
q^2\,\phi_{i}\circ\phi_{i}^*- q^{m-2i+1}\,[m-2i]\,\id_{\ce_i}
\big\|_{h_i}^2
\label{YMHBog}\end{align}
where 
$$
\Top_1(\ce_i,\omega):=\big(F_{\nabla_i}^\omega\,,\,\id_{\ce_i} \big)_{h_i} \quad 
\textup{and} \quad \Top_2(\ce_i,\omega):=-\Big(F_{\nabla_i} \,,\,\star
\big(F_{\nabla_i}\wedge
\frac{\omega^{d-2}}{(d-2)!}\big) \Big)_{h_i} \ .
$$
\end{theo}
\begin{proof}
We will apply a Bogomol'nyi-type transformation to the action
functional (\ref{YMHdef}). Firstly, we have
\begin{multline}
\big\|F_{\nabla_i}\big\|_{h_i}^2
+\big(q^2+1\big)\,\big\|\nabla_{i-1,i}(\phi_i)\big\|^2_{h_{i-1,i}} \\ = 
4 \big\|F_{\nabla_i}^{0,2}
\big\|_{h_i}^2 + \big\|F^\omega_{\nabla_i}\big\|_{h_i}^2  
+\, 2 \big(q^2+1\big)\,\big\|\nabla_{i-1,i}^{\dolb}
(\phi_i)\big\|^2_{h_{i-1,i}} -\Top_2(\ce_i,\omega)
\label{Fphidecomp}
\end{multline}
for each $i=0,1,\dots,m$ (see e.g.~\cite[\S4]{A-CG-P2}). For
$(\nabla^{\dolb},\phi^*)\in\scrc(\ceX)^{1,1}$, this is equal to
$$
\big\|F^\omega_{\nabla_i}\big\|_{h_i}^2-\Top_2(\ce_i,\omega) \ . 
$$
We now combine
$\|F^\omega_{\nabla_i}\|_{h_i}^2$ with the last set of norms in
(\ref{YMHdef}). For this, we expand out the last set of inner products
in (\ref{YMHBog}) to get
\begin{multline*}
  \big\|F_{\nabla_i}^\omega+ \phi^*_{i+1}\circ\phi_{i+1}-
q^2\,\phi_{i}\circ\phi_{i}^*- q^{m-2i+1}\,[m-2i]\,\id
\big\|_{h_i}^2 \\  \qquad \=
\big\|F^\omega_{\nabla_i}\big\|_{h_i}^2 + \big\|
\phi^*_{i+1}\circ\phi_{i+1}- 
q^2\,\phi_{i}\circ\phi_{i}^*- q^{m-2i+1}\,[m-2i]\,\id
\big\|_{h_i}^2 \\   \qquad \qquad \quad +\, 2\big( F^\omega_{\nabla_i} \,,\,
\phi^*_{i+1}\circ\phi_{i+1}-
q^2\,\phi_{i}\circ\phi_{i}^*- q^{m-2i+1}\,[m-2i]\,\id \big)_{h_i}
\ .
\end{multline*}
The last term in the inner product here gives
$-2q^{m-2i+1}\,[m-2i]\, \Top_1(\ce_i,\omega)$. The first two terms in
this inner product can be evaluated by using
(\ref{indconn}) and the graded commutator to deduce that
the curvature of the induced connection $\nabla_{i-1,i}$ on
$\Hom_{\ca(M)}(\ce_{i-1},\ce_i)$ is  
$$
F_{\nabla_{i-1,i}}(\phi_i):=\nabla_{i-1,i}\circ
\nabla_{i-1,i}(\phi_i)= \phi_i\circ F_{\nabla_{i-1}}
-F_{\nabla_i}\circ \phi_i \ .
$$
Since $F_{\nabla_i}^{0,2}=0$ for each $i=0,1,\dots,m$, we can use
standard K\"ahler identities~\cite[eq.~(4.10)]{A-CG-P2} for the induced holomorphic structures on
the $\ca(M)$-bimodules $\Hom_{\ca(M)}(\ce_{i-1},\ce_i)$ to
get
$$
\big(F^\omega_{\nabla_{i-1,i}}(\phi_i)\,,\,\phi_i\big)_{h_{i-1,i}} =
\big\|\nabla_{i-1,i}^{\dol} 
(\phi_i)\big\|^2_{h_{i-1,i}}- \big\|\nabla_{i-1,i}^{\dolb}
(\phi_i)\big\|^2_{h_{i-1,i}} \ ,
$$
which vanishes for $(\nabla^{\dolb},\phi^*)\in\scrc(\ceX)^{1,1}$ by Proposition~\ref{brholconn}. It
follows that
\begin{eqnarray*}
&& \sum_{i=0}^m\, \big( F^\omega_{\nabla_i} \,,\,
\phi^*_{i+1}\circ\phi_{i+1}-
q^2\,\phi_i\circ\phi_i^*\big)_{h_i} = \sum_{i=1}^m\,
\big(\phi_i\circ
F^\omega_{\nabla_{i-1}}- q^2\, F^\omega_{\nabla_i} \circ \phi_i\,,\,
\phi_i\big)_{h_{i-1,i}} \\[4pt]  & & \qquad\qquad\qquad =
\sum_{i=1}^m\, \Big[
\big(F^\omega_{\nabla_{i-1,i}}(\phi_i)\,,\,\phi_i\big)_{h_{i-1,i}} 
+\, \big(1-q^2\big)\,\big(F_{\nabla_{i}}^\omega\circ\phi_i
\,,\, \phi_i\big)_{h_{i-1,i}}\, \Big] \\[4pt] 
& & \qquad\qquad =
\big(1-q^2\big)\, \|\phi\|_{\scrr(\ceX)}^2 \ .
\end{eqnarray*}
Putting everything together yields (\ref{YMHBog}).
\end{proof}

\begin{coro}\label{coro:BPSvort}
The minima of the Yang--Mills--Higgs action functional $\YMH_{q,m}$ on
$\scrc(\ceX)$, having values in $[0, \infty )$, are given by elements
$(\nabla^{\dolb},\phi^*)\in\scrc(\ceX)^{1,1}$ which satisfy the
holomorphic chain $q$-vortex equations (\ref{qchaineqs}), and for
which the curvature projections $F_{\nabla_i}^\omega$ in
$\End_{\ca(M)}(\ce_i)$ for each $i=1,\dots,m$ are non-negative with
\beq
\|\phi\|_{\scrr(\ceX)}^2\leq \frac1{2\big(1-q^2\big)}\, \sum_{i=0}^m\,
\big(2q^{m-2i+1}\,[m-2i]\, \Top_1(\ce_i,\omega)
-\Top_2(\ce_i,\omega)\big) \ .
\label{phibound}\eeq
When equality holds in (\ref{phibound}), the minima achieve the
infemum $\YMH_{q,m}(\nabla^{\dolb},\phi^*)=0$.
\end{coro}
\begin{proof}
From (\ref{Fphidecomp}) it follows that the action functional
(\ref{YMHdef}) is minimized by taking
$(\nabla^{\dolb},\phi^*)\in\scrc(\ceX)^{1,1}$. From (\ref{YMHBog}) one 
then gets that there is a Bogomol'nyi-type inequality
$$
\YMH_{q,m}(\nabla,\phi)\geq \sum_{i=0}^m\,
\big(2q^{m-2i+1}\,[m-2i]\,\Top_1(\ce_i,\omega) 
-\Top_2(\ce_i,\omega)\big) - 2\big(1-q^2\big)\, \|\phi\|_{\scrr(\ceX)}^2 \ ,
$$
with $\Top_1(\ce_i,\omega)$ and $\Top_2(\ce_i,\omega)$ not dependent on
the choice of connection $\nabla_i$ on the $\ca(M)$-module
$\ce_i$. This inequality is saturated by solutions to the holomorphic
$q$-vortex equations, with the bound (\ref{phibound}) since
$\YMH_{q,m}(\nabla,\phi)\geq0$ and $0<q<1$.
\end{proof}

\begin{coro}
A stable $q$-quiver bundle on $M$ has
characteristic classes constrained by the Bogomolov--Gieseker-type
inequality
$$
2\, \sum_{i=0}^m\, q^{m-2i}\,[m-2i]\, \Top_1(\ce_i,\omega) \geq
\sum_{i=0}^m\, \Top_2(\ce_i,\omega) \ .
$$
If equality holds, then the connections $\nabla_i$ are flat and
\beq
\phi^*_{i+1}\circ\phi_{i+1}-
q^2\,\phi_i\circ\phi_i^* = q^{m-2i+1}\,[m-2i]\,\id_{\ce_i}
\label{flateqs}\eeq
for each $i=0,1,\dots,m$.
\end{coro}
\begin{proof}
The first statement follows from the inequality (\ref{phibound})
together with $\|\phi\|_{\scrr(\ceX)}^2\geq0$ and $0<q<1$. For the second
statement, we use (\ref{YMHBog}) to get
$$
\YMH_{q,m}(\nabla,\phi)=-2\big(1-q^2\big)\, \|\phi\|^2_{\scrr(\ceX)}\leq
0 \ . 
$$
But from its definition (\ref{YMHdef}) the Yang--Mills--Higgs
functional is a sum of non-negative terms, whence
$\YMH_{q,m}(\nabla,\phi)=0$ and thus $F_{\nabla_i}=0$ for each
$i=0,1,\dots,m$.
\end{proof}

Let us demonstrate explicitly that the vacuum moduli
space of the Yang--Mills--Higgs functional is generically non-empty.
\begin{prop}
Suppose that the finitely-generated projective $\ca(M)$-bimodules
$\ce_i$ have the same rank for all $i=0,1,\dots,m$. Then a class of
explicit solutions to the vacuum equations (\ref{flateqs}) is given by
\beq
\phi_i=\phi_i{}^0:=\sqrt{\lambda_i}~u_i \ , \qquad i=1,\dots,m
\ ,
\label{phivac}\eeq
where $u_i\in\Hom_{\ca(M)}(\ce_{i-1},\ce_{i})$ are arbitrary
holomorphic morphisms  unitary with respect to the hermitian structures $h_{i-1}$ on
$\ce_{i-1}$ and $h_i$ on $\ce_i$, and the induced
connection $\nabla_{i-1,i}^{\dolb}$, and $\lambda_i$ are the
$q$-numbers
\beq
\lambda_i=\frac{q^{m-2i+3}}{\big(1-q^2\big)\,\big(1-q^6 \big)}\, \Big(
[m-2i+2]-q^4\,[m-2i] -q^{4i}\,\big( [m+2]-q^4\, [m]\big) \Big) \ .
\label{lambdai}\eeq
\label{prop:phivac}\end{prop}
\begin{proof}
Substituting (\ref{phivac}) into (\ref{flateqs}) gives the linear
recursion relation
\beq
\lambda_{i+1}-q^2\, \lambda_i=q^{m-2i+1}\,[m-2i] \ .
\label{lambdarec}\eeq
With $\lambda_0=0=\lambda_{m+1}$, the solution of (\ref{lambdarec}) is
given by
$$
\lambda_i=\sum_{k=0}^{i-1}\, q^{m-2(i-2k)+3}\,\big[ m-2(i-k-1)\big] \
.
$$
We now use the definition of the $q$-integers $[m-2(i-k-1)]$ given in (\ref{eq:q-integer}) and
perform the sums over $k$ using
$$
\sum_{k=0}^{i-1}\, x^k=\frac{1-x^i}{1-x} \ ,
$$
with $x=q^6$ and $x=q^2$. This gives
\begin{eqnarray*}
\lambda_i&=&\frac{q^{m-2i+3}}{\big(q-q^{-1}\big)\, \big(1-q^6\big) \,
  \big(1-q^2\big)}\, \Big(\big(1-q^{6i}\big)\,
\big(q^{m-2i+2}-q^{m-2i+4}\big) \\ && \hspace{6cm} -\, \big(1-q^{2i}\big)\,
\big(q^{-m+2i-2}-q^{-m+2i+4}\big) \Big) \ ,
\end{eqnarray*}
which is easily manipulated into the form
(\ref{lambdai}).
\end{proof}

\medskip

\subsection{Stability conditions}~\\[5pt]
We can also derive topological obstructions to the existence of
solutions to the $q$-vortex equations (\ref{qchaineqs}). For this, we
suppose that $M$ is compact, and define the degree of a hermitian
finitely-generated projective $\ca(M)$-module $(\ce,h)$ by
$$
\deg(\ce)=\frac{\Top_1(\ce,\omega)}{\vol_\omega(M)}
$$
where $\vol_\omega(M)=\int_M\, \frac{\omega^d}{d!}$ is the K\"ahler
volume of $M$. The degree depends on the cohomology class of
$\omega$. The slope of $\ce$ is the number
$$
\mu(\ce)=\frac{\deg(\ce)}{\rank(\ce)} \ , 
$$
with the rank, $\rank(\ce)$, defined as the trace of the identity endomorphism acting on $\ce$. 

Analogously to~\cite[\S3]{A-CG-P1}, we define the $(q,m)$-degree of a finitely-generated
$\SU$-equivariant projective bimodule $\ceX$ over the algebra
$\ca(\man)$, with equivariant decomposition as in (\ref{decomp}),
by
\beq
\deg_{q,m}(\ceX)= \sum_{i=0}^m\,\big( \deg(\ce_i)-
q^{m-2i+1}\,[m-2i]\,\rank(\ce_i) \big)
\label{degqm}\eeq
and its $(q,m)$-slope by
$$
\mu_{q,m}(\ceX)=\frac{\deg_{q,m}(\ceX)}{\rank(\ceX)}
$$
where $\rank(\ceX)=\sum_i\,\rank(\ce_i)$. The definition (\ref{degqm}) has a natural topological meaning, when we recall from \S\ref{se:wn} that the $q$-integers $[m-2i]$ label classes in the $\SU$-equivariant K-theory $\K^{\su}_0(\pq)$. Since the quantum group $\SU$ acts trivially on the manifold $M$, the $(q,m)$-degree labels classes in the $\SU$-equivariant K-theory of $\man=\pq\times M$. Thus the standard assignment~\cite{PS,LPS1} of D-brane charges in equivariant K-theory to quiver vortices extends to our $q$-vortices as well.

The parameters $q,m$ and the
topology of the bundles $\ce_i$ over $M$ are then constrained by the
following (semi-)stability criteria.
\begin{prop}\label{prop:stability}
A stable $q$-quiver bundle on $M$ has slopes constrained by the
inequalities:
\begin{itemize}
\item[(a)] \ $\mu(\ce_0)\leq q^{m+1}\,[m]$, with equality if and only
  if $\ce_0$ admits a holomorphic connection $\nabla_0$ solving the
  hermitian Yang--Mills equation
  $F_{\nabla_0}^\omega=q^{m+1}\,[m]\,\id_{\ce_0}$.
\\
\item[(b)] \ $\mu(\ce_m)\geq -q^{-m+1}\,[m]$, with equality if and only
  if $\ce_m$ admits a holomorphic connection $\nabla_m$ solving the
  hermitian Yang--Mills equation
  $F_{\nabla_m}^\omega=-q^{-m+1}\,[m]\,\id_{\ce_m}$.
  \\
\item[(c)] \ $\mu_{q,m}(\ceX)\leq0$, with equality if and only
  if $\ce_i$ admits a holomorphic connection $\nabla_i$ solving the
  hermitian Yang--Mills equation
  $F_{\nabla_i}^\omega=q^{m-2i+1}\,[m-2i]\,\id_{\ce_i}$ for each $i=0,1,\dots,m$.
\end{itemize}
\end{prop}
\begin{proof}
Point (a) follows from the $q$-vortex equation (\ref{qchaineqs}) for $i=0$ after taking
inner products on both sides with $\id_{\ce_0}$, and using
$(\id_{\ce_0},\id_{\ce_0})_{h_0}= \rank(\ce_0)\, \vol_\omega(M)$
together with $\phi_0:=0$ and $\|\phi_1\|_{h_{0,1}}^2\geq0$. Point (b)
follows similarly from (\ref{qchaineqs}) with $i=m$ and
$\phi_{m+1}:=0$. For point (c), we take inner products on both
sides of (\ref{qchaineqs}) with $\id_{\ce_i}$ and sum over
$i=0,1,\dots,m$ to get the constraint
$$
\sum_{i=0}^m\, \deg(\ce_i)= \sum_{i=0}^m\, q^{m-2i+1}\,[m-2i]\, \rank(\ce_i)
+\frac{q^2-1}{\vol_\omega(M)}\, \sum_{i=1}^m\,
\|\phi_i\|_{h_{i-1,i}}^2 \ ,
$$
and the result follows since $0<q<1$ and
$\|\phi_i\|_{h_{i-1,i}}^2\geq0$ (with $\|\phi_i\|_{h_{i-1,i}}^2=0$ if
and only if $\phi_i=0$).
\end{proof}

\section{Examples}

In this final section we will briefly consider some explicit examples
of the $q$-vortex equations~(\ref{qchaineqs}). In particular, we will
describe how the $q$-deformations affect stability conditions for the
existence of solutions to these equations and the structure of the
corresponding moduli spaces. These considerations provide the first
step to formulating a general algebro-geometric notion of stability for
$\SU$-equivariant modules over $\ca(\man)$ and the corresponding
$q$-quiver bundles over $M$.

\medskip

\subsection{Deformations of holomorphic triples and stable pairs}~\\[5pt]
A holomorphic triple $(\ce_0,\ce_1,\phi)$ on a compact K\"ahler manifold
$(M,\omega)$ consists of a pair of holomorphic vector bundles
$\ce_0,\ce_1$ over $M$ and a holomorphic morphism
$$
\ce_0~\xrightarrow{\phi}~ \ce_1
$$
which together obey coupled vortex equations~\cite{GP1,BG-P}. For $m=1$,
our $q$-vortex equations (\ref{qchaineqs}) provide a $q$-deformation
of such triples. In this case the equations (\ref{qchaineqs}) read
\beqa
F_{\nabla_0}^\omega&=& q^2\, \big(\id_{\ce_0}-q^{-2}\,\phi\circ\phi^*
\big) \ , \nonumber \\[4pt]
F_{\nabla_1}^\omega&=& -\big(\id_{\ce_1}-q^{2}\,\phi^* \circ\phi
\big) \ ,
\label{m1eqs}\eeqa
where $\phi:=\phi_1$. The topological stability conditions on these
triples are governed by Proposition~\ref{prop:stability} for
$m=1$. Additionally, by taking inner products on both sides of the equations (\ref{m1eqs})
with $\id_{\ce_0}$ and $\id_{\ce_1}$ respectively, and summing shows
that the degrees of the bundles are related by
$$
\deg(\ce_0)+q^{-2}\,\deg(\ce_1)=q^2\,\rank(\ce_0)-
q^{-2}\,\rank(\ce_1) \ .
$$
Substituting this into the formula for the $(q,1)$-degree of the
quiver bundle then yields
$$
\deg_{q,1}(\ceX)=\big(1-q^{-2}\big)\,\big(\deg(\ce_1)
+\rank(\ce_1)\big) \ .
$$
These criteria are much more stringent than the stability condition
of~\cite{BG-P}.

Let us now denote $\ce:=\ce_0$, $\nabla:=\nabla_0$, and take $\ce_1$
to be the $\ca(M)$-module of sections of the trivial holomorphic
line bundle over $M$, i.e. $\ce_1=\IC\otimes\ca(M)\simeq \ca(M)$. Then
$F_{\nabla_1}^\omega=0$, and $\Hom_{\ca(M)}(\ce_0,\ce_1)\simeq \ce_0$
so that the Higgs field $\phi$ can be regarded as an element of $\ce$,
i.e. as a holomorphic section. We set $\phi:=q^{-1}\,\varphi$ with
$\varphi\in\ce$, so that
$\varphi^*\otimes_{\ca(M)}\varphi=\id_{\ca(M)}$ by the second equation
of (\ref{m1eqs}). The first equation of (\ref{m1eqs}) can then written
as
\beq
F_\nabla^\omega+q^{-2}\,\varphi\otimes_{\ca(M)} \varphi^*=q^2\,\id_\ce
\ .
\label{qpair}\eeq
Thus in this case the triple describes a $q$-deformation of the stable
pairs $(\ce,\varphi)$ considered in~\cite{Bradlow}. The stability
condition reads
$$
\deg(\ce)=q^2\,\rank(\ce)-q^{-2}
$$
which is much more restrictive than the undeformed one of~\cite{Bradlow,GP1}.

\medskip

\subsection{$q$-vortices on Riemann surfaces}~\\[5pt]
Let $M$ be a compact oriented Riemann surface of genus $g$. Then the
equations (\ref{m1eqs}) describe $q$-deformations of non-abelian
vortices on $M$. For $g\neq1$, the area of $M$ is  
$$
\vol_\omega(M)=\frac{8\pi}\kappa\, (1-g)
$$
by the Gauss--Bonnet theorem, where $\kappa$ is the scalar curvature
of $M$ with respect to the K\"ahler metric corresponding to
$\omega$. Let us again consider a particular case. If $\ce:=\ce_0$,
$\nabla:=\nabla_0$ and $\ce_1\simeq \IC^r\otimes\ca(M)$ with
$r=\rank(\ce)$, then the Higgs field $\phi=q^{-1}\,\varphi$ can be regarded as an
element of $\IC^r\otimes \ce$. The characteristic class
$\frac1{2\pi}\, \Top_1(M,\omega)$ is the first Chern class $c_1(\ce)$
of the bundle $\ce$. A non-empty moduli space of solutions to the
$q$-vortex equations, formally the same as in (\ref{qpair}), is
ensured in this case by the stability condition
$$
c_1(\ce)=\frac{4r}{\kappa}\, \big(q^2-q^{-2}\big)\, (1-g)
$$
for $g\neq1$. Since $0<q<1$, this degree satisfies the bound
$c_1(\ce)<\frac{4q^2}\kappa\,(1-g)$. Hence the pair $(\ce,\varphi)$ is
$\tau$-stable in the sense of~\cite{BDG-PW}, and by the Hitchin--Kobayashi
correspondence it is gauge equivalent to a solution of the non-abelian $q$-vortex
equations. The corresponding moduli space of solutions is described
explicitly in~\cite{Baptista}. For abelian vortices, $r=1$, this
moduli space coincides with the $|n|$-th symmetric product orbifold of
$M$, i.e. the space of effective divisors on $M$ of degree
$n=c_1(\ce)$.

By taking $\ce_1$ to be a (generically non-trivial) holomorphic line
bundle, one also obtains from (\ref{m1eqs}) a $q$-deformation of the
non-abelian vortex equations studied in~\cite{Popov}. However, contrary to the
$q=1$ case, wherein the reduction $r=1$, $\ce_0\simeq \ce_1$ and
$\nabla_0=-\nabla_1$ would lead to the standard abelian BPS vortex
equations on $M$, such an abelian reduction in (\ref{m1eqs}) is not
consistent for $q\neq1$. Indeed, as we first witnessed in point~(c) of
Proposition~\ref{prop:stability}, the $q$-deformation generically
imposes very stringent constraints on the allowed stable quiver
bundles. Moreover, the $q$-vortices do not
exist on the complex plane $M=\IC$, wherein the formal limit
$\vol_\omega(M)=\infty$ would necessitate infinite vortex number and
action. The features spelled out in this section are generic properties of the $q$-vortex equations (\ref{m1eqs}).

\medskip

\subsection{$q$-instantons}~\\[5pt]
Let $(M,\omega)$ be a K\"ahler surface. Set $\ce_0\simeq\ce_1=:\ce$, with $r=\rank(\ce)$, and $\phi=\id_{\ce}$. Then since $\phi$ is a holomorphic section, $\nabla^{\dolb}_{0,1}(\phi)=0$, from (\ref{indconn}) we have $\nabla_0=\nabla_1=:\nabla$ and both equations in (\ref{m1eqs}) simplify to
\beq
F_\nabla^\omega=\big(q^2-1\big)\, \id_\ce \ .
\label{HYM4d}\eeq
For bundles $\ce$ of vanishing degree and $q^2\neq1$, this equation gives a deformation of the hermitian Yang--Mills equation on $M$, and hence of the standard anti-selfduality equations $\star F_\nabla=-F_\nabla$. Gauge equivalence classes of solutions to (\ref{HYM4d}) are thus called $q$-instantons, and their moduli spaces can be described explicitly in the following way.

The fixed points on the space $\cc(\ce)$ of unitary connections on $\ce$ under the action of the group of gauge transformations $\cu(\ce)$ are given by integrable connections $\nabla\in\cc(\ce)^{1,1}$. The natural $\cu(\ce)$-invariant symplectic form $\omega_{\cc}$ on $\cc(\ce)$ is thus given by
$$
\omega_{\cc}(\alpha,\alpha'\,)= \frac{1}{2}\ \int_M\,\tr\big(\alpha^*\wedge
\alpha'\, \big)^\omega ~  \omega^2 
$$
for $\alpha,\alpha'\in\Hom^a_{\ca(M)}(\ce,\Omega^1(\ce))$. We implicitly use the inclusion of the Lie algebra of $\cu(\ce)$ in its dual space by means of the hermitian structure $h$ on $\ce$. Then the corresponding moment map $\mu_\cc:\cc(\ce)\to (\Lie\,\cu(\ce))^*$ is given by
$$
\mu_\cc(\nabla)=F_\nabla^\omega \ .
$$
The moduli space of $q$-instantons on $M$ is thus realized as the symplectic quotient
$$
\mu_\cc^{-1}\big((q^2-1)\,\id_\ce\big) \, \big/ \, \cu(\ce) \ ,
$$
and hence the $q$-vortices in this case correspond to points of $\mu_\cc^{-1}\big((q^2-1)\,\id_\ce\big)$ which lie inside the K\"ahler submanifold $\cc(\ce)^{1,1}$ (outside the singularities).

When $M=\IC^2$, the constant shift in the moment map condition from $\mu_\cc=0$ to $\mu_\cc=(q^2-1)\,\id_\ce$ induces a shift in the corresponding real ADHM equation. The effect of this shift is to 
augment~\cite{Nakajima} the moduli space of holomorphic instanton bundles to the moduli space of torsion free sheaves on the projective plane $\IC\mathrm{P}^2$ with a trivialization on a fixed projective line $\IC\mathrm{P}^1\subset \IC\mathrm{P}^2$. This resolves the small instanton singularities and turns the instanton moduli space into a hyper-K\"ahler manifold of complex dimension $4r\,k$, where $k=\frac1{8\pi^2}\, \Top_2(M,\omega)=c_2(\ce)$. It is well-known~\cite{NS} that this modification arises explicitly in the equations which determine instantons on a certain noncommutative deformation of $\IR^4$. Here we have shown that the same sort of resolution of instanton moduli space is achieved via our $q$-deformed dimensional reduction procedure over the quantum projective line $\pq$. The essential feature behind such resolutions, provided here by the deformation in (\ref{HYM4d}), lies in the content of point~(c) of Proposition~\ref{prop:stability}.

\section*{Final remarks}

In this paper we have shown that the formalism of ${\rm
  SU}(2)$-equivariant dimensional reduction over the sphere has a
natural $\SU$ Hopf algebraic generalization to reductions over the
quantum sphere. This was achieved by recasting the standard
dimensional reduction procedure into a purely algebraic framework and
using the fact that much of the geometry of the projective line
survives $q$-deformation to the quantum projective line (the quantum
sphere with additional structure). We obtained a $q$-deformed
Yang--Mills--Higgs theory from the reduction of Yang--Mills theory,
and also $q$-deformations of quiver chain vortex equations from the
reduction of natural first order gauge theory equations. We
demonstrated that the moduli spaces of solutions to these $q$-vortex
equations are more constrained but generically better behaved than their $q\to1$ limits. In some instances, the vacuum moduli space can be described as a symplectic or even hyper-K\"ahler quotient. It would be interesting to explore
  whether the generic $q$-vortex equations admit such a moment map interpretation, as they do in the $q=1$ case from the action of a unitary group on a representation space of quiver modules (see~\cite[\S2.2]{A-CG-P2}). This presumably involves interpreting the $q$-commutator terms as moment map equations for a sort of quantum group action on the space of quiver gauge connections. This may also help fill an important gap in our construction, namely the proper formulation of stability conditions and the ensuing Hitchin--Kobayashi-type correspondence which relates the existence of solutions to the gauge equations with a stability criterion. This problem appears to lie in the general realm of extending noncommutative geometry into the algebro-geometric setting, which is not yet fully developed.
It would be interesting to see if the $q$-deformations of the Yang--Mills--Higgs models derived in this paper improve the phenomenological viability of the models constructed in~\cite{BS}. The somewhat intricate $q$-dependence of the vacuum Higgs field configurations described by Proposition~\ref{prop:phivac} may drastically alter the dynamical mass generation in these models. It would also be interesting to extend our constructions to Hopf algebraic equivariant dimensional reduction over other quantum homogeneous spaces.


\begin{thebibliography}{99}

\bibitem{A-CG-P1}
L.~\'{A}lvarez-C\'onsul, O.~Garc\'{\i}a-Prada,
{\it Dimensional reduction, $\slc$-equivariant bundles
  and stable holomorphic chains},
Int.~J. Math. {12} (2001) 159--201.

\bibitem{A-CG-P3}
L.~\'{A}lvarez-C\'onsul, O.~Garc\'{\i}a-Prada,
{\it Dimensional reduction and quiver bundles},
J. Reine Angew. Math. 556 (2003) 1--46.

\bibitem{A-CG-P2}
L.~\'{A}lvarez-C\'onsul, O.~Garc\'{\i}a-Prada,
{\it Hitchin--Kobayashi correspondence, quivers and vortices},
Comm. Math. Phys. 238 (2003) 1--33.

\bibitem{Aschieri1}
  P.~Aschieri, T.~Grammatikopoulos, H.~Steinacker, G.~Zoupanos,
  {\it Dynamical generation of fuzzy extra dimensions, dimensional reduction and
  symmetry breaking},
  J. High Energy Phys. {0609} (2006) 026.

\bibitem{Aschieri2}
  P.~Aschieri, J.~Madore, P.~Manousselis, G.~Zoupanos,
  {\it Dimensional reduction over fuzzy coset spaces},
  J. High Energy Phys. {0404} (2004) 034.

\bibitem{Baptista}
J.M.~Baptista,
{\it Non-abelian vortices on compact Riemann surfaces},
Comm. Math. Phys. 291 (2009) 799--812.

\bibitem{Bradlow}
S.B.~Bradlow,
{\it Special metrics and stability for holomorphic bundles with global
  sections},
J. Diff. Geom. 33 (1991) 169--213.

\bibitem{BG-P}
S.B.~Bradlow, O.~Garc\'{\i}a-Prada,
{\it Stable triples, equivariant bundles and dimensional reduction},
Math. Ann. 304 (1996) 225--252.

\bibitem{BDG-PW}
S.B.~Bradlow, G.~Daskalopoulos, O.~Garc\'{\i}a-Prada, R.~Wentworth,
{\it Stable augmented bundles over Riemann surfaces},
London Math. Soc. Lect. Notes Ser. 208 (1995) 15--67.

\bibitem{BM93}
T. Brzezi\'nski, S. Majid, {\it Quantum group gauge theory on quantum
spaces}, Comm. Math. Phys. 157 (1993) 591--638 [Erratum \emph{ibid.}
167 (1995) 235].

\bibitem{BM97}
T. Brzezi\'nski, S. Majid, {\it Quantum differential and the
  $q$-monopole revisited}, Acta Appl. Math. 54 (1998) 185--233.

\bibitem{BS}
  B.P.~Dolan, R.J.~Szabo,
  {\it Dimensional reduction, monopoles and dynamical symmetry breaking},
  J. High Energy Phys. {0903} (2009) 059.

\bibitem{GP1} 
O.~Garc\'{\i}a-Prada,
{\it Dimensional reduction of stable bundles, vortices and stable
  pairs},
Int. J. Math. {5} (1994) 1--52.

\bibitem{H00}
P.M. Hajac, {\it Bundles over quantum sphere and noncommutative
index theorem}, K-Theory 21 (2000) 141--150.

\bibitem{HM99}
P.M. Hajac, S. Majid, 
{\it Projective module description of the q-monopole},
Commun. Math. Phys. 206 (1999) 247--264.

\bibitem{Harland}
  D.~Harland, S.~K\"urk\c{c}\"uo\v{g}lu,
  {\it Equivariant reduction of Yang--Mills theory over the fuzzy sphere and the
  emergent vortices},
  Nucl.\ Phys.\  B {821} (2009) 380--398.

\bibitem{KLvS} M. Khalkhali, G. Landi, W.D. van Suijlekom, {\it
    Holomorphic structures on the quantum projective line},
Int.  Math. Res. Not. 4 (2010) 851--884.  

\bibitem{KS97}
A.~Klimyk and K.~Schm{\"u}dgen, \emph{\sl Quantum Groups and their
  Representations} (Springer, 1997).

\bibitem{LRZ} G. Landi, C. Reina, A. Zampini, 
{\it Gauged laplacians on quantum Hopf bundles},  
Comm. Math. Phys. 287 (2009)  179--209.

\bibitem{LPS1}
  O.~Lechtenfeld, A.D.~Popov, R.J.~Szabo,
  {\it Rank two quiver gauge theory, graded connections and noncommutative
  vortices},
  J. High Energy Phys. {0609} (2006) 054.

\bibitem{LPS2}
  O.~Lechtenfeld, A.D.~Popov, R.J.~Szabo,
  {\it Quiver gauge theory and noncommutative vortices},
  Prog.\ Theor.\ Phys.\ Suppl.\  {171} (2007) 258--268.

\bibitem{maj05} S. Majid, {\it Noncommutative riemannian and spin
geometry of the standard $q$-sphere}, Comm. Math. Phys. 256 (2005)
255--285.

\bibitem{MNW} T. Masuda, Y. Nakagami, J. Watanabe,
\emph{Noncommutative differential geometry on the quantum two-sphere
  of P.~Podle{\'s}. I: An algebraic viewpoint}, K-Theory 5 (1991)
151--175.

\bibitem{maetal} T. Masuda, K. Mimachi, Y. Nakagami, M. Noumi,
  K. Ueno, {\it Representations of the quantum group $\SU$ and the
    little $q$-Jacobi polynomials}, J. Funct. Anal. 99 (1991)
  357--387.

\bibitem{Nakajima}
H.~Nakajima,
{\it Heisenberg algebra and Hilbert schemes of points on projective surfaces},
Ann. Math. 145 (1997) 379--388.

\bibitem{NS}
N.A.~Nekrasov, A.S.~Schwarz,
{\it Instantons on noncommutative $\IR^4$ and $(2,0)$ superconformal six-dimensional theory},
Comm. Math. Phys. 198 (1998) 689--703.

\bibitem{NT}
S. Neshveyev, L. Tuset, {\it A local index formula for the quantum
sphere}, Comm. Math. Phys. 254 (2005) 323--341.

\bibitem{Po87} P. Podle\'s, {\it Quantum spheres},
  Lett. Math. Phys. 14 (1987) 193--202.

\bibitem{pod89} P. Podle\'s, {\it Differential calculus on quantum
spheres}, Lett. Math. Phys. 18 (1989) 107--119.

\bibitem{Popov}
A.D.~Popov,
{\it Non-abelian vortices on Riemann surfaces: An integrable case},
Lett. Math. Phys. 84 (2008) 139--148.

\bibitem{PS}
A.D.~Popov, R.J.~Szabo,
  {\it Quiver gauge theory of non-abelian vortices and noncommutative instantons
  in higher dimensions},
  J.\ Math.\ Phys.\  {47} (2006) 012306.

\bibitem{SW04}
K. Schm\"udgen, E. Wagner, {\it Dirac operator and a twisted cyclic
cocycle on the standard Podle\'s quantum sphere}, J. Reine Angew.
Math. 574 (2004) 219--235.

\bibitem{SWPod}
K. Schm\"udgen, E. Wagner, {\it Representations of cross product
  algebras of Podle\'s quantum spheres}, 
J. Lie Theory 17 (2007) 751--790.

\bibitem{Tian}
G.~Tian,
{\it Gauge theory and calibrated geometry. I},
Ann. Math. 151 (2000) 193--268.

\bibitem{wa07} E. Wagner,
{\it On the noncommutative spin geometry of the standard Podle\'s
  sphere and index computations}, J. Geom. Phys. 59 (2009) 998--1016.  

\bibitem{wor87} S.L. Woronowicz, {\it Twisted $\SU$ group. An
example of a noncommutative differential calculus},
Publ. Rest. Inst. Math. Sci., Kyoto Univ. 23 (1987) 117--181.

\end{thebibliography}
\end{document}